\documentclass[%
 reprint,
superscriptaddress,
 amsmath,amssymb,
 aps,
]{revtex4-2}

\usepackage{graphicx}
\usepackage{dcolumn}
\usepackage{bm}
\usepackage{soul}

\usepackage[counterclockwise, figuresleft]{rotating}
\usepackage{multirow}
\usepackage{longtable}
\usepackage{enumitem}

\begin{document}

\preprint{APS/123-QED}

\title{Search for gravitational wave signals from known pulsars in LIGO-Virgo O3 data using the 5\textbf{\textit{n}}-vector ensemble method}

\author{Luca D'Onofrio} \email{ldonofrio@roma1.infn.it}
\affiliation{INFN, Sezione di Roma, I-00185 Roma, Italy} 
\author{Rosario De Rosa}
\affiliation{Università di Napoli “Federico II”, I-80126 Napoli, 
Italy}\affiliation{INFN, Sezione di Napoli, I-80126 Napoli, Italy} 
\author{Cristiano Palomba}
\affiliation{INFN, Sezione di Roma, I-00185 Roma, Italy}
\author{Paola Leaci}\affiliation{INFN, Sezione di Roma, I-00185 Roma, Italy}
\affiliation{Università di Roma "Sapienza", I-00185 Roma, Italy}
\author{Ornella J. Piccinni}\affiliation{Institut de Fisica d’Altes Energies (IFAE), The Barcelona Institute of Science and Technology, Campus UAB, 08193 Bellaterra (Barcelona) Spain}
\author{Valeria Sequino}\affiliation{Università di Napoli “Federico II”, I-80126 Napoli, 
Italy}\affiliation{INFN, Sezione di Napoli, I-80126 Napoli, Italy} 
\author{Luciano Errico}\affiliation{Università di Napoli “Federico II”, I-80126 Napoli, 
Italy}\affiliation{INFN, Sezione di Napoli, I-80126 Napoli, Italy} 
\author{Lucia Trozzo}
\affiliation{INFN, Sezione di Napoli, I-80126 Napoli, Italy} 
\author{Jim Palfreyman}\affiliation{University of Tasmania}
\author{James W. McKee}\affiliation{E.A. Milne Centre for Astrophysics, University of Hull, Cottingham Road, Kingston-upon-Hull, HU6 7RX, UK}\affiliation{Centre of Excellence for Data Science, Artificial Intelligence and Modelling (DAIM), University of Hull, Cottingham Road, Kingston-upon-Hull, HU6 7RX, UK}
\author{Bradley W. Meyers}\affiliation{International Centre for Radio Astronomy Research (ICRAR), Curtin University, Bentley WA 6102 Australia}
\author{Ingrid Stairs}\affiliation{Dept. of Physics and Astronomy, University of British Columbia, 6224 Agricultural Road, Vancouver, BC V6J 2B1 Canada}
\author{Lucas Guillemot}\affiliation{Laboratoire de Physique et Chimie de l’Environnement et de l’Espace, Universit\'e d’Orl\'eans/CNRS, F-45071 Orl\'eans Cedex 02, France}\affiliation{Observatoire Radioastronomique de Nan{\c c}ay, Observatoire de Paris, Universit{\'e} PSL, Universit{\'e} d’Orl{\'e}ans, CNRS, 18330 Nan{\c c}ay, France}
\author{Ismaël Cognard}\affiliation{Laboratoire de Physique et Chimie de l’Environnement et de l’Espace, Universit\'e d’Orl\'eans/CNRS, F-45071 Orl\'eans Cedex 02, France}\affiliation{Observatoire Radioastronomique de Nan{\c c}ay, Observatoire de Paris, Universit{\'e} PSL, Universit{\'e} d’Orl{\'e}ans, CNRS, 18330 Nan{\c c}ay, France}
\author{Gilles Theureau}\affiliation{Laboratoire de Physique et Chimie de l’Environnement et de l’Espace, Universit\'e d’Orl\'eans/CNRS, F-45071 Orl\'eans Cedex 02, France}\affiliation{Observatoire Radioastronomique de Nan{\c c}ay, Observatoire de Paris, Universit{\'e} PSL, Universit{\'e} d’Orl{\'e}ans, CNRS, 18330 Nan{\c c}ay, France}\affiliation{LUTH, Observatoire de Paris, PSL Research University, Meudon, France}
\author{Michael J. Keith}\affiliation{Jodrell Bank Centre for Astrophysics, Department of Physics and Astronomy, The University of Manchester, Manchester M13 9PL, UK}
\author{Andrew Lyne}\affiliation{Jodrell Bank Centre for Astrophysics, Department of Physics and Astronomy, The University of Manchester, Manchester M13 9PL, UK}
\author{Chris Flynn}\affiliation{OzGrav-Swinburne University of Technology, VIC 3166, Australia}
\author{Ben Stappers}\affiliation{Jodrell Bank Centre for Astrophysics, Department of Physics and Astronomy, The University of Manchester, Manchester M13 9PL, UK}

\date{\today}

\begin{abstract}
The 5\textit{n}-vector ensemble method is a multiple test for the targeted search of continuous gravitational waves from an ensemble of known pulsars. This method can improve the detection probability combining the results from individually undetectable pulsars if few signals are near the detection threshold. In this paper, we apply the 5\textit{n}-vector ensemble method to the O3 data set from the LIGO and Virgo detectors considering an ensemble of 201 known pulsars. We find no evidence for a signal from the ensemble and set a 95\% credible upper limit on the mean ellipticity assuming a common exponential distribution for the pulsars' ellipticities. Using two independent hierarchical Bayesian procedures, we find upper limits of $1.2 \times 10^{-9}$  and $2.5 \times 10^{-9}$  on the mean ellipticity for the 201 analyzed pulsars.
\end{abstract}
\maketitle

\section{Introduction}
Continuous gravitational waves (CWs) signals are promising targets for the LIGO \cite{LIGO} and Virgo \cite{virgo} detectors defined by persistence and near-monochromaticity over long time scales. Among the most plausible sources of this type of signal there are spinning neutron stars with a non-axisymmetric mass distribution \cite{1era} due to deformations from shear strains in the crust, or from magnetic stresses \cite{review}. 

The signal strain amplitude depends on the ellipticity, the physical parameter that quantifies the mass distribution asymmetry with respect to the rotation axis \cite{JKS}. Different neutron star equations of state can allow different sizes of deformations, i.e. of ellipticities, to be sustained in the star. The detection of a CW signal could allow us to make physical inference about the equation of state, and/or about the internal magnetic field strength \cite{cwemission}.

According to the knowledge of the source parameters (sky position, rotation parameters), different strategies can be adopted to search for a CW signal \cite{ornella}. The simplest, but also the most sensitive strategy, is the \textit{targeted} search \cite{firstsearch}: using information about the spin evolution, astrometry, and binary orbit gained from continued pulsar monitoring, it is possible to infer with high accuracy the sky position and the rotation parameters including the source spin frequency and the spin-down. Indeed, most pulsars are observed to steadily spin-down due to the emission of electromagnetic waves (and, possibly, also CWs). From the pulsar spin-down, it is possible to set a theoretical upper limit to the strain amplitude, the so-called spin-down limit, assuming that all of the rotation energy lost by the pulsar is converted to gravitational wave (GW) energy \cite{O3targ}.

At the source reference frame, the CW signal is almost monochromatic with frequency proportional to the rotation frequency of the source, through a factor which depends on the considered emission model \cite{review}. At the detector, the CW signal has a frequency and phase modulation mainly due to the Doppler effect for the relative motion between the source and the Earth. 

De-modulation techniques are useful for removing the phase modulation caused by the spin-down/Doppler shift and hence to precisely  unwind  the apparent  phase  evolution  of the source.
Once these corrections have been applied, a CW signal would become monochromatic except for the sidereal modulation due to the antenna patterns. This modulation depends on the detector position and orientation on the Earth and it is used by the \textit{5n-vector method} to define a detection statistic as well as to estimate the signal parameters \cite{2010,2014}. 

So far, no evidence for a CW signal has been found in LIGO and Virgo data. The last targeted search from the LIGO-Virgo-Kagra (LVK) Collaboration \cite{O3targ}, using the data from the observing run O3 and analyzing 236 known pulsars, set 95\% credible upper limits on the amplitude and the ellipticity of each pulsar. 

To improve the detection probability for the targeted search of CWs, different strategies can be used considering an ensemble of individually undetectable pulsars as in \cite{ensbayes,Fstat2,mio}. In this paper, we use \textit{the 5n-vector ensemble method}, described in \cite{mio2}. This method defines an ensemble statistic, $T(k)$, as the partial sum of the statistics of single pulsar, ranked for increasing $p$-values. 

The ensemble procedure is a rank truncation method that selects the top ranking sources according to the significance of the corresponding individual test.  Using the statistics $T(k)$, we define a $p$-value as a function of $k$, that is a $p$-value for the overall hypothesis of the presence of CWs from the ensemble.

In case of no detection, we can set 95\% credible upper limit on the global parameter $\Lambda$ (defined in \cite{mio2}) that describes the ensemble, and on the mean ellipticity that returns information about the population properties using two independent hierarchical procedures.

The paper is organized as follows. In Sec.~\ref{sec:intro}, we review the 5\textit{n}-vector ensemble method and describe the procedures to set upper limits in case of no detection from the ensemble. In Sec.~\ref{sec:dataset}, we describe the data set used for the analysis while in Sec.~\ref{sec:res_sing}, we show the results of the targeted search for single pulsar using the 5\textit{n}-vector method with the first application to pulsars in binary system. In Sec.~\ref{sec:res_ens}, we report the results of the ensemble procedure on three different sets. Since there is no evidence of a signal from the ensemble, we set the upper limits, that are compared with the results of previous similar searches. Conclusions are presented in Sec.~\ref{sec:end}. 

\section{The 5\textit{n}-vector ensemble method}\label{sec:intro}
The 5\textit{n}-vector ensemble method \cite{mio2} is a multiple test that combines the effects of individually undetectable pulsars using the results of the single pulsar analysis through the 5\textit{n}-vector method.

The 5\textit{n}-vector method \cite{2014} is a frequentist pipeline used in the LVK Collaboration for CW searches. It is based on the splitting at the detector of the expected GW frequency $f_{\text{gw}}$ in five components due to the sidereal modulation of the Earth. The expected signal $h(t)$ can be written using the polarization ellipse \cite{2010}:
\begin{equation}\label{compsign}
h(t)=H_0(H_+ A_+(t) + H_\times A_\times(t))e^{i\Phi(t)}
\end{equation}
where $H_{+/\times}$ are the polarization functions for the plus and cross components:
\begin{equation}
H_+=\frac{\cos(2\psi)-j\eta \sin(2\psi)}{\sqrt{1+\eta^2}}   
\end{equation}
\begin{equation}
 H_\times=\frac{\sin(2\psi)+j\eta \cos(2\psi)}{\sqrt{1+\eta^2}} \,.
\end{equation} $\eta$ and $\psi$ are the polarization parameters \cite{2010}, $A_{+/\times}$ are linked to the antenna patterns while $\Phi(t)$ is the GW phase. The parameter $\eta$ is defined as:
\begin{equation}\label{iota}
    \eta=-\frac{2\cos(\iota)}{1+\cos^2(\iota)}
\end{equation}
where $\iota$ is the angle between the star
rotation axis and the line of sight. The amplitude $H_0$ is linked to the classic amplitude $h_0$, defined for example in \cite{O3targ}, by:
\begin{equation}
 H_0=h_0\sqrt{\frac{1+6\cos^2(\iota)+\cos^4(\iota)}{4}} \,.
\end{equation}
 
 The phase $\Phi(t)$ in Eq.~\ref{compsign} shows a time dependence due to different phenomena that modulate in time the received signal frequency. Heterodine corrections \cite{2019} are applied to account for the pulsar spin-down and  Doppler effect due to the Earth motion. The residual modulation is in the functions $A_{+/\times}$, i.e. in the detector response, that depend on the 5 frequencies $f_{\text{gw}},f_{\text{gw}} \pm \Omega_\oplus,f_{\text{gw}} \pm 2 \Omega_\oplus $ where $\Omega_\oplus$ is the Earth's sidereal angular frequency. 

The data 5-vector $\textbf{X}$ and the signal template 5-vectors  $\textbf{A}^{+/\times}$ are defined as the Fourier transforms of the data and of the template functions $A_{+/\times}$ at the 5 frequencies where the signal power is split. The data/template 5-vectors from each of the $n$ considered detectors are then combined as described in Appendix \ref{app:5n} for a multidetector analysis. 

The 5\textit{n}-vector defines two matched filters between the data $\textbf{X}$ and the signal templates $\textbf{A}^{+/\times}$ vectors, used in order to maximize the signal-to-noise ratio:
\begin{equation}\label{estim}
\hat{H}_+=\frac{\textbf{X}\cdot \textbf{A}^+}{|\textbf{A}^+|^2} \qquad \text{and} \qquad \hat{H}_\times=\frac{\textbf{X}\cdot \textbf{A}^\times}{|\textbf{A}^\times|^2}
\end{equation} The two matched filters are estimators \cite{2010} of the signal plus and cross amplitudes.
They are used to construct the detection statistic $S$:
\begin{equation}\label{statS}\small
S= \frac{|\textbf{A}^{+}|^4 }{\sum_{j=1}^n \sigma_j^2 \, T_{j} \, |\textbf{A}^{+}_{j}|^2} |\hat{H}_+|^2 +  \frac{|\textbf{A}^{\times}|^4 }{\sum_{k=1}^n \sigma_k^2 \, T_{k} \, |\textbf{A}^{\times}_{k}|^2} |\hat{H}_\times|^2 \,.
\end{equation}
where $\sigma_j^2$ and $T_j$ are the variance of the data distribution in a frequency band around $f_{gw}$ (usually few tenths of Hz wide) and the observation time\footnote{\textit{Observation time} refers to the amount of data; that is, the time span not considering the periods where the detector was in NO-science mode} in the $j^{\text{th}}$ detector, respectively. In the case of Gaussian noise, the distribution of the statistic $S$ in Eq.~\eqref{statS} is a Gamma distribution $\Gamma(x;2,1)$.

Fixing the number $N$ of analyzed pulsars, the 5\textit{n}-vector ensemble method ranks these pulsars according to the $p$-values $P_i$ (with $i=1,...,N$),
\begin{equation}
P_i=P(S_i \ge \overline{S}_i|\text{noise})
\end{equation} computed from the $S_i$ noise distribution and the measured value $\overline{S}_i$ for the $i^{\text{th}}$ pulsar.
Then, the ensemble statistic $T(k)$ is defined as the partial sum of the first $k$ largest order statistics (statistics with the $k$ smallest p-values):
\begin{equation}
    T(k)=\sum_{j=N-k+1}^N S_{(j)} \,.
\end{equation}

From the statistics $T(k)$, we can compute the $p$-value of ensemble $P(k)$ as a function of $k$:
\begin{equation}\label{ptk}
    P(k)=P(T(k) \ge \overline{T}(k)|\text{noise}) 
\end{equation}with 
\begin{equation}
     \overline{T}(k) =\sum_{j=N-k+1}^N \overline{S}_{(j)} \,,
\end{equation} and $\overline{S}_{(j)}$ is the $j^{\text{th}}$ largest measured value for the detection statistic for a certain pulsar in the ensemble. For example, for an ensemble of $N$ pulsars with $P_{j}\le P_i\,\, \forall i$ and $P_{k} \ge P_i \,\, \forall i$, the smallest and the largest order statistics values are $\overline{S}_{(1)}\equiv \overline{S}_{k}$ and $\overline{S}_{(N)} \equiv \overline{S}_{j}$, respectively.

We note that the  p-value  refers  only  to  the null  hypothesis  and  does  not  make  reference  to  or  allow conclusions about any other hypotheses, such as the alternative hypothesis.
A small p-value (for example, below the 1\%) can be used to recognize interesting ensemble of pulsars as starting point for the detection.

The noise and the signal distribution of the statistics $T(k)$ can be inferred using a Monte Carlo procedure starting from the $N$ experimental noise and signal distributions for the statistics of single pulsar, as described in \cite{mio2}. 

The $T(N)$ statistic coincides with the simple sum of the statistics $S_i$; in case of Gaussian noise, $T(N)$ follows a $\Gamma(x;2N,1)$ distribution. The $T(N)$ signal distribution can be also expressed in closed-form and it is proportional to a non-central $\chi^2$ distribution with $4N$ degrees of freedom and non-centrality parameter $\Lambda$ \cite{mio2}:
\begin{equation}\label{Lambda}
    \Lambda=\sum_{i=1}^N H_{0,i}^2\, f_i(\eta_i,\psi_i)
\end{equation}
where the coefficient $f_i$ is:
\begin{equation}\small{
    f_i(\psi_i,\eta_i) = 2 \left( \frac{|\textbf{A}^{+}_i|^4\, |H_{+,i}|^2}{\sum_{j=1}^n \sigma_j^2 T_{j} |\textbf{A}^{+}_{j,i}|^2}+\frac{|\textbf{A}^{\times}_i|^4\, |H_{\times,i}|^2}{\sum_{k=1}^n \sigma_k^2 T_{k} |\textbf{A}^{\times}_{k,i}|^2} \right)} \,.
\end{equation}

Even though we have defined a $p$-value of ensemble as a function of $k$, the statistical inference concerns the entire set of considered pulsars. The claim of a rank truncation method is the same of the Fisher's combination method where all the individual tests are combined together: i.e. there are some signals among all the $N$ tests, regardless that a certain $k<N$ led to a rejection.

\subsection{Upper limit procedures}
For the single pulsar analysis, we can set the 95\% credible upper limits on the amplitude assuming uninformative uniform priors on $\psi$ (with $-\pi/2\leq \psi \leq \pi/2$) and $\cos(\iota)$ (with $-1\leq \cos(\iota)\leq 1$) that defines $\eta$ (see Eq.~\ref{iota}). Knowing the distance and assuming a fiducial moment of inertia, we set the 95\% credible upper limits on the ellipticity.

The upper limit for the ensemble procedure must be set on population parameters since using a multiple test, we can not infer statistical information about the single pulsar parameters. 
\\Considering the statistic $T(N)$, i.e. the entire set of considered pulsars, two different approaches can be used. 

First, we can constrain the parameter $\Lambda$, defined in Eq.~\eqref{Lambda}, that fixes the $T(N)$ distribution considering a mixed Bayesian-frequentist approach as in the case of single pulsar \cite{targO2}. 
\\ For the Bayes' theorem, the posterior on $\Lambda$ is:
\begin{equation}
    P(\Lambda|\overline{T}(N))\varpropto L(\overline{T}(N)|\Lambda)\Pi(\Lambda)
\end{equation}where $\overline{T}(N)$ is the measured value of the ensemble statistic, $\Pi(\Lambda)$ is the prior while $L(\overline{T}(N)|\Lambda)$ is the likelihood, that is estimated from the theoretical $T(N)$ signal distribution evaluated at the value $\overline{T}(N)$ for different $\Lambda$ values.

The upper limit $\Lambda^{95\%}$ on the parameter $\Lambda$ is:
\begin{equation}
    \Lambda^{95\%}\, : \, \int_0^{\Lambda^{95\%}} P(\Lambda|\overline{T}(N)) d \Lambda=0.95 \,.
\end{equation}$\Lambda$ is a global parameter for the ensemble and $\Lambda^{95\%}$ can be used to show the improvement in the method sensitivity analyzing future runs.

Secondly, assuming a common distribution for the ellipticities, we can use a hierarchical Bayesian framework to constrain the value of the hyperparameter that fixes the assumed common distribution. Indeed, hierarchical Bayesian inference allows to study population properties for the analyzed ensemble of pulsars \cite{hierarc}.

Let us consider an ensemble of $N$ pulsars and a common distribution fixed by the hyperparameter $\mu_\epsilon$ for the set of ellipticities ${\epsilon_i}$ (with $i=1,..,N$). 

We can constrain the value of the hyperparameter $\mu_\epsilon$ considering the posterior probability density function inferred from a Bayesian procedure:
\begin{equation}\label{mulam}
    P(\mu_\epsilon|\overline{T}(N)) \varpropto L(\overline{T}(N)|\mu_\epsilon) \Pi(\mu_\epsilon)
\end{equation}
where $\Pi(\mu_\epsilon)$ is the prior distribution on the hyperparameter, while the likelihood $L(\overline{T}(N)|\mu_\epsilon)$ can be defined as the value of the $T(N)$ signal distribution evaluated at $\overline{T}(N)$, if the common distribution of the ellipticities is fixed by the hyperparameter $\mu_\epsilon$.  The $T(N)$ signal distribution and hence, the likelihood, is fixed by $\Lambda$ in case of Gaussian noise. It follows that:
\begin{equation}
    L(\overline{T}(N)|\mu_\epsilon)=\int L(\overline{T}(N)|\Lambda)\Pi(\Lambda|\mu_\epsilon) d \Lambda \,.
\end{equation}

From the definition in Eq.~\eqref{Lambda}, deducing the analytical prior on $\Lambda$ for an assumed distribution of the ellipticities with hyperparameter $\mu_\epsilon$ is not straightforward. For a large ensemble, $N \gg 1$, $\Pi(\Lambda|\mu_\epsilon)$ can be approximated with a Gaussian distribution according to the Central Limit Theorem. The mean and variance of the Gaussian distribution depend on the assumed distribution since it is the weighted sum of the mean and variance for the squared $\epsilon_i$ random variables.

The upper limit $\mu^{(\Lambda)}$ is:
\begin{equation}
 \mu^{(\Lambda)} \, : \, \int_0^{\mu^{(\Lambda)}} P(\mu_\epsilon|\overline{T}(N)) d \mu_\epsilon=0.95   \,.
\end{equation}
Differently, we can also consider a classic hierarchical procedure combining the results of the single pulsar analysis as in \cite{ensbayes}. In this case, the posterior probability density function on the hyperparameter $\mu_\epsilon$ is:
\begin{equation}\label{uleps}
P(\mu_\epsilon|\overline{S}_1,..,\overline{S}_N) \varpropto \left( \prod_{i=1}^N \int L(\overline{S}_i|\epsilon_i) \Pi(\epsilon_i|\mu_\epsilon) d \epsilon_i \right) \Pi(\mu_\epsilon) \,.
\end{equation}

This posterior is independent of the ensemble procedure (i.e. of the statistic $T(N)$) and depends on the results of the single pulsar analysis through the likelihoods $L(\overline{S}_i|\epsilon_i)$. In addition, it does not depend on the Gaussian noise hypothesis and it can be used to constrain the hyperparameter $\mu_\epsilon$ using the likelihood $L(\overline{S}_i|\epsilon_i)$ computed for the $i^{\text{th}}$ pulsar analysis.
\\The upper limit $\mu^{(S)}$ is
\begin{equation}
 \mu^{(S)} \, : \, \int_0^{\mu^{(S)}} P(\mu_\epsilon|\overline{S}_1,..,\overline{S}_N) d \mu_\epsilon=0.95   \,.
\end{equation}

\section{Data set}\label{sec:dataset}
In this section, we describe the data set used for the analysis. It consists of the GW data set from the LIGO and Virgo detectors, and the electromagnetic data set from the different observatories used to determine the pulsar ephemerides.
\subsection{Gravitational wave data}
The O3 observing run started on 2019 April 1 (MJD: 58574.625) and ended on 2020  March  27  (MJD:  58935.708) for both the LIGO and Virgo detectors. O3  consisted  of  two   parts,  separated  by  a one month-long  commissioning  break.   O3a  ran  from  2019 April 1,  15:00 UTC until 2019 October 1,  15:00 UTC. O3b  ran  from  2019  November  1,  15:00  UTC,  to  2020 March  27,  17:00  UTC.   

The duty cycles for this run were 76\%, 71\%, and 76\% for LIGO Livingston (LLO), LIGO Hanford (LHO), and Virgo, respectively. The O3 data set is publicly available via the Gravitational Wave Open Science Center \cite{gwdata}.

In this paper, we use the Band Sampled Data (BSD) framework \cite{2019} that consists of band-limited, down-sampled time series of the detector calibrated data, called BSD files. This framework can be described as a database of sub-databases where the BSD file is a complex time series that covers 10$\,$Hz and spans 1 month of the original data. Using the BSD libraries, it is possible to extract frequency sub-bands or time sub-periods in a flexible way, reducing the computational cost of the analysis.
\subsection{Electromagnetic data}
The set of analyzed pulsars is in Table \ref{tab:long}. We consider the ephemerides used in \cite{O3targ}.  

The  observatories  which  have  contributed  to the data set \footnote{The ephemerides and the rotational/orbital parameters may be obtainable on request at the discretion of the individual EM pulsar groups from the above-mentioned observatories.} are: the Canadian Hydrogen Intensity Mapping Experiment (CHIME) \cite{2021ApJS..255....5A}, the Mount Pleasant Observatory 26-m telescope, the 42-ft telescope and Lovell telescope at Jodrell Bank, MeerKAT \cite{MeerTime}, the Nan\c{c}ay Decimetric Radio Telescope, the Neutron Star Interior Composition Explorer (NICER) and the Molonglo Observatory Synthesis Telescope (as part of the UTMOST pulsar timing program) \cite{2019MNRAS.484.3691J, 2020MNRAS.494..228L}.
Pulsar timing solutions were determined from this data using {\sc Tempo} \cite{2015ascl.soft09002N} or {\sc Tempo2} \cite{tempo22, tempo21, tempo23} to fit the model parameters. The ephemerides have been created using pulse time-of-arrival observations that mainly overlapped with O3 observing period.

To estimate the theoretical spin-down limit or the upper limit for the fiducial ellipticity for a certain source, the pulsars' distances are needed \cite{O3targ}. For many pulsars, the distance can be found in the ATNF Pulsar Catalog \cite{atnf}. These are distances mostly based on the observed dispersion measure and estimated using the Galactic electron density distribution model YMW16 \cite{pulsarYao} or the NE2001 model \cite{NE2001}, although others are  based  on  parallax  measurements,  or  inferred  from associations with other objects or flux measurements. 

The distances used for each pulsar are given in Table \ref{tab:long}. The distances based on the dispersion measure are estimates with uncertainties that can amount up to a factor of two \cite{pulsarYao}.  
Nearby pulsars, for which parallax measurements are possible, have smaller distance uncertainties.

We select pulsars whose rotation frequency is greater than  10$\,$Hz  to match  the  sensitivity  band  of the  GW  detectors.   This  leads  to  primarily  targeting millisecond pulsars and fast spinning young pulsars.  In the analyzed set, there are 165 millisecond pulsars with expected CW frequencies above 100$\,$Hz.  

Pulsars can experience glitches during the observation time (as for the Crab pulsar during O3 \cite{O3targ}), i.e. sudden, impulsive increases in their spin \cite{glitch}.  The general procedure for glitching pulsars is to analyze each inter-glitch period independently and then to sum the resulting statistics \cite{O3targ}. In this analysis, we have not considered glitching pulsars; a dedicated study is needed to optimize the detection probability when glitching pulsars are present in the ensemble. 
\begin{figure}[t] 
\centering
\includegraphics[clip,trim=10 0 0 10,scale=0.61]{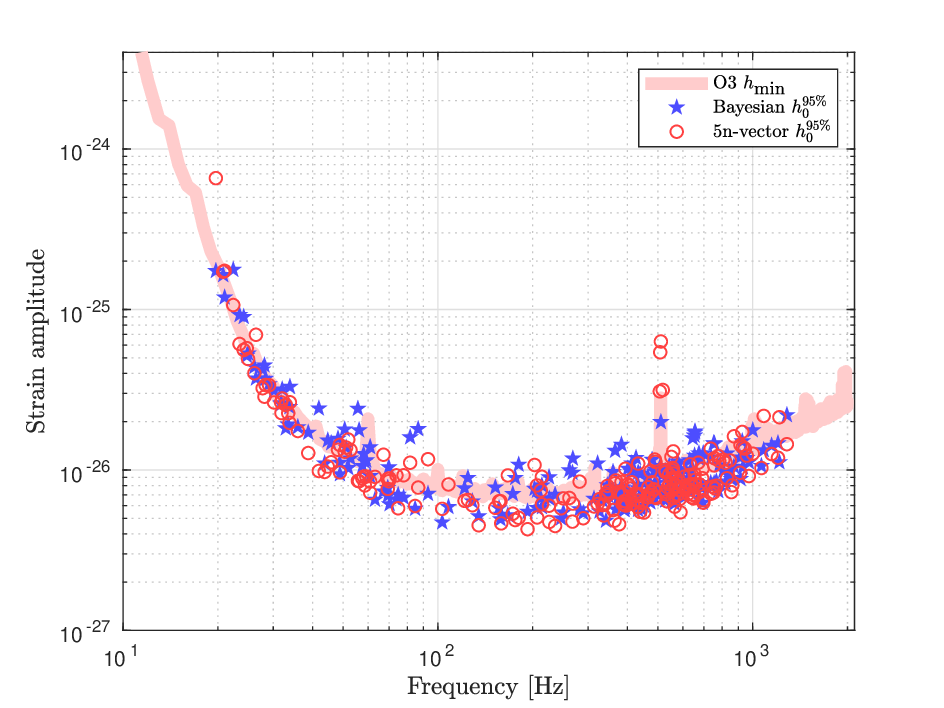}
\caption{\small{Strain amplitude as a function of the GW frequency for O3 data of the LIGO and Virgo detectors. Blue stars are the upper limits inferred in \cite{O3targ} using the Bayesian method for the 223 pulsars in Table \ref{tab:long} while the red circles are the upper limits for the 5n-vector method. The continuous pink line is the minimum detectable amplitude for the O3 run considering a multidetector analysis.}}
\label{fig:res_singlep}
\end{figure}
\section{Single pulsar analysis}\label{sec:res_sing}
In this section, we describe the results of the analysis of single pulsars using the 5\textit{n}-vector method. Differently from the analysis in \cite{O3targ}, we use the normalized single pulsar statistic $S$ defined in Eq.~\eqref{statS} using the weighted multidetector extension described in Appendix \ref{app:5n}. For the first time using the 5\textit{n}-vectors, we also consider pulsars in binary systems.

The results, reported in Table \ref{tab:long}, are obtained considering a single harmonic emission, i.e. the expected GW frequency is assumed exactly twice the rotation frequency of the source, and analyzing O3 data from the two LIGO and Virgo detectors.

For each pulsar, we consider a small frequency band (at least 0.2$\,$Hz and at most 0.5$\,$Hz wide) around the expected GW frequency. The frequency band extracted from the BSD files is chosen considering the spin-down of the source and in case of binary systems, considering also the Doppler effect due to the orbital motion. 

Using the heterodyne correction \cite{2019}, we remove the spin-down and the Doppler frequency modulation for the expected GW signal. The Doppler correction for pulsars in binary systems follows the model described in \cite{deltabin}. We do not take into account the uncertainty estimates in the binary parameters; the possible effects of these uncertainties on the heterodyne correction is described in the analysis in Appendix \ref{app:bin_par}.

The experimental noise distribution is inferred considering a range of off-source frequencies as described in \cite{2014}. The measured value of the statistics is compared with the noise distribution computing the $p$-value. Using the BSD framework, the computational cost of the analysis is reduced to a few CPU-minutes per source per detector.

In agreement with \cite{O3targ}, there is no evidence of a CW signal from any pulsar in O3 data. We set 95\% credible upper limit on the amplitude $h_0$ for each pulsar. 

For the 5\textit{n}-vector method since we are considering a different statistic, the results do not exactly coincide with the results in \cite{O3targ}  but there is a very good agreement. For example, the upper limit on the amplitude for the pulsar J0711-6830 in \cite{O3targ} is $5.0\times 10^{-27}$, while $4.8\times 10^{-27}$ in this work.

As shown in Fig.~\ref{fig:res_singlep}, the upper limits from the 5\textit{n}-vector method (red circles) are in good agreement with the results of the Bayesian method (blue stars as in \cite{O3targ}). The differences in the results can be  due to the data frameworks that are used in the two pipelines that entail different analyzed frequency bands for each pulsar and different pre-processing of the data.
\begin{figure*}[t]
\includegraphics[clip,trim=100 0 50 10,scale=0.43]{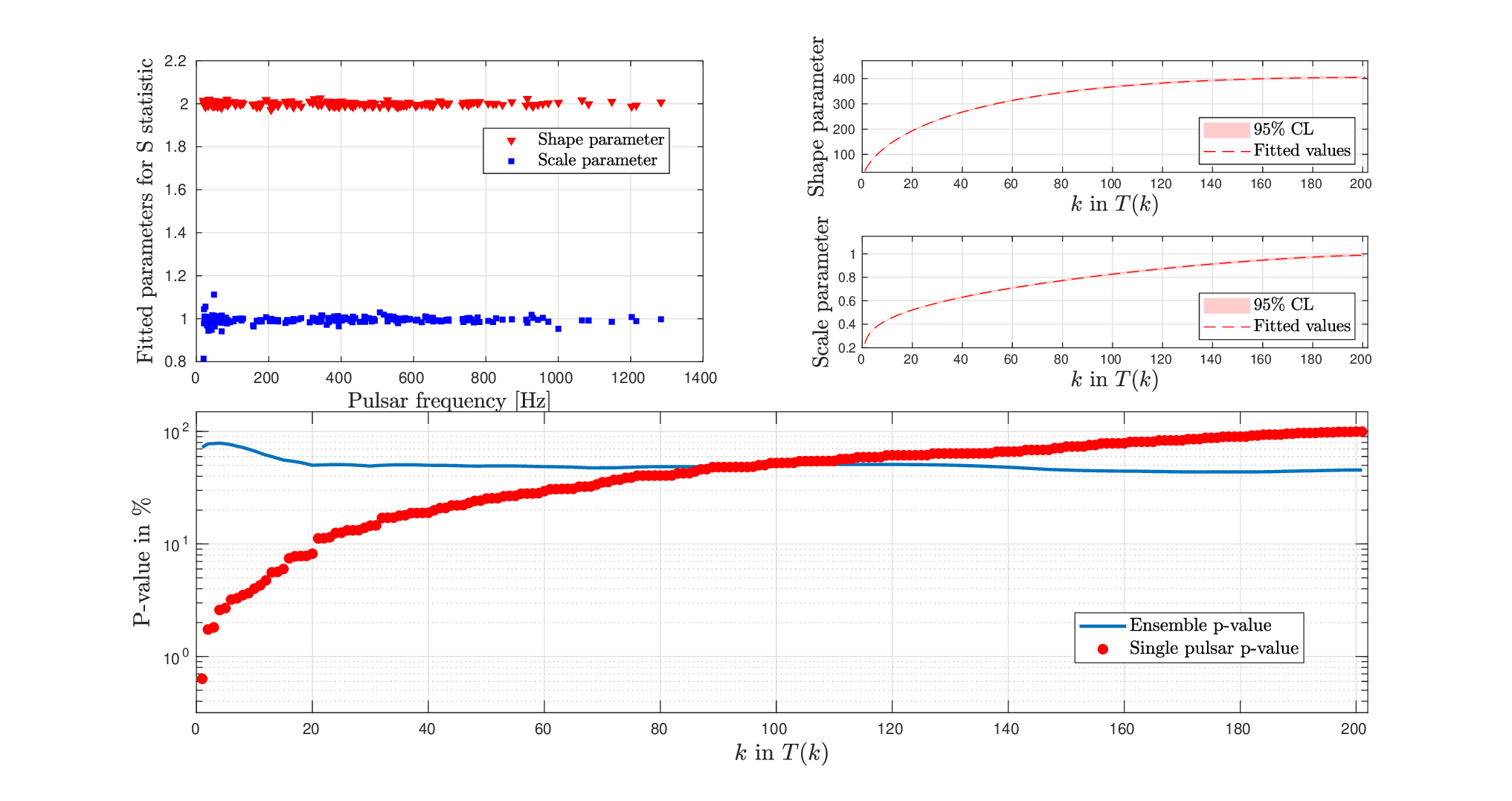}
\caption{\small{Summary plot of the results for the ensemble over 201 pulsars considering O3 data of LLO, LHO and Virgo detectors. \textit{Top-left plot} Fitted shape (red triangles) and scale parameters (blue squares) using a Gamma distribution to the experimental noise distribution for each pulsar as a function of the emission frequency of the pulsar. \textit{Top-right plots} Fitted shape and scale parameters (dashed lines) with $95\%$ confidence level (coloured area) for the $T(k)$ noise distributions for each $k$ inferred from the Monte Carlo procedure. \textit{Bottom plot} $p$-value of ensemble (blue line)  compared with the single pulsar $p$-values (red dots), ranked for increasing values.}}\label{fig:res_all}
\end{figure*}
\section{Ensemble analysis} \label{sec:res_ens}
In this section, we show the results of the 5\textit{n}-vector ensemble method  applied to O3 data. From the set of pulsars in Table \ref{tab:long}, we select 201 pulsars excluding 22 binary pulsars according to the selection criteria in Eq.~\ref{thr_bin} described in Appendix \ref{app:bin_par} based on the uncertainties of the binary parameters.

As described in \cite{mio2}, the improvement in the detection probability for the 5\textit{n}-vector ensemble method depends on the number of analyzed pulsars $N$ and on the power of the combined individual tests. In a real search, we can not optimize $N$ since we do not know the strength of the signals a priori.

We fixed three different ensembles from the set in Table \ref{tab:long}: the entire set for the three detectors, the set of millisecond pulsars ($f_{\text{gw}}>100$$\,$Hz) for the LIGO detectors and the high-value pulsars set for the LIGO detectors. High-value pulsars are the analyzed pulsars whose ratio between the experimental upper limit on the amplitude $h_0^{95\%}$ and the theoretical spin-down limit $h_{\text{sd}}$  (the so-called spin-down ratio) is:
\begin{equation}
    h_0^{95\%}/h_{\text{sd}}\lesssim 1 \,.
\end{equation}

We chose the three ensembles according to physical observations. Indeed, millisecond pulsars are most likely to have undergone an accretion phase, which could alter the structure of their crust and magnetic field compared to non-recycled young pulsars \cite{ppdot}, while high-value pulsars are promising targets that have surpassed their spin-down limits.

The three ensembles correspond to fix $N=201$ for the first set, $N=146$ for the millisecond pulsars set and $N=20$ for the high-value pulsars set.

After analyzing each pulsar singularly, we reconstruct the statistics $T(k)$ and compute a $p$-value of ensemble as a function of $k$. The results are shown in the form of summary plot; the top-left plot shows the fitted shape and scale parameters of a Gamma distribution for the experimental noise distribution for the statistic $S$ as a function of the GW frequency, the top-right plots show the fitted shape and scale parameters to the $T(k)$ noise distributions as a function of $k$ while the bottom-plot compares the ordered single pulsar $p$-values (red dots) with the ensemble $p$-values from the $T(k)$ statistics (blue line).

Since there is no evidence of signal from the ensemble, we set 95\% credible upper limit on the global parameter $\Lambda$ and on the hyperparameter $\mu_\epsilon$, that is the mean value of the assumed exponential distribution for the ellipticities.
\begin{figure*}[t]
\includegraphics[clip,trim=100 0 50 10,scale=0.43]{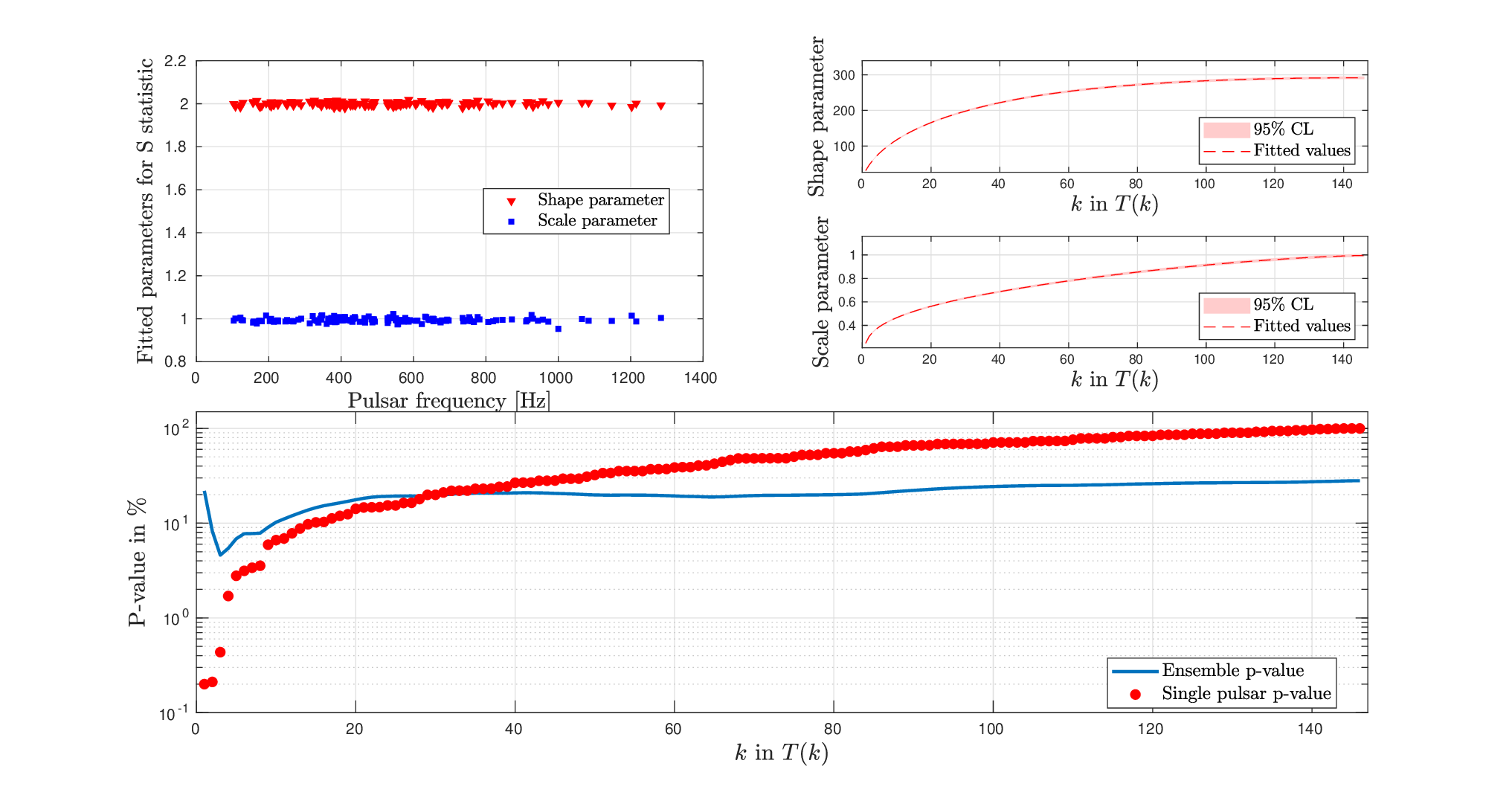}
\caption{\small{Summary plot of the ensemble composed by the millisecond pulsars in Table \ref{tab:long}, considering O3 data and the two LIGO detectors. \textit{Top-left plot} Fitted shape (red triangles) and scale parameters (blue squares) using a Gamma distribution to the experimental noise distribution for each pulsar as a function of the emission frequency for the considered pulsar. \textit{Top-right plots} Fitted shape and scale parameters (dashed lines) with $95\%$ confidence level (coloured area) for the $T(k)$ noise distributions for each $k$ inferred from the Monte Carlo procedure. \textit{Bottom plot} $p$-value of ensemble (blue line)  compared with the single pulsar $p$-value (red dots), ranked for increasing values.}}\label{fig:res_MS}
\end{figure*}
\subsection{All pulsars, all detectors}
The summary plot for the entire set of analyzed pulsars is shown in Fig.~\ref{fig:res_all}. For this analysis, we consider the two LIGO and Virgo detectors. 

A criterion based on noise is used to select the detectors for the multi-detector analysis of single pulsar. Since the normalized statistic $S$, in the hypothesis of Gaussian noise, is distributed according to a $\Gamma(x;2,1)$, from the fitted parameters to the experimental noise $S$ distribution (top-left plot in Fig.~\ref{fig:res_all}) we can easily check the Gaussianity in the selected frequency band for each pulsar. For the multi-detector analysis, we consider only the detectors that have fitted shape and scale parameters respectively in the range $[1.6,2.4]$ and $[0.8,1.2]$ (that is a maximum difference of $20\%$ from the expected values for Gaussian noise). For each pulsar, this procedure allows to exclude the detectors with large noise disturbances in the analyzed frequency band.  Indeed, experimental noise $S$ distributions that are very different from the $\Gamma(x;2,1)$ can influence the Monte Carlo procedure for the $T(k)$ noise distributions.

From the bottom-plot of Fig.~\ref{fig:res_all}, there is no evidence of a signal form the ensemble. One pulsar has single pulsar $p$-value (red dot) below $1\%$ as expected for a set of 201 pulsars. The $p$-value computed for the statistics $T(k)$ are consistent with the noise hypothesis.
\subsection{Millisecond pulsars, LIGO detectors}
The summary plot for the millisecond pulsars is shown in Fig.~\ref{fig:res_MS}. For this analysis, we consider only the two LIGO detectors with the same noise criteria introduced in the previous subsection, not considering the Virgo detector that has a higher noise level. As shown in Appendix \ref{app:5n}, a multidetector analysis does not necessarily entail a better sensitivity compared to the most sensitive detector's analysis.

As shown in the bottom plot in Fig.~\ref{fig:res_MS}, there are \textbf{three} pulsars -  J0824+0028, J1544+4937, J1551$-$0658 - below the $1\%$ threshold. There is no  evidence of a signal for any of these pulsars (as for the Bayesian results in \cite{O3targ}); the combination of the two LIGO detectors produces lower $p$-values with respect to the three detectors case. 

For example considering the pulsar J0824+0028, the $p$-values for the single-detector single-pulsar analysis are 0.01, 0.04, 0.37  for LLO, LHO and Virgo respectively. The multidetector analysis considering all the three detectors entails a $p$-value of 0.01 while considering only the two LIGO detectors, the $p$-value is 0.002. 

The ensemble $p$-value has a minimum of almost $3\%$ for $k=4$ with no evidence of a signal from the ensemble. The strong decrease for the first values of $k$ is due to the presence of the three pulsars with $p$-values below 1\%. 
\begin{figure*} [t]
\includegraphics[clip,trim=100 0 50 10,scale=0.43]{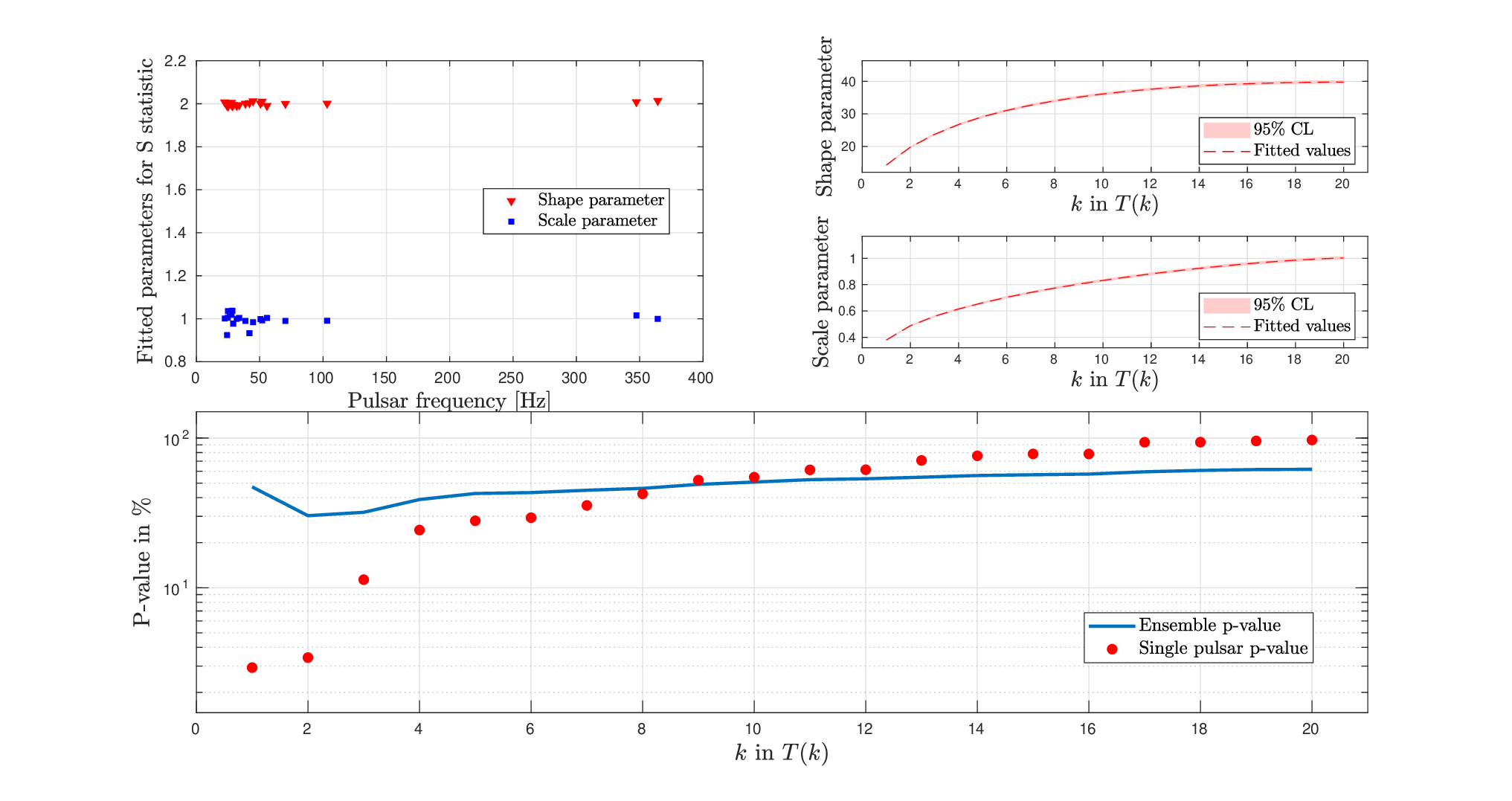}
\caption{\small{Summary plot of the ensemble composed by the high-value pulsars with spin-down ratio $\lesssim 1$ considering O3 data and the two LIGO detectors. \textit{Top-left plot} Fitted shape (red triangles) and scale parameters (blue squares) using a Gamma distribution to the experimental noise distribution for each pulsar as a function of the frequency for the considered pulsar. \textit{Top-right plots} Fitted shape and scale parameters (dashed lines) with $95\%$ confidence level (coloured area) for the $T(k)$ noise distributions for each $k$ inferred from the Monte Carlo procedure. \textit{Bottom plot} $p$-value of ensemble (blue line)  compared with the single pulsar $p$-value (red dots), ranked for increasing values.}}\label{fig:res_HV}
\end{figure*}
\subsection{High-value pulsars, LIGO detectors}
There are 20 high-value pulsars in our dataset, that is there are 20 pulsars with spin-down ratio $\lesssim 1$ (as in \cite{O3targ}, excluding the glitching pulsars). 

The summary plot for the high-value pulsars is shown in Fig.~\ref{fig:res_HV}. For this analysis, we consider only the two LIGO detectors with the same noise criteria introduced in the previous subsection.

As shown by the the blue line in the bottom plot of Fig.~\ref{fig:res_HV}, the $p$-value of the ensemble is fully consistent with the noise hypothesis.

\subsection{Upper limits}
For the upper limit computation, we consider the set of known pulsars analyzed in Fig.~\ref{fig:res_all} and the O3 dataset for the LIGO and Virgo detectors.

\begin{figure}[t] 
\centering
\includegraphics[clip,trim=31 0 35 20,scale=0.65]{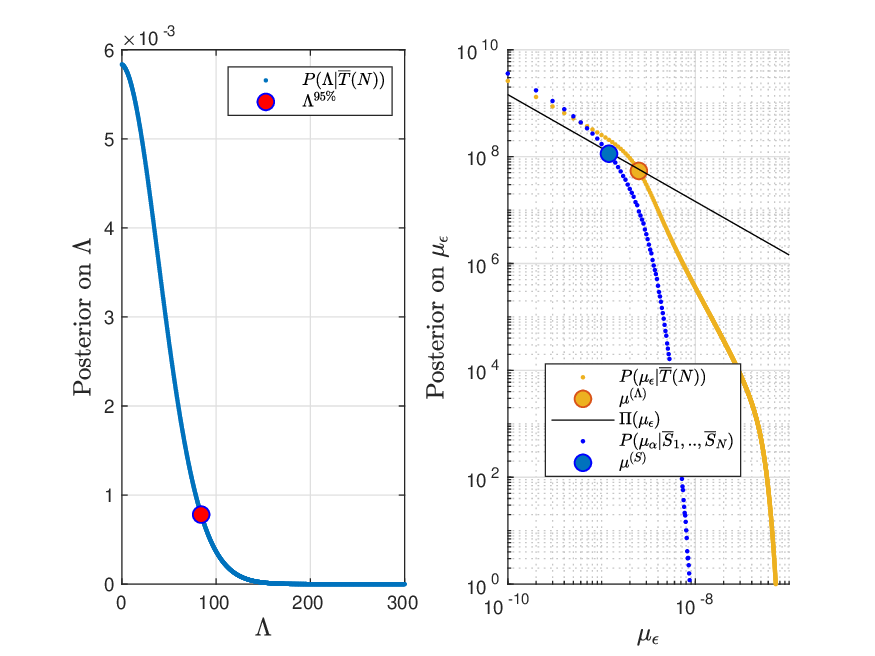}
\caption{\small{Upper limits inferred from the ensemble procedure considering the set of 201 pulsars in Figure \ref{fig:res_all} and considering the O3 data for the LIGO and Virgo detectors. \textit{Left plot:} posterior distribution for the global parameter $\Lambda$ (defined in Eq.~\ref{Lambda}). The red point indicates the 95\% credible upper limit for $\Lambda$. \textit{Right plot:} posterior distribution for the two independent hierarchical procedures to set the upper limits on the mean $\mu_\epsilon$ of the assumed exponential distribution. The continuous line is the log-uniform hyperprior on $\mu_\epsilon$. The blue and yellow points indicate the 95\% credible upper limit for $\mu_\epsilon$.}}
\label{fig:res_ulensALL}
\end{figure}

Using the statistic $T(N)$, that is the simple sum of the statistics for the pulsars in the analyzed ensemble, we set 95\% credible upper limit on the parameter $\Lambda$ defined in Eq~\ref{Lambda}. The prior on $\Lambda$ is chosen uniform while the likelihood is the value of the $T(N)$ signal distribution evaluated at the measured value $\overline{T}(N)$. The posterior $P(\Lambda|\overline{T}(N))$, shown in the left plot of Fig.~\ref{fig:res_ulensALL}, is compatible with zero as expected for the noise-only case. The 95\% credible upper limit is:
\begin{equation}
\Lambda^{95\%}=84.1    \,.
\end{equation}

It is not possible to infer any information for the single pulsars from the upper limit $\Lambda^{95\%}$ since $\Lambda$ is a global parameter for the ensemble. $\Lambda^{95\%}$ can be used to constrain different global parameters based on some assumptions for the single pulsar amplitudes. For example, by assuming that each pulsar in the ensemble emits a CW signal with the same amplitude $\overline{h}$, from the upper limit  $\Lambda^{95\%}$ it follows that $\overline{h}=1.9 \times 10^{-27}$.

For hierarchical procedures, following the analysis in \cite{ensbayes} and the stochastic searches in \cite{delillo,deep}, we assume a common exponential distribution for the ellipticities with the hyperprior $\Pi(\mu_\epsilon)$ that is log-uniform between $10^{-10}$ and $10^{-7}$. The posteriors $P(\mu_\epsilon|\overline{T}(N))$ and $P(\mu_\epsilon|\overline{S}_1,..,\overline{S}_N)$ are shown in the right plot of Fig.~\ref{fig:res_ulensALL}.

The upper limit $\mu_\epsilon^{(\Lambda)}$ for the hierarchical procedure using the $\Lambda$ parameter is 
\begin{equation}
    \mu_\epsilon^{(\Lambda)}=2.5\times 10^{-9} \,.
\end{equation}

This upper limit procedure considers the theoretical signal distribution to evaluate the likelihood. The theoretical distribution can be used in the hypothesis of Gaussian noise in the analyzed frequency band for each pulsar. This can be checked looking at the fitted Gamma parameters for the $T(N)$ statistic in the top-right plot of Fig.~\ref{fig:res_all}.

The upper limit $\mu_\epsilon^{(S)}$ for the hierarchical procedure using the single pulsar results is
\begin{equation}
    \mu_\epsilon^{(S)}=1.2\times 10^{-9} \,.
\end{equation}

The results and also the posteriors  in Fig.~\ref{fig:res_ulensALL} can be compared with Figure 8 in \cite{ensbayes}, where the authors analyzed 92 pulsars using data from the LIGO S6 science run showing the 90\% upper limit on the mean of the assumed exponential distribution of $3.8 \times 10^{-8}$. 

\subsection{Discussion}\label{subsec:disc}
As shown in Fig.~\ref{fig:res_all}-\ref{fig:res_HV} there is no evidence for a CW signal from the three different considered ensemble. In the following, we report some comments on the results found in this work. 
\begin{enumerate}[noitemsep,nosep,topsep=0pt,partopsep=0pt,itemindent=0.6cm,labelwidth=0cm,listparindent=0cm,leftmargin=0pt]
\item The upper limits set in this paper on the mean of the exponential distribution for the set of analyzed pulsars are more than one order of magnitude below the value in \cite{ensbayes}. The combined O3 data for the LIGO and Virgo detectors are largely more sensitive than the LIGO S6 science run data and the analyzed ensemble is larger. A more reliable comparison between the two pipelines should consider the same ensemble, i.e. the same pulsars, and is outside the scope of this paper.
\item The lowest limit on the ellipticity from the targeted search in \cite{O3targ} is $3.2 \times 10^{-9}$ for J0636+5129 while from the all-sky search in \cite{O3allsky} is $1.4 \times 10^{-9}$ for a neutron star at $10\,$pc and $2047.5\,$Hz. The upper limits on the mean ellipticity set in this work are consistent with these values. We also note that these results approach the  minimum  limit  on  the ellipticity for millisecond pulsars of $\approx 10^{-9}$ supposed in \cite{min_ell}. 
\item The upper limits in \cite{delillo,deep} on average ellipticity for the neutron star population are $\mathcal{O}(10^{-8})$ from  cross-correlation-based searches of a stochastic gravitational wave background. These limits can not be directly compared with the limits in this work since we focused on known pulsars that represent only a small subset of the entire set of neutron stars in our Galaxy. 
\item The upper limits on the hyperparameter for the exponential distribution do not consider the uncertainties on the distance. The distance should be included as a variable assuming, for example following \cite{ensbayes}, a Gaussian prior with a mean given by each pulsar’s best fit distance, and a standard deviation of 20\% of that value with a hard cutoff at zero.
\item The hierarchical procedures assume that all the analyzed pulsar ellipticities are drawn from a common distribution that can be too simplistic to describe the true $\epsilon$ distribution. For example, the population of young pulsars  and  old  recycled  millisecond  pulsars have  undergone different evolution and so have a different distribution. These populations should be treated independently, or a more complex  distribution  that  allows  separation  of  the  two distributions should be used \cite{ensbayes}. Therefore, a bimodal distribution, or two independent exponential distributions with different mean values, could be more appropriate and worth of study.
\item Using the multidetector procedure, the results of single pulsar analysis in Fig.~\ref{fig:res_singlep} are in agreement with \cite{O3targ}. However, it is still not clear when a multidetector analysis outperforms the most sensitive detector analysis. To choose which detectors should be used in the multidetector case, criteria based on the source sky position, on the detectors' sensitivities and observation times will be studied in the next future.
\end{enumerate}

\section{Conclusion}\label{sec:end}
In this paper, we have described the application of the 5\textit{n}-vector ensemble method to O3 data for the LIGO and Virgo detectors considering a set of 201 known pulsars.

The single pulsar analysis can be seen as an hypothesis test where the null hypothesis of pure noise is tested against the alternative hypothesis of the presence of a CW signal in the data. To improve the detection probability, we apply the 5\textit{n}-vector ensemble method, a multiple test that combines multi-detector single pulsar statistics defined through the 5\textit{n}-vector method. 

For the analysis of single pulsar, the single harmonic search (i.e. gravitational wave frequency exactly twice the rotation frequency of the source) on the 223 known pulsars shows good agreement with the Bayesian results in \cite{O3targ}. For the first time, the 5\textit{n}-vector method is applied to binary systems for the targeted search and compared with the Bayesian results. The dual harmonic search (GW frequency at both once and twice the rotation frequency) is not implemented for the 5\textit{n}-vector method. In Appendix \ref{app:bin_par}, we describe a preliminary analysis to quantify the effect of the binary parameters uncertainties on the detection sensitivity. We exclude from the ensemble analysis 22 binary pulsars whose combined uncertainties on $t_p$ and $\omega$ do not respect the selection criteria in Eq.~\ref{thr_bin}.

The ensemble statistics $T(k)$ for the rank truncation method is defined as the partial sum of the $k$ largest order statistics to control the look-elsewhere effect. Reconstructing the $T(k)$ noise distributions for each $k$, using a Monte Carlo procedure, we compute the $p$-value of ensemble as a function of $k$, that is a $p$-value for the overall hypothesis of the presence of CWs from the ensemble.

In case of no detection, we propose different procedures to set upper limits using the ensemble procedure and the statistic $T(N)$, i.e. the statistic for the entire ensemble. Using a mixed frequentist-Bayesian procedure, we can constrain the value of the global parameter $\Lambda$ that fixes the $T(N)$ signal distribution. Assuming a common exponential distribution for the ellipticities, we propose two independent hierarchical procedures to set upper limit on the mean $\mu_\epsilon$ of the assumed distribution. 

Section \ref{sec:res_ens} shows the application of the ensemble procedure to the set of 223 pulsars in Table \ref{tab:long} considering O3 data. Glitching pulsars, as the Crab, are not considered in the single pulsar analysis and therefore, also in the ensemble analysis. In future searches, glitching pulsars can be included in the ensemble procedure considering the resulting statistic for the single pulsar analysis, or considering each inter-glitch period as an independent pulsar. 

Three different ensembles are analyzed: the full set for the LIGO and Virgo detectors, the millisecond pulsar set for the two LIGO detectors and the high-value pulsars set for the two LIGO detectors. The results are shown in the summary plots in Fig.~\ref{fig:res_all}, Fig.~\ref{fig:res_MS} and Fig.~\ref{fig:res_HV}.

There is no evidence of a CW signal from the ensemble; the $p$-values as a function of $k$ are well above the assumed 1\% threshold. 

The upper limits procedures are applied to the entire set of pulsars considering the LIGO and Virgo detectors and O3 data. The posterior on the $\Lambda$ parameter and on $\mu_\epsilon$ are shown in Fig.~\ref{fig:res_ulensALL}. The upper limit set on $\Lambda$ is $\Lambda^{95\%}=84.1$ while the upper limit on $\mu_\epsilon$ is $2.5 \times 10^{-9}$ for the hierarchical procedure using the $\Lambda$ parameter, and $1.2 \times 10^{-9}$ for the hierarchical procedure using the single pulsar results. These results are more than one order of magnitude below the upper limit in \cite{ensbayes}, where the authors considered a classic hierarchical Bayesian procedure on an ensemble of 92 pulsars and data from the LIGO S6 science run.

The 5\textit{n}-vector ensemble method improves the detection probability for the targeted search of CWs from known pulsars. Application of this procedure on the next observing runs will improve the possibility to detect a CW signal from known pulsars for the first time.

\section*{Acknowledgement}
This research has made use of data obtained from the Gravitational Wave Open Science Center (https://www.gw-openscience.org/ ), a service of LIGO Laboratory, the LIGO Scientific Collaboration and the Virgo Collaboration. LIGO Laboratory and Advanced LIGO are funded by the United States National Science Foundation (NSF) as well as the Science and Technology Facilities Council (STFC) of the United Kingdom, the Max-Planck-Society (MPS), and the State of Niedersachsen/Germany for support of the construction of Advanced LIGO and construction and operation of the GEO600 detector. Additional support for Advanced LIGO was provided by the Australian Research Council. Virgo is funded, through the European Gravitational Observatory (EGO), by the French Centre National de Recherche Scientifique (CNRS), the Italian Istituto Nazionale di Fisica Nucleare (INFN) and the Dutch Nikhef, with contributions by institutions from Belgium, Germany, Greece, Hungary, Ireland, Japan, Monaco, Poland, Portugal, Spain.
\\JWM gratefully acknowledges support by the Natural Sciences and Engineering
Research Council of Canada (NSERC), [funding reference \#CITA 490888-16].
\\Pulsar research at UBC is supported by an NSERC Discovery Grant and by the Canadian Institute for Advanced Research.
\\Pulsar research at UoM is supported by a Consolidated Grant from STFC. 
\\The Nan\c{c}ay Radio Observatory is operated by the Paris Observatory, associated with the French Centre National de la Recherche Scientifique (CNRS). We acknowledge financial support from ``Programme National de Cosmologie et Galaxies'' (PNCG) and ``Programme National Hautes Energies'' (PNHE) of CNRS/INSU, France.
\\We acknowledge that CHIME is located on the traditional, ancestral, and 
unceded territory of the Syilx/Okanagan people. We are grateful to the 
staff of the Dominion Radio Astrophysical Observatory, which is operated 
by the National Research Council of Canada.  CHIME is funded by a grant 
from the Canada Foundation for Innovation (CFI) 2012 Leading Edge Fund 
(Project 31170) and by contributions from the provinces of British 
Columbia, Qu\'{e}bec and Ontario. The CHIME/FRB Project, which enabled 
development in common with the CHIME/Pulsar instrument, is funded by a 
grant from the CFI 2015 Innovation Fund (Project 33213) and by 
contributions from the provinces of British Columbia and Qu\'{e}bec, and 
by the Dunlap Institute for Astronomy and Astrophysics at the University 
of Toronto. Additional support was provided by the Canadian Institute 
for Advanced Research (CIFAR), McGill University and the McGill Space 
Institute thanks to the Trottier Family Foundation, and the University 
of British Columbia. The CHIME/Pulsar instrument hardware was funded by 
NSERC RTI-1 grant EQPEQ 458893-2014. This research was enabled in part 
by support provided by WestGrid (www.westgrid.ca) and Compute Canada 
(www.computecanada.ca).


\appendix

\section{Weighted multidetector extension}\label{app:5n}
Considering $n$ detectors, the classic 5\textit{n}-vector definition, described in \cite{5n}, combines together the 5-vectors $\mathbf X_i$ from the $i^{\text{th}}$ detector:
\begin{equation}
\mathbf X=[\mathbf X_1,..., \mathbf X_n] \,.
\end{equation}

Since the detectors can have different sensitivities, the statistic based on the 5\textit{n}-vector can reduce the signal to noise ratio compared to the statistic based only on the 5-vector of the most sensitive detector.

In this paper, we define the weighted data 5\textit{n}-vector as:
\begin{equation}\label{5nvec_weight}
\mathbf X=[c_1\mathbf X_1,..., c_n\mathbf X_n] 
\end{equation} where the weights $c_j$ are defined as:
\begin{equation}
c_j = \frac{\sqrt{n}}{\sqrt{\sum\limits_{i=1}^n\left( \frac{T_i}{S_i}  \right)} }\sqrt{\frac{T_j}{S_j}}= \sqrt{\mathcal{H}} \, \sqrt{\frac{T_j}{S_j}} \,.
\end{equation}$S_j$ and $T_j$ are the variance of the data in a small frequency band around the expected GW frequency and the observation time in the $j^{\text{th}}$ detector, respectively. $\mathcal{H}$ is  the harmonic mean of the time-weighted variances. 

The matched filter is (similarly for $\hat{H}_\times$):
\begin{equation}\begin{split}
\hat{H}_+ &=\frac{\textbf{X} \cdot \textbf{A}^{+}}{|\textbf{A}^{+}|^2}=\frac{\sum\limits_{j=1}^{n}c_j \textbf{X}_j \, (\textbf{A}_j^{+})^*}{\sum\limits_{k=1}^{n}\textbf{A}_k^{+}\,(\textbf{A}_k^{+})^*}= \\
&= \frac{1}{|\textbf{A}^{+}|^2} \left(c_1 |\textbf{A}_1^{+}|^2 \hat{H}_{+,1}+...+ c_n |\textbf{A}_n^{+}|^2 \hat{H}_{+,n} \right) \,.
\end{split}
\end{equation}

In the hypothesis of Gaussian noise for the $j^{\text{th}}$ detector with variance $\sigma_j^2$, the two complex estimators $\hat{H}_{+/\times}$ have also Gaussian distributions,
\begin{equation}\label{variance}
\hat{H}_{+/\times}\sim Gauss\left(x;\, 0,\sigma^2_{+/\times}\right)
\end{equation}\text{with} 
\begin{equation}
\sigma^2_{+/\times}=\sum\limits_{j=1}^{n}\frac{(c_j\sigma_j)^2\, T_j \,  |\textbf{A}_j^{+/\times}|^2}{|\textbf{A}^{+/\times}|^4} \,.
\end{equation}
  Using the $c_j$, we re-define the noise variance in each detector:
\begin{equation}
(c_j\sigma_j)^2 \varpropto c_j^2 \, S_j = \frac{n \, T_j}{\sum\limits_{i=1}^n\left( \frac{T_i}{S_i} \right) } \,.
\end{equation}
 
 If the observation time was the same ($\forall \, j,\,T_j=t $), this corresponds to "equalize" the noise in each detector. In this case, it follows that:\begin{equation}\begin{split}
\sigma^2_{+/\times}&=\sum\limits_{j=1}^{n}\frac{(c_j\sigma_j)^2\, T_j \, |\textbf{A}_j^{+/\times}|^2}{|\textbf{A}^{+/\times}|^4}=\sum\limits_{j=1}^{n}\frac{n\, t \, |\textbf{A}_j^{+/\times}|^2}{\sum\limits_{i=1}^{n} \left(\frac{1}{\sigma_i^2}\right)|\textbf{A}^{+/\times}|^4}=\\
&=\frac{n \, t}{\sum\limits_{i=1}^{n} \left(\frac{1}{\sigma_i^2}\right)|\textbf{A}^{+/\times}|^2}=\frac{n \, t}{\sum\limits_{i=1}^{n} \left(\frac{1}{\sigma_i^2}\right)\sum\limits_{k=1}^{n} |\textbf{A}_k^{+/\times}|^2} \,.
\end{split}\end{equation}

Let us consider the toy case of $n$ co-located detectors where $\,|\textbf{A}_k^{+/\times}|^2=|\textbf{A}^{+/\times}_1|^2 \,,\forall \, k$; the variances are 
\begin{equation}
\sigma^2_{+/\times}=\frac{t}{\sum\limits_{i=1}^{n} \left(\frac{1}{\sigma_i^2}\right) |\textbf{A}^{+/\times}_1|^2} \,.
\end{equation}This corresponds to the case of one detector with observation time $t$ and variance $V^2$:
\begin{equation}
V^2=\frac{1}{\sum\limits_{i=1}^{n} \left(\frac{1}{\sigma_i^2}\right)}=\frac{\mathcal{H}}{n}
\end{equation} 
where $\mathcal{H}$ is the harmonic mean of the variances. Since there is the condition:
\begin{equation}
min\{\sigma_1^2,...,\sigma^2_n\}\leq \mathcal{H} \leq  n\, min\{\sigma_1^2,...,\sigma^2_n\}
\end{equation}this means that:
\begin{equation}\label{imp}
\frac{min\{\sigma_1^2,...,\sigma^2_n\}}{n}\leq \frac{\mathcal{H}}{n} \leq  min\{\sigma_1^2,...,\sigma^2_n\} \,.
\end{equation}

It follows that for $n$ co-located detectors with the same observation time, we always have an improvement in the detection sensitivity using the coefficients $c_j$. 

For example, let us consider the case of two co-located detectors $n=2$ (equal to consider different datasets of the same detector), with $\sigma_2^2=C\, \sigma_1^2$ and $C>1$:
\begin{equation}
\sigma^2_{+/\times}=\frac{C\, \sigma_1^2\, t}{(C+1)\,  |\textbf{A}^{+/\times}_1|^2} \,.
\end{equation} 
The minimum detectable signal is $h_{min}$:\begin{equation}
h_{min}\varpropto \sqrt{\frac{C}{C+1}\frac{\sigma_1^2}{t}} \,.
\end{equation}

It is clear that in the general case (different detectors' locations and observation times), the multi-detector analysis not necessarily outperforms the most sensitive detectors. 

It is important to briefly describe the signal distribution for $|\hat{H}_{+/\times}|^2$, see also \cite{2014}.
If a CW signal is present into the analyzed data, the distributions of the two complex estimators $\hat{H}_{+/\times}$ are: 
\begin{equation}
\hat{H}_{+/\times}\sim Gauss\left(x\,;\, H_{0}\, e^{j\gamma}\, H_{+/\times}\, M_{+/\times}\,,\,\sigma^2_{+/\times}\right)
\end{equation}where $H_{+/\times}$ are the polarization functions, $H_{0}$ is the amplitude, $\gamma$ the phase and \begin{equation}
M_{+/\times}=\sum\limits_{j=1}^{n}\frac{c_j \, |\textbf{A}_j^{+/\times}|^2}{|\textbf{A}^{+/\times}|^2}
\end{equation}is a known factor for the targeted search.
\\$|\hat{H}_{+/\times}|^2$ have a non central-$\chi^2$ distribution (apart from the factor $k_{+/\times}$):
\begin{equation}
|\hat{H}_{+/\times}|^2\sim \frac{k_{+/\times}}{2}e^{-\frac{k_{+/\times} x+\lambda_{+/\times}}{2}}I_0(\sqrt{k_{+/\times}\lambda_{+/\times} x})
\end{equation}where $I_0$ is the modified Bessel function of the first kind, and\begin{equation}
 k_{+/\times}=\frac{2}{\sigma^2_{+/\times}} \qquad  \lambda_{+/\times}=k_{+/\times}\, |H_{+/\times}|^2\, H_0^2\, M_{+/\times} \,.
\end{equation}

With respect to the classic definition of the 5\textit{n}-vector, the distributions do not change.  There is just a re-definition of the variances for the detectors noise. This means that using the coefficients $c_j$, it is changed how $\lambda_{+/\times}$ depends on the detectors' sensitivity.

\section{Test of the binary correction}\label{app:HI}
Hardware injections are simulated signals added to the detectors data physically displacing the detectors’ test masses \cite{HI}. Differential displacement of the test masses mimics the detectors’ response to a GW signal. Continuous hardware injections can be used to test CW analysis methods. 

During the O3 run, $17$ hardware injections that simulate different CW signals from spinning neutron stars were added in the two LIGO detectors. The list of injected pulsar parameters can be found in \cite{HIlist}. 

In this section, we consider two different hardware injections to test the 5\textit{n}-vector pipeline. We report the results for the injected pulsar P03 (SNR(1 yr) $\approx$ 30) that simulates an isolated spinning neutron stars with $f_{\text{gw}}\simeq 108.8$$\,$Hz and for P16 (SNR(1 yr) $\approx$ 68), a neutron star in binary system with $f_{\text{gw}}\simeq 234.5$$\,$Hz.
\begin{figure} [t]
\centering
\includegraphics[scale=0.6]{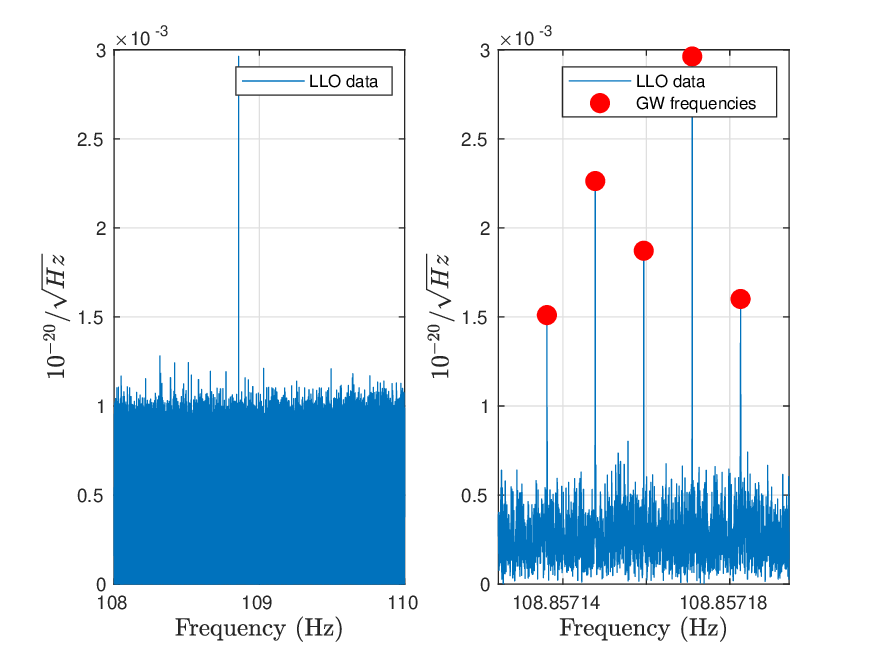}
\caption{\small{Amplitude spectral density for P03 and LLO data after the corrections for Doppler and spin-down effect. The right plot is the zoom around the GW frequency; the red dots show the expected 5 frequencies due to the Earth sidereal motion. Central peak frequency is the expected GW frequency.}}
\label{fig:HI3ASDL}
\end{figure}
\begin{figure}[t]
\centering
\includegraphics[scale=0.6]{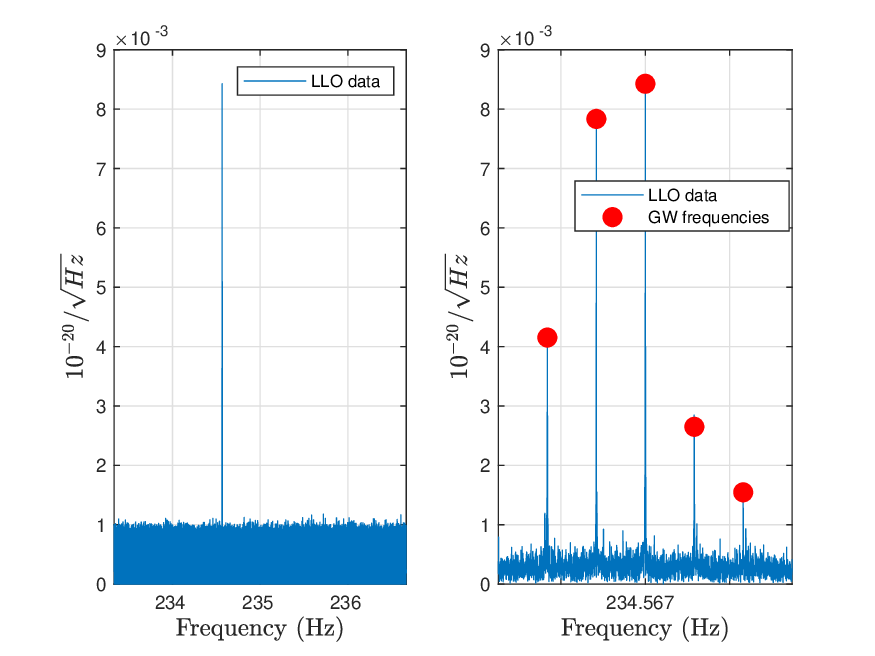}
\caption{\small{Amplitude spectral density for P16 and LLO data after the corrections for Doppler and spin-down effect. The right plot is the zoom around the GW frequency; the red dots show the expected 5 frequencies due to the Earth sidereal motion. Central peak frequency is the expected GW frequency.}}
\label{fig:HI16ASDL}
\end{figure}
\begin{table}[htpb]
\setlength{\arrayrulewidth}{0.12mm}
\setlength{\tabcolsep}{8pt}
\renewcommand{\arraystretch}{1.5}
  \begin{center}
    	\begin{tabular}{c c c c c}
    	\hline\hline
HI & Detector & $\epsilon_{H_0}$     & $\epsilon_\eta$      & $\epsilon_\psi$ \\
    	\hline
\multirow{2}{*}{P03}        & LLO   & 0.93  & 0.54\% &  1.49\%     \\
                            & LHO   & 0.95 & 1.27\%  & 0.39\%      \\
    	\hline
\multirow{2}{*}{P16}        & LLO   & 0.91  & 0.66\% &  -     \\
                            & LHO   & 0.93 & 0.90\% & -     \\
    	\hline                            
\end{tabular}
\end{center}
  \caption{ Table of the parameters mismatch for the hardware injections (HI) P03 and P16 analyzing O3 data from the two LIGO detectors. $\epsilon_{H_0}$ is the ratio between the estimated and injected amplitude, while  $\epsilon_{\eta}$ and  $\epsilon_\psi$ are the normalized relative errors for the polarization parameters}
  \label{tab:result}
\end{table}
Figures \ref{fig:HI3ASDL} and \ref{fig:HI16ASDL} show the amplitude spectral density for LLO detector in the frequency band around the expected GW frequency for P03 and P16  after the Doppler and spin-down corrections using the BSD heterodyne method. The red dots are the theoretical expected 5 frequency peaks for the GW signal due to the sidereal modulation. The different "heights" of corresponding peaks in the two detectors depend on the antenna pattern, i.e. on the signal 5-vector templates $\textbf{A}^{+/\times}$. It is important to note that the formalism used to construct the hardware injections' signal is independent from the 5-vector method. 

In Table \ref{tab:result}, there are the results obtained for the two hardware injections considering the O3 datasets of the two LIGO detectors. The estimation of the intrinsic parameters (see \cite{2010}) is not influenced by the choice of the detection statistic since it depends only on the estimators $\hat{H}_{+/\times}$.

The small discrepancies (below 10\%) in Table \ref{tab:result} for the amplitude  fall within the uncertainties of the actuation system used for the injections. For P16, the $\psi$ parameter is not well defined, since $\eta \approx 1$ and the signal is nearly circularly polarized. 
\begin{figure*} [t]
\centering
\includegraphics[clip,trim=60 0 50 10,scale=0.5]{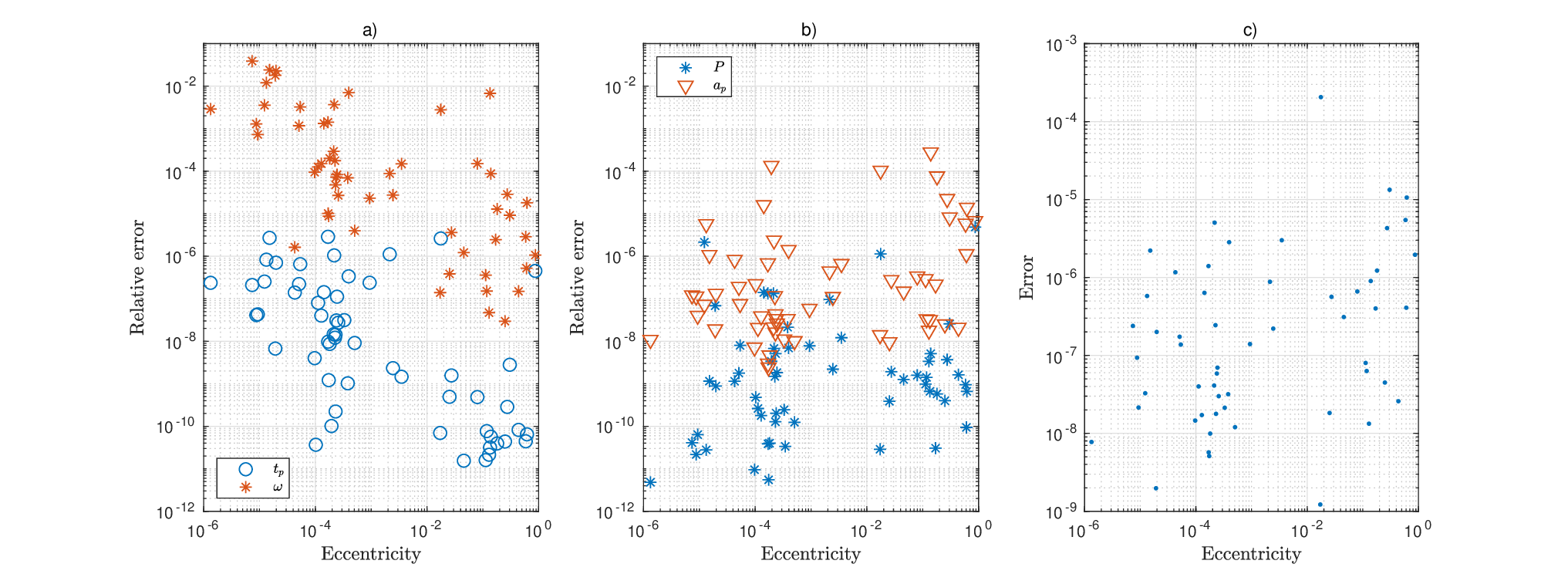}
\caption{\small{Uncertainties for the orbital parameters provided by the T2 model. \textit{a)} Relative errors on $t_p$ and $\omega$ as a function of the eccentricity. \textit{b)} Relative errors on $P$ and $a_p$ as a function of the eccentricity. \textit{c)} 1-$\sigma$ errors on the eccentricity as a function of the eccentricity values.}}\label{fig:rel_err}
\end{figure*}
\section{Orbital parameters uncertainties}\label{app:bin_par}
In this Appendix, we describe a preliminary analysis on the possible effects of the uncertainties on the heterodine de-modulation correction in the 5-vector context. We also describe the selection criteria used in this paper to include pulsars in binary systems for the ensemble procedure.
\\Indeed, large orbital parameters uncertainties could affect the sensitivity of targeted searches since  large mismatches on the binary parameters could reduce the signal-to-noise ratio.
\\The heterodine correction is applied by multiplying the data by the exponential factor $e^{-i\Phi_{corr}(t)}$. The total signal phase correction $\Phi_{corr}(t)$ can be written as the sum of the spin-down and the Doppler contribution. The Doppler contribution is the sum of the terms due to the Earth motion and due to the pulsar orbital motion $\Phi_{bin}(t)$. For the linear phase model and the small-eccentricity limit, $e\ll 1$ (details in \cite{deltabin}): 
\begin{equation}\small{
    \Phi_{bin}(t)\approx 2\pi f \left[ - a_p \left( \sin \psi(t) + \frac{k}{2} \sin 2\psi(t) -\frac{\beta}{2}\cos 2\psi(t)  \right) \right]}
\end{equation}where $a_p$ is the semi-major axis whike $k$ and $\beta$ are the Laplace-Lagrange parameters:
\begin{equation}
    k\equiv e\cos(\omega) \qquad \beta\equiv e\sin(\omega)
\end{equation} The mean orbital phase is
\begin{equation}
    \psi(t)=\Omega (t-t_{asc})
\end{equation}measured from the time of ascending nodes $t_{asc}$. For small eccentricity and mean orbital angular velocity $\Omega=\frac{2\pi}{P}$ ($P$ is the orbital period), $t_{asc}$ is:
\begin{equation}
    t_{asc}=t_p-\frac{\omega}{\Omega}
\end{equation}  $t_p$ and $\omega$ are the time and the argument of periapse, respectively.
There are two main models for the orbital parameters \cite{tempo23}: the T2 model that returns the parameters ($P$, $a_p$, $t_p$, $\omega$, $e$) and the ELL1 model that returns the parameters ($P$, $a_p$, $t_{asc}$, $k$, $\beta$). Indeed in the low-eccentricity limit,  the T2 model returns highly correlated values for $t_p$ and $\omega$, and the ELL1 model was developed for pulsars in such orbits.
\\In our preliminary analysis on the effects of the uncertainties, we focus on the T2 model since the ELL1 model provides better constrained parameters. For the ensemble of pulsars considered in our manuscript, almost 80 pulsars follows the T2 model. The parameters with the larger relative uncertainties are $t_p$ and $\omega$  as shown in Figure \ref{fig:rel_err} (particularly when the orbital eccentricity is below 0.01 as expected). These two parameters are also expected to be more relevant since they contribute directly to the orbital phase $\psi(t)$. We note that for 15 pulsars from the T2 model, we do not have an estimation of $\Delta t_p$ while 11 pulsars are in circular orbits with $e=\omega=0$.
\\Using the P16, we have computed the coherence\footnote{ The coherence is a number between 0 and 1 that measures the resemblance between the shape of the expected signal and the data. Differently form the p-value, the noise distribution of the coherence is not flat between 0 and 1 (see \cite{2010}). It is a supplemental parameter to show the significance of the detection.} for a wide range of the uncertainties $\Delta t_p$ and $\Delta \omega$. The results show minima in the coherence corresponding to certain values of the combination:
\begin{equation}
    \Delta \psi=\Omega \Delta t_p +  \Delta \omega
\end{equation}The values of $\Delta \psi$ corresponding to the minima can be analytically computed from the phase $\Phi_{bin}$ considering a mismatch $\Delta t_p$ and $\Delta \omega$.  
\\We fix a threshold using the value of $\Delta \psi$ associated to the first minimum:
\begin{equation}\label{thr_bin}
    \Delta \psi \sim F\cdot \frac{3}{8fa_p}
\end{equation}
We can take the factor $F=5$, which seems reasonable considering that the first minima are very sharp. \\Considering the 223 pulsars analyzed in Table \ref{tab:long}, 80 pulsars have parameters form the T2 model; 22 pulsars (J0218$+$4232, J1017$-$7156, J1125$-$5825, J1312$+$0051, J1529$-$3828, J1545$-$4550, J1603$-$7202, J1614$-$2230, J1630$+$3734, J1708$-$3506, J1709$+$2313, J1737$-$0811, J1750$-$2536, J1751$-$2857, J1835$-$0114, J1840$-$0643, J1904$+$0412, J1946$+$3417, J1949$+$3106, J2019$+$2425, J2053$+$4650, J2129$-$5721) have a $\Delta \psi$ above the threshold in Eq.~$\ref{thr_bin}$. We have excluded these pulsars from the ensemble analysis.  
\\We stress that the uncertainties provided by the binary model are 1$\sigma$ Gaussian errors while in our relation, $\Delta t_p$ and $\Delta \omega$ are considered as an absolute error.
\\The preliminary results described in this response consider only the $t_p$ and $\omega$ parameters; it remains unclear how the uncertainties of the entire set of parameters combine together in the heterodyne correction. A precise and independent analysis is still required and falls beyond the scope of our manuscript.

\clearpage

\setlength{\tabcolsep}{10pt}
\begin{longtable*}{lrccccc}
\caption{Table of the parameters (name, GW frequency in Hz, distance in kpc, spin-down limit $h_{\text{sd}}$) and the results inferred using the 5\textit{n}-vector method (upper limit on the amplitude $h_0^{95\%}$, upper limit on the ellipticity $\epsilon^{95\%}$, $p$-value) for the 223 analyzed pulsars. Some pulsars don't have a value for $h_{\text{sd}}$, and $\epsilon^{95\%}$ since the corresponding distance value (or the frequency derivative value) is not available in the ATNF Catalogue \cite{atnf}. More details about the references for the ephemerides and distance estimation for the analyzed pulsars are in \cite{O3targ}.} \label{tab:long} \\

\hline \hline \rule{0pt}{3ex}  
Pulsar name & $f_{\text{gw}}$ & Distance  &
$h_{\text{sd}}$ &
$h_0^{\, 95\%}$ &
$\epsilon^{95\%}$ &
P-value\\ 
(J2000)  & (Hz) & (kpc) &  &  & &  \\   [1ex]  
\hline  \rule{0pt}{0.2ex}
\endfirsthead

\hline \hline \rule{0pt}{3ex}  
Pulsar name & $f_{\text{gw}}$ & Distance  &
$h_{\text{sd}}$ &
$h_0^{\, 95\%}$ &
$\epsilon^{95\%}$ &
P-value \\ 
(J2000)  & (Hz) & (kpc) &  &  &  & \\ [1ex]  
\hline  \rule{0pt}{0.2ex}
\endhead

\hline
\endfoot

J0023+0923  & $655.7$ & $1.1$ & $1.3 \cdot 10^{-27} $ & $ $8.2$ \cdot 10^{-27} $ & $2.0 \cdot 10^{-8}$ & $0.68$ \\ 
J0030+0451  & $411.1$ & $0.3$ & $3.6 \cdot 10^{-27} $ & $ $6.5$ \cdot 10^{-27} $ & $1.2 \cdot 10^{-8}$ & $0.71$ \\ 
J0034$-$0534  & $1065.4$ & $1.4$ & $8.9 \cdot 10^{-28} $ & $ $1.0$ \cdot 10^{-26} $ & $1.2 \cdot 10^{-8}$ & $0.93$ \\ 
J0101$-$6422  & $777.3$ & $1.0$ & $9.7 \cdot 10^{-28} $ & $ $1.2$ \cdot 10^{-26} $ & $1.9 \cdot 10^{-8}$ & $0.11$ \\ 
J0102+4839  & $674.7$ & $2.3$ & $6.8 \cdot 10^{-28} $ & $ $7.3$ \cdot 10^{-27} $ & $3.5 \cdot 10^{-8}$ & $0.81$ \\ 
J0117+5914  & $19.7$ & $1.8$ & $1.1 \cdot 10^{-25} $ & $ $6.6$ \cdot 10^{-25} $ & $2.8 \cdot 10^{-3}$ & $0.66$ \\ 
J0125$-$2327  & $544.2$ & $0.9$ & $2.0 \cdot 10^{-27} $ & $ $8.1$ \cdot 10^{-27} $ & $2.4 \cdot 10^{-8}$ & $0.48$ \\ 
J0154+1833  & $845.8$ & $1.6$ & $3.4 \cdot 10^{-28} $ & $ $1.1$ \cdot 10^{-26} $ & $2.5 \cdot 10^{-8}$ & $0.4$ \\ 
J0218+4232  & $860.9$ & $3.1$ & $1.5 \cdot 10^{-27} $ & $ $8.0$ \cdot 10^{-27} $ & $3.2 \cdot 10^{-8}$ & $1$ \\ 
J0340+4130  & $606.2$ & $1.6$ & $7.2 \cdot 10^{-28} $ & $ $9.1$ \cdot 10^{-27} $ & $3.7 \cdot 10^{-8}$ & $0.33$ \\ 
J0348+0432  & $51.1$ & $2.1$ & $9.3 \cdot 10^{-28} $ & $ $1.4$ \cdot 10^{-26} $ & $1.0 \cdot 10^{-5}$ & $0.2$ \\ 
J0406+3039  & $766.7$ & - & - & $8.0 \cdot 10^{-27}$  & -  & $0.85$ \\ 
J0407+1607  & $77.8$ & $1.3$ & $1.1 \cdot 10^{-27} $ & $ $9.3$ \cdot 10^{-27} $ & $1.9 \cdot 10^{-6}$ & $0.2$ \\ 
J0437$-$4715  & $347.4$ & $0.2$ & $8.0 \cdot 10^{-27} $ & $ $7.4$ \cdot 10^{-27} $ & $9.3 \cdot 10^{-9}$ & $0.24$ \\ 
J0453+1559  & $43.7$ & $0.5$ & $3.1 \cdot 10^{-27} $ & $ $9.7$ \cdot 10^{-27} $ & $2.5 \cdot 10^{-6}$ & $1$ \\ 
J0509+0856  & $493.1$ & $0.8$ & $9.6 \cdot 10^{-28} $ & $ $6.7$ \cdot 10^{-27} $ & $2.1 \cdot 10^{-8}$ & $0.83$ \\ 
J0509+3801  & $26.1$ & $1.6$ & $5.3 \cdot 10^{-27} $ & $ $4.0$ \cdot 10^{-26} $ & $8.7 \cdot 10^{-5}$ & $0.7$ \\ 
J0557+1550  & $782.4$ & $1.8$ & $7.5 \cdot 10^{-28} $ & $ $1.1$ \cdot 10^{-26} $ & $3.2 \cdot 10^{-8}$ & $0.33$ \\ 
J0557$-$2948  & $45.8$ & $4.3$ & $2.4 \cdot 10^{-28} $ & $ $1.1$ \cdot 10^{-26} $ & $2.1 \cdot 10^{-5}$ & $0.65$ \\ 
J0609+2130  & $35.9$ & $0.6$ & $2.9 \cdot 10^{-27} $ & $ $1.8$ \cdot 10^{-26} $ & $7.3 \cdot 10^{-6}$ & $0.73$ \\ 
J0610$-$2100  & $518.0$ & $3.3$ & $1.3 \cdot 10^{-28} $ & $ $3.2$ \cdot 10^{-26} $ & $3.6 \cdot 10^{-7}$ & $0.63$ \\ 
J0613$-$0200  & $653.2$ & $0.6$ & $2.2 \cdot 10^{-27} $ & $ $6.9$ \cdot 10^{-27} $ & $9.2 \cdot 10^{-9}$ & $0.96$ \\ 
J0614$-$3329  & $635.2$ & $0.6$ & $3.0 \cdot 10^{-27} $ & $ $8.2$ \cdot 10^{-27} $ & $1.2 \cdot 10^{-8}$ & $0.56$ \\ 
J0621+1002  & $69.3$ & $0.4$ & $2.4 \cdot 10^{-27} $ & $ $8.8$ \cdot 10^{-27} $ & $7.3 \cdot 10^{-7}$ & $0.34$ \\ 
J0636$-$3044  & $506.9$ & $0.7$ & $2.6 \cdot 10^{-27} $ & $ $3.1$ \cdot 10^{-26} $ & $8.1 \cdot 10^{-8}$ & $0.71$ \\ 
J0636+5129  & $697.1$ & $0.2$ & $4.2 \cdot 10^{-27} $ & $ $6.3$ \cdot 10^{-27} $ & $2.6 \cdot 10^{-9}$ & $1$ \\ 
J0645+5158  & $225.9$ & $1.2$ & $3.7 \cdot 10^{-28} $ & $ $4.7$ \cdot 10^{-27} $ & $1.1 \cdot 10^{-7}$ & $0.86$ \\ 
J0709+0458  & $58.1$ & $1.2$ & $2.2 \cdot 10^{-27} $ & $ $9.1$ \cdot 10^{-27} $ & $3.1 \cdot 10^{-6}$ & $0.78$ \\ 
J0711$-$6830  & $364.2$ & $0.1$ & $1.2 \cdot 10^{-26} $ & $ $4.8$ \cdot 10^{-27} $ & $3.8 \cdot 10^{-9}$ & $0.91$ \\ 
J0721$-$2038  & $128.7$ & $2.7$ & $5.1 \cdot 10^{-28} $ & $ $6.0$ \cdot 10^{-27} $ & $9.2 \cdot 10^{-7}$ & $0.59$ \\ 
J0732+2314  & $489.0$ & $1.1$ & $8.5 \cdot 10^{-28} $ & $ $6.8$ \cdot 10^{-27} $ & $3.1 \cdot 10^{-8}$ & $0.76$ \\ 
J0740+6620  & $693.1$ & $1.1$ & $1.1 \cdot 10^{-27} $ & $ $1.0$ \cdot 10^{-26} $ & $2.3 \cdot 10^{-8}$ & $0.21$ \\ 
J0751+1807  & $574.9$ & $0.6$ & $1.9 \cdot 10^{-27} $ & $ $8.4$ \cdot 10^{-27} $ & $1.4 \cdot 10^{-8}$ & $0.62$ \\ 
J0824+0028  & $202.8$ & $1.7$ & $1.8 \cdot 10^{-27} $ & $ $1.1$ \cdot 10^{-26} $ & $4.2 \cdot 10^{-7}$ & $0.01$ \\ 
J0835$-$4510  & $22.4$ & $0.3$ & $3.4 \cdot 10^{-24} $ & $ $1.1$ \cdot 10^{-25} $ & $5.6 \cdot 10^{-5}$ & $0.26$ \\ 
J0921$-$5202  & $206.6$ & $0.4$ & $2.9 \cdot 10^{-27} $ & $ $5.1$ \cdot 10^{-27} $ & $4.5 \cdot 10^{-8}$ & $0.73$ \\ 
J0931$-$1902  & $431.2$ & $3.7$ & $1.8 \cdot 10^{-28} $ & $ $5.9$ \cdot 10^{-27} $ & $1.1 \cdot 10^{-7}$ & $0.84$ \\ 
J0955$-$6150  & $1000.3$ & $2.2$ & $9.9 \cdot 10^{-28} $ & $ $1.3$ \cdot 10^{-26} $ & $2.6 \cdot 10^{-8}$ & $0.51$ \\ 
J1012$-$4235  & $644.9$ & $0.4$ & $3.1 \cdot 10^{-27} $ & $ $7.9$ \cdot 10^{-27} $ & $6.7 \cdot 10^{-9}$ & $0.61$ \\ 
J1012+5307  & $380.5$ & $0.7$ & $1.6 \cdot 10^{-27} $ & $ $7.2$ \cdot 10^{-27} $ & $3.3 \cdot 10^{-8}$ & $0.3$ \\ 
J1017$-$7156  & $855.2$ & $3.5$ & $2.5 \cdot 10^{-28} $ & $ $7.3$ \cdot 10^{-27} $ & $3.3 \cdot 10^{-8}$ & $0.97$ \\ 
J1022+1001  & $121.6$ & $0.6$ & $1.8 \cdot 10^{-27} $ & $ $6.4$ \cdot 10^{-27} $ & $2.6 \cdot 10^{-7}$ & $0.56$ \\ 
J1024$-$0719  & $387.4$ & $1.2$ & - &  $6.1 \cdot 10^{-27}$ & $4.6 \cdot 10^{-8}$ & $0.75$ \\ 
J1035$-$6720  & $696.4$ & $1.5$ & $2.2 \cdot 10^{-27} $ & $ $6.5$ \cdot 10^{-27} $ & $1.8 \cdot 10^{-8}$ & $0.96$ \\ 
J1036$-$8317  & $586.7$ & $0.9$ & $2.6 \cdot 10^{-27} $ & $ $6.6$ \cdot 10^{-27} $ & $1.7 \cdot 10^{-8}$ & $0.74$ \\ 
J1038+0032  & $69.3$ & $6.0$ & $2.1 \cdot 10^{-28} $ & $ $7.6$ \cdot 10^{-27} $ & $8.9 \cdot 10^{-6}$ & $0.59$ \\ 
J1045$-$4509  & $267.6$ & $0.6$ & $2.1 \cdot 10^{-27} $ & $ $4.8$ \cdot 10^{-27} $ & $3.7 \cdot 10^{-8}$ & $0.91$ \\ 
J1101$-$6101  & $31.8$ & $7.0$ & $4.2 \cdot 10^{-26} $ & $ $2.3$ \cdot 10^{-26} $ & $1.5 \cdot 10^{-4}$ & $0.62$ \\ 
J1101$-$6424  & $391.4$ & $2.2$ & $2.2 \cdot 10^{-28} $ & $ $6.1$ \cdot 10^{-27} $ & $8.1 \cdot 10^{-8}$ & $0.6$ \\ 
J1103$-$5403  & $589.5$ & $1.7$ & $5.0 \cdot 10^{-28} $ & $ $5.4$ \cdot 10^{-27} $ & $2.5 \cdot 10^{-8}$ & $1$ \\ 
J1125$-$5825  & $644.7$ & $1.7$ & $2.0 \cdot 10^{-27} $ & $ $7.0$ \cdot 10^{-27} $ & $2.8 \cdot 10^{-8}$ & $0.78$ \\ 
J1125$-$6014  & $760.3$ & $1.4$ & $7.1 \cdot 10^{-28} $ & $ $7.7$ \cdot 10^{-27} $ & $1.8 \cdot 10^{-8}$ & $0.78$ \\ 
J1125+7819  & $476.0$ & $0.9$ & - &  $7.3 \cdot 10^{-27}$ & $2.7 \cdot 10^{-8}$ & $0.48$ \\ 
J1142+0119  & $394.1$ & $2.2$ & - &  $1.0 \cdot 10^{-26}$ & $1.3 \cdot 10^{-7}$ & $0.067$ \\ 
J1207$-$5050  & $413.0$ & $1.3$ & $6.9 \cdot 10^{-28} $ & $ $8.4$ \cdot 10^{-27} $ & $5.9 \cdot 10^{-8}$ & $0.17$ \\ 
J1216$-$6410  & $565.1$ & $1.1$ & $5.0 \cdot 10^{-28} $ & $ $5.9$ \cdot 10^{-27} $ & $1.9 \cdot 10^{-8}$ & $0.92$ \\ 
J1231$-$1411  & $542.9$ & $0.4$ & $2.9 \cdot 10^{-27} $ & $ $9.4$ \cdot 10^{-27} $ & $1.3 \cdot 10^{-8}$ & $0.29$ \\ 
J1300+1240  & $321.6$ & $0.6$ & - &  $6.1 \cdot 10^{-27}$ & $3.4 \cdot 10^{-8}$ & $0.64$ \\ 
J1302$-$3258  & $530.4$ & $1.4$ & $7.4 \cdot 10^{-28} $ & $ $7.1$ \cdot 10^{-27} $ & $3.4 \cdot 10^{-8}$ & $0.67$ \\ 
J1302$-$6350  & $41.9$ & $2.3$ & $7.7 \cdot 10^{-26} $ & $ $9.9$ \cdot 10^{-27} $ & $1.2 \cdot 10^{-5}$ & $0.97$ \\ 
J1312+0051  & $473.0$ & $1.5$ & $7.6 \cdot 10^{-28} $ & $ $9.8$ \cdot 10^{-27} $ & $6.1 \cdot 10^{-8}$ & $0.21$ \\ 
J1327$-$0755  & $746.8$ & $25.0$ & - &  $7.9 \cdot 10^{-27}$ & $3.3 \cdot 10^{-7}$ & $0.9$ \\ 
J1327+3423  & $48.2$ & - & - & $1.3 \cdot 10^{-26}$  & -  & $0.5$ \\ 
J1337$-$6423  & $212.2$ & $5.9$ & $1.9 \cdot 10^{-28} $ & $ $7.5$ \cdot 10^{-27} $ & $9.3 \cdot 10^{-7}$ & $0.13$ \\ 
J1400$-$1431  & $648.5$ & $0.3$ & $5.1 \cdot 10^{-28} $ & $ $6.7$ \cdot 10^{-27} $ & $4.0 \cdot 10^{-9}$ & $0.98$ \\ 
J1411+2551  & $32.0$ & $1.1$ & $8.5 \cdot 10^{-28} $ & $ $2.8$ \cdot 10^{-26} $ & $2.9 \cdot 10^{-5}$ & $0.3$ \\ 
J1412+7922  & $33.8$ & $2.0$ & $9.5 \cdot 10^{-26} $ & $ $2.0$ \cdot 10^{-26} $ & $3.3 \cdot 10^{-5}$ & $0.48$ \\ 
J1420$-$5625  & $58.6$ & $1.3$ & $8.5 \cdot 10^{-28} $ & $ $8.0$ \cdot 10^{-27} $ & $2.9 \cdot 10^{-6}$ & $0.88$ \\ 
J1421$-$4409  & $313.2$ & $2.1$ & $3.8 \cdot 10^{-28} $ & $ $5.9$ \cdot 10^{-27} $ & $1.2 \cdot 10^{-7}$ & $0.59$ \\ 
J1431$-$5740  & $486.6$ & $3.5$ & $2.8 \cdot 10^{-28} $ & $ $7.6$ \cdot 10^{-27} $ & $1.1 \cdot 10^{-7}$ & $0.44$ \\ 
J1435$-$6100  & $214.0$ & $2.8$ & $4.6 \cdot 10^{-28} $ & $ $9.1$ \cdot 10^{-27} $ & $5.3 \cdot 10^{-7}$ & $0.024$ \\ 
J1439$-$5501  & $69.8$ & $0.7$ & $2.7 \cdot 10^{-27} $ & $ $8.5$ \cdot 10^{-27} $ & $1.1 \cdot 10^{-6}$ & $0.21$ \\ 
J1446$-$4701  & $911.3$ & $1.5$ & $1.1 \cdot 10^{-27} $ & $ $9.6$ \cdot 10^{-27} $ & $1.6 \cdot 10^{-8}$ & $0.69$ \\ 
J1453+1902  & $345.3$ & $1.3$ & $8.0 \cdot 10^{-28} $ & $ $6.9$ \cdot 10^{-27} $ & $7.0 \cdot 10^{-8}$ & $0.44$ \\ 
J1455$-$3330  & $250.4$ & $1.0$ & $1.3 \cdot 10^{-27} $ & $ $6.7$ \cdot 10^{-27} $ & $1.0 \cdot 10^{-7}$ & $0.41$ \\ 
J1502$-$6752  & $74.8$ & $7.7$ & $3.0 \cdot 10^{-28} $ & $ $5.8$ \cdot 10^{-27} $ & $7.5 \cdot 10^{-6}$ & $0.84$ \\ 
J1513$-$2550  & $943.8$ & $4.0$ & $6.5 \cdot 10^{-28} $ & $ $1.4$ \cdot 10^{-26} $ & $5.7 \cdot 10^{-8}$ & $0.26$ \\ 
J1514$-$4946  & $557.2$ & $0.9$ & $1.6 \cdot 10^{-27} $ & $ $1.0$ \cdot 10^{-26} $ & $2.9 \cdot 10^{-8}$ & $0.19$ \\ 
J1518+0204A & $360.1$ & $8.0$ & $2.8 \cdot 10^{-28} $ & $ $5.6$ \cdot 10^{-27} $ & $3.3 \cdot 10^{-7}$ & $1$ \\ 
J1518+4904  & $48.9$ & $1.0$ & $6.3 \cdot 10^{-28} $ & $ $9.7$ \cdot 10^{-27} $ & $3.7 \cdot 10^{-6}$ & $0.97$ \\ 
J1525$-$5545  & $176.1$ & $3.1$ & $8.7 \cdot 10^{-28} $ & $ $4.9$ \cdot 10^{-27} $ & $4.7 \cdot 10^{-7}$ & $0.81$ \\ 
J1528$-$3146  & $32.9$ & $0.8$ & $2.1 \cdot 10^{-27} $ & $ $2.5$ \cdot 10^{-26} $ & $1.7 \cdot 10^{-5}$ & $0.41$ \\ 
J1529$-$3828  & $235.7$ & $4.3$ & $3.4 \cdot 10^{-28} $ & $ $4.5$ \cdot 10^{-27} $ & $3.3 \cdot 10^{-7}$ & $0.97$ \\ 
J1537+1155  & $52.8$ & $1.1$ & $6.0 \cdot 10^{-27} $ & $ $1.3$ \cdot 10^{-26} $ & $4.8 \cdot 10^{-6}$ & $0.17$ \\ 
J1543$-$5149  & $972.3$ & $1.1$ & $1.9 \cdot 10^{-27} $ & $ $9.2$ \cdot 10^{-27} $ & $1.1 \cdot 10^{-8}$ & $0.85$ \\ 
J1544+4937  & $926.2$ & $3.0$ & $3.1 \cdot 10^{-28} $ & $ $1.7$ \cdot 10^{-26} $ & $5.7 \cdot 10^{-8}$ & $0.019$ \\ 
J1545$-$4550  & $559.4$ & $2.2$ & $1.4 \cdot 10^{-27} $ & $ $9.5$ \cdot 10^{-27} $ & $6.3 \cdot 10^{-8}$ & $0.32$ \\ 
J1547$-$5709  & $466.1$ & $2.7$ & $3.9 \cdot 10^{-28} $ & $ $7.8$ \cdot 10^{-27} $ & $9.2 \cdot 10^{-8}$ & $0.32$ \\ 
J1551$-$0658  & $281.9$ & $1.3$ & $1.0 \cdot 10^{-27} $ & $ $8.5$ \cdot 10^{-27} $ & $1.3 \cdot 10^{-7}$ & $0.15$ \\ 
J1600$-$3053  & $555.9$ & $3.0$ & $4.0 \cdot 10^{-28} $ & $ $9.9$ \cdot 10^{-27} $ & $9.1 \cdot 10^{-8}$ & $0.25$ \\ 
J1603$-$7202  & $134.8$ & $3.4$ & $2.5 \cdot 10^{-28} $ & $ $4.5$ \cdot 10^{-27} $ & $8.0 \cdot 10^{-7}$ & $0.92$ \\ 
J1614$-$2230  & $634.8$ & $0.7$ & $1.2 \cdot 10^{-27} $ & $ $7.6$ \cdot 10^{-27} $ & $1.2 \cdot 10^{-8}$ & $0.76$ \\ 
J1618$-$3921  & $166.8$ & $5.5$ & $3.1 \cdot 10^{-28} $ & $ $9.3$ \cdot 10^{-27} $ & $1.7 \cdot 10^{-6}$ & $0.036$ \\ 
J1618$-$4624  & $337.2$ & $3.0$ & $1.9 \cdot 10^{-28} $ & $ $6.1$ \cdot 10^{-27} $ & $1.5 \cdot 10^{-7}$ & $0.56$ \\ 
J1622$-$6617  & $84.7$ & $4.0$ & $2.9 \cdot 10^{-28} $ & $ $6.0$ \cdot 10^{-27} $ & $3.2 \cdot 10^{-6}$ & $0.69$ \\ 
J1623$-$2631  & $180.6$ & $1.8$ & $1.3 \cdot 10^{-27} $ & $ $5.0$ \cdot 10^{-27} $ & $2.6 \cdot 10^{-7}$ & $0.92$ \\ 
J1628$-$3205  & $622.7$ & $1.2$ & - &  $7.1 \cdot 10^{-27}$ & $2.1 \cdot 10^{-8}$ & $0.79$ \\ 
J1629$-$6902  & $333.3$ & $1.0$ & $1.1 \cdot 10^{-27} $ & $ $6.0$ \cdot 10^{-27} $ & $4.9 \cdot 10^{-8}$ & $0.79$ \\ 
J1630+3734  & $602.8$ & $1.2$ & $1.1 \cdot 10^{-27} $ & $ $7.9$ \cdot 10^{-27} $ & $2.5 \cdot 10^{-8}$ & $0.53$ \\ 
J1640+2224  & $632.2$ & $1.5$ & $3.4 \cdot 10^{-28} $ & $ $1.1$ \cdot 10^{-26} $ & $3.9 \cdot 10^{-8}$ & $0.17$ \\ 
J1641+3627A & $192.7$ & $7.1$ & $3.2 \cdot 10^{-28} $ & $ $4.3$ \cdot 10^{-27} $ & $7.7 \cdot 10^{-7}$ & $1$ \\ 
J1643$-$1224  & $432.7$ & $1.2$ & $1.3 \cdot 10^{-27} $ & $ $1.1$ \cdot 10^{-26} $ & $6.6 \cdot 10^{-8}$ & $0.045$ \\ 
J1652$-$4838  & $528.4$ & - & - & $9.7 \cdot 10^{-27}$  & -  & $0.14$ \\ 
J1653$-$2054  & $484.4$ & $2.6$ & $5.0 \cdot 10^{-28} $ & $ $7.5$ \cdot 10^{-27} $ & $8.0 \cdot 10^{-8}$ & $0.6$ \\ 
J1658$-$5324  & $819.9$ & $0.9$ & $1.9 \cdot 10^{-27} $ & $ $1.1$ \cdot 10^{-26} $ & $1.4 \cdot 10^{-8}$ & $0.3$ \\ 
J1705$-$1903  & $806.4$ & $2.4$ & $9.8 \cdot 10^{-28} $ & $ $1.2$ \cdot 10^{-26} $ & $4.0 \cdot 10^{-8}$ & $0.33$ \\ 
J1708$-$3506  & $443.9$ & $3.3$ & $3.4 \cdot 10^{-28} $ & $ $6.2$ \cdot 10^{-27} $ & $9.9 \cdot 10^{-8}$ & $0.74$ \\ 
J1709+2313  & $431.9$ & $2.2$ & $1.9 \cdot 10^{-28} $ & $ $6.4$ \cdot 10^{-27} $ & $7.1 \cdot 10^{-8}$ & $0.68$ \\ 
J1713+0747  & $437.6$ & $1.0$ & $1.1 \cdot 10^{-27} $ & $ $7.0$ \cdot 10^{-27} $ & $3.5 \cdot 10^{-8}$ & $0.59$ \\ 
J1719$-$1438  & $345.4$ & $0.3$ & $2.6 \cdot 10^{-27} $ & $ $8.0$ \cdot 10^{-27} $ & $2.1 \cdot 10^{-8}$ & $0.25$ \\ 
J1721$-$2457  & $572.0$ & $1.4$ & - &  $8.0 \cdot 10^{-27}$ & $3.1 \cdot 10^{-8}$ & $0.59$ \\ 
J1727$-$2946  & $73.8$ & $1.9$ & $1.3 \cdot 10^{-27} $ & $ $9.3$ \cdot 10^{-27} $ & $3.0 \cdot 10^{-6}$ & $0.2$ \\ 
J1729$-$2117  & $30.2$ & $1.0$ & $1.3 \cdot 10^{-27} $ & $ $2.6$ \cdot 10^{-26} $ & $2.7 \cdot 10^{-5}$ & $0.83$ \\ 
J1730$-$2304  & $246.2$ & $0.5$ & $2.0 \cdot 10^{-27} $ & $ $5.8$ \cdot 10^{-27} $ & $4.2 \cdot 10^{-8}$ & $0.64$ \\ 
J1732$-$5049  & $376.5$ & $1.9$ & $6.2 \cdot 10^{-28} $ & $ $4.6$ \cdot 10^{-27} $ & $5.8 \cdot 10^{-8}$ & $1$ \\ 
J1737$-$0811  & $479.0$ & $0.2$ & $5.4 \cdot 10^{-27} $ & $ $8.4$ \cdot 10^{-27} $ & $7.3 \cdot 10^{-9}$ & $0.44$ \\ 
J1738+0333  & $341.9$ & $1.5$ & $1.1 \cdot 10^{-27} $ & $ $7.1$ \cdot 10^{-27} $ & $8.5 \cdot 10^{-8}$ & $0.44$ \\ 
J1741+1351  & $533.7$ & $1.1$ & $2.1 \cdot 10^{-27} $ & $ $6.4$ \cdot 10^{-27} $ & $2.3 \cdot 10^{-8}$ & $0.9$ \\ 
J1744$-$1134  & $490.9$ & $0.4$ & $2.6 \cdot 10^{-27} $ & $ $1.1$ \cdot 10^{-26} $ & $1.8 \cdot 10^{-8}$ & $0.087$ \\ 
J1745$-$0952  & $103.2$ & $0.2$ & $7.5 \cdot 10^{-27} $ & $ $5.7$ \cdot 10^{-27} $ & $1.2 \cdot 10^{-7}$ & $0.82$ \\ 
J1745+1017  & $754.1$ & $1.2$ & $6.0 \cdot 10^{-28} $ & $ $1.1$ \cdot 10^{-26} $ & $2.2 \cdot 10^{-8}$ & $0.38$ \\ 
J1747$-$4036  & $1215.4$ & $7.2$ & $2.9 \cdot 10^{-28} $ & $ $2.1$ \cdot 10^{-26} $ & $9.8 \cdot 10^{-8}$ & $0.036$ \\ 
J1748$-$2446A & $173.0$ & $6.9$ & $3.3 \cdot 10^{-28} $ & $ $5.3$ \cdot 10^{-27} $ & $1.2 \cdot 10^{-6}$ & $0.76$ \\ 
J1748$-$3009  & $206.5$ & $5.0$ & - &  $8.0 \cdot 10^{-27}$ & $9.0 \cdot 10^{-7}$ & $0.13$ \\ 
J1750$-$2536  & $57.6$ & $3.2$ & $5.1 \cdot 10^{-28} $ & $ $8.9$ \cdot 10^{-27} $ & $8.2 \cdot 10^{-6}$ & $0.74$ \\ 
J1751$-$2857  & $510.9$ & $1.1$ & $1.2 \cdot 10^{-27} $ & $ $6.3$ \cdot 10^{-26} $ & $2.5 \cdot 10^{-7}$ & $0.088$ \\ 
J1753$-$1914  & $31.8$ & $2.9$ & $1.6 \cdot 10^{-27} $ & $ $2.6$ \cdot 10^{-26} $ & $7.2 \cdot 10^{-5}$ & $0.56$ \\ 
J1753$-$2240  & $21.0$ & $3.2$ & $8.0 \cdot 10^{-28} $ & $ $1.7$ \cdot 10^{-25} $ & $1.2 \cdot 10^{-3}$ & $0.33$ \\ 
J1755$-$3716  & $156.4$ & $8.2$ & $1.5 \cdot 10^{-28} $ & $ $6.4$ \cdot 10^{-27} $ & $2.0 \cdot 10^{-6}$ & $0.41$ \\ 
J1756$-$2251  & $70.3$ & $0.7$ & $6.6 \cdot 10^{-27} $ & $ $8.8$ \cdot 10^{-27} $ & $1.2 \cdot 10^{-6}$ & $0.41$ \\ 
J1757$-$1854  & $93.0$ & $19.6$ & $4.6 \cdot 10^{-28} $ & $ $1.2$ \cdot 10^{-26} $ & $2.5 \cdot 10^{-5}$ & $0.018$ \\ 
J1757$-$5322  & $225.5$ & $0.9$ & $1.5 \cdot 10^{-27} $ & $ $6.0$ \cdot 10^{-27} $ & $1.0 \cdot 10^{-7}$ & $0.45$ \\ 
J1801$-$1417  & $551.7$ & $1.1$ & $7.4 \cdot 10^{-28} $ & $ $8.9$ \cdot 10^{-27} $ & $3.0 \cdot 10^{-8}$ & $0.41$ \\ 
J1801$-$3210  & $268.3$ & $6.1$ & - &  $6.2 \cdot 10^{-27}$ & $5.0 \cdot 10^{-7}$ & $0.52$ \\ 
J1802$-$2124  & $158.1$ & $0.8$ & $2.5 \cdot 10^{-27} $ & $ $6.4$ \cdot 10^{-27} $ & $1.8 \cdot 10^{-7}$ & $0.49$ \\ 
J1804$-$0735  & $86.6$ & $7.8$ & $2.9 \cdot 10^{-28} $ & $ $7.8$ \cdot 10^{-27} $ & $7.7 \cdot 10^{-6}$ & $0.39$ \\ 
J1804$-$2717  & $214.1$ & $0.8$ & $1.9 \cdot 10^{-27} $ & $ $6.1$ \cdot 10^{-27} $ & $1.0 \cdot 10^{-7}$ & $0.51$ \\ 
J1807$-$2459A & $653.7$ & $3.0$ & $7.6 \cdot 10^{-28} $ & $ $9.4$ \cdot 10^{-27} $ & $6.3 \cdot 10^{-8}$ & $0.41$ \\ 
J1809$-$1917  & $24.2$ & $3.3$ & $1.4 \cdot 10^{-25} $ & $ $5.6$ \cdot 10^{-26} $ & $3.0 \cdot 10^{-4}$ & $0.85$ \\ 
J1810+1744  & $1202.8$ & $2.4$ & $5.5 \cdot 10^{-28} $ & $ $1.2$ \cdot 10^{-26} $ & $1.8 \cdot 10^{-8}$ & $0.82$ \\ 
J1810$-$2005  & $60.9$ & $3.5$ & $2.9 \cdot 10^{-28} $ & $ $8.5$ \cdot 10^{-27} $ & $7.6 \cdot 10^{-6}$ & $0.69$ \\ 
J1811$-$2405  & $751.7$ & $1.8$ & $9.9 \cdot 10^{-28} $ & $ $7.2$ \cdot 10^{-27} $ & $2.2 \cdot 10^{-8}$ & $0.98$ \\ 
J1813$-$1749  & $44.7$ & $6.2$ & $2.2 \cdot 10^{-25} $ & $ $1.1$ \cdot 10^{-26} $ & $3.1 \cdot 10^{-5}$ & $0.89$ \\ 
J1813$-$2621  & $451.5$ & $3.2$ & - &  $5.4 \cdot 10^{-27}$ & $8.0 \cdot 10^{-8}$ & $0.98$ \\ 
J1821+0155  & $59.2$ & $1.7$ & $4.3 \cdot 10^{-28} $ & $ $9.1$ \cdot 10^{-27} $ & $4.2 \cdot 10^{-6}$ & $0.92$ \\ 
J1823$-$3021A & $367.6$ & $7.9$ & $2.9 \cdot 10^{-28} $ & $ $8.6$ \cdot 10^{-27} $ & $4.8 \cdot 10^{-7}$ & $0.15$ \\ 
J1824$-$2452A & $654.8$ & $5.5$ & $4.1 \cdot 10^{-28} $ & $ $8.7$ \cdot 10^{-27} $ & $1.1 \cdot 10^{-7}$ & $0.55$ \\ 
J1825$-$0319  & $439.2$ & $3.9$ & $2.6 \cdot 10^{-28} $ & $ $5.5$ \cdot 10^{-27} $ & $1.0 \cdot 10^{-7}$ & $0.98$ \\ 
J1826$-$2415  & $425.9$ & $2.7$ & $5.7 \cdot 10^{-28} $ & $ $7.0$ \cdot 10^{-27} $ & $10.0 \cdot 10^{-8}$ & $0.53$ \\ 
J1828$-$1101  & $27.8$ & $4.8$ & $7.7 \cdot 10^{-26} $ & $ $3.2$ \cdot 10^{-26} $ & $1.9 \cdot 10^{-4}$ & $0.91$ \\ 
J1829+2456  & $48.8$ & $0.9$ & $10.0 \cdot 10^{-28} $ & $ $1.4$ \cdot 10^{-26} $ & $5.2 \cdot 10^{-6}$ & $0.57$ \\ 
J1832$-$0836  & $735.5$ & $1.6$ & - &  $1.4 \cdot 10^{-26}$ & $3.8 \cdot 10^{-8}$ & $0.093$ \\ 
J1833$-$0827  & $23.4$ & $4.5$ & $5.9 \cdot 10^{-26} $ & $ $6.1$ \cdot 10^{-26} $ & $4.7 \cdot 10^{-4}$ & $0.91$ \\ 
J1835$-$0114  & $390.9$ & $3.5$ & $2.7 \cdot 10^{-28} $ & $ $6.4$ \cdot 10^{-27} $ & $1.4 \cdot 10^{-7}$ & $0.71$ \\ 
J1838$-$0655  & $28.4$ & $6.6$ & $1.0 \cdot 10^{-25} $ & $ $3.4$ \cdot 10^{-26} $ & $2.6 \cdot 10^{-4}$ & $0.72$ \\ 
J1840$-$0643  & $56.2$ & $5.0$ & $2.6 \cdot 10^{-28} $ & $ $8.5$ \cdot 10^{-27} $ & $1.3 \cdot 10^{-5}$ & $0.82$ \\ 
J1841+0130  & $67.2$ & $4.2$ & $3.2 \cdot 10^{-27} $ & $ $1.2$ \cdot 10^{-26} $ & $1.1 \cdot 10^{-5}$ & $0.037$ \\ 
J1843$-$1113  & $1083.6$ & $1.3$ & $1.4 \cdot 10^{-27} $ & $ $2.2$ \cdot 10^{-26} $ & $2.2 \cdot 10^{-8}$ & $0.019$ \\ 
J1843$-$1448  & $365.5$ & $3.5$ & - &  $8.3 \cdot 10^{-27}$ & $2.0 \cdot 10^{-7}$ & $0.23$ \\ 
J1849$-$0001  & $51.9$ & $7.0$ & $7.0 \cdot 10^{-26} $ & $ $1.5$ \cdot 10^{-26} $ & $3.8 \cdot 10^{-5}$ & $0.08$ \\ 
J1853+1303  & $488.8$ & $1.3$ & $8.9 \cdot 10^{-28} $ & $ $1.2$ \cdot 10^{-26} $ & $6.1 \cdot 10^{-8}$ & $0.058$ \\ 
J1856+0245  & $24.7$ & $6.3$ & $1.1 \cdot 10^{-25} $ & $ $5.7$ \cdot 10^{-26} $ & $5.6 \cdot 10^{-4}$ & $0.69$ \\ 
J1857+0943  & $373.0$ & $1.2$ & $1.2 \cdot 10^{-27} $ & $ $7.2$ \cdot 10^{-27} $ & $5.8 \cdot 10^{-8}$ & $0.48$ \\ 
J1902$-$5105  & $1147.8$ & $1.6$ & $1.1 \cdot 10^{-27} $ & $ $1.3$ \cdot 10^{-26} $ & $1.5 \cdot 10^{-8}$ & $0.59$ \\ 
J1903+0327  & $930.3$ & $6.1$ & $3.9 \cdot 10^{-28} $ & $ $1.4$ \cdot 10^{-26} $ & $9.3 \cdot 10^{-8}$ & $0.24$ \\ 
J1903$-$7051  & $555.9$ & $0.9$ & $1.3 \cdot 10^{-27} $ & $ $6.7$ \cdot 10^{-27} $ & $1.9 \cdot 10^{-8}$ & $0.77$ \\ 
J1904+0412  & $28.1$ & $4.6$ & $2.2 \cdot 10^{-28} $ & $ $2.9$ \cdot 10^{-26} $ & $1.6 \cdot 10^{-4}$ & $0.98$ \\ 
J1905+0400  & $528.5$ & $1.1$ & $8.0 \cdot 10^{-28} $ & $ $7.3$ \cdot 10^{-27} $ & $2.6 \cdot 10^{-8}$ & $0.68$ \\ 
J1909$-$3744  & $678.6$ & $1.1$ & $6.6 \cdot 10^{-28} $ & $ $8.1$ \cdot 10^{-27} $ & $1.9 \cdot 10^{-8}$ & $0.64$ \\ 
J1910+1256  & $401.3$ & $1.5$ & $7.2 \cdot 10^{-28} $ & $ $8.1$ \cdot 10^{-27} $ & $7.2 \cdot 10^{-8}$ & $0.29$ \\ 
J1911$-$1114  & $551.6$ & $1.1$ & $1.3 \cdot 10^{-27} $ & $ $1.2$ \cdot 10^{-26} $ & $4.0 \cdot 10^{-8}$ & $0.067$ \\ 
J1911+1347  & $432.3$ & $1.4$ & $1.1 \cdot 10^{-27} $ & $ $7.1$ \cdot 10^{-27} $ & $4.9 \cdot 10^{-8}$ & $0.54$ \\ 
J1913+1011  & $55.7$ & $4.6$ & $5.3 \cdot 10^{-26} $ & $ $8.6$ \cdot 10^{-27} $ & $1.2 \cdot 10^{-5}$ & $0.82$ \\ 
J1914+0659  & $108.0$ & $8.5$ & $1.2 \cdot 10^{-28} $ & $ $8.1$ \cdot 10^{-27} $ & $5.6 \cdot 10^{-6}$ & $0.23$ \\ 
J1915+1606  & $33.9$ & $5.2$ & $1.9 \cdot 10^{-27} $ & $ $2.6$ \cdot 10^{-26} $ & $1.1 \cdot 10^{-4}$ & $0.21$ \\ 
J1918$-$0642  & $261.6$ & $1.1$ & $1.3 \cdot 10^{-27} $ & $ $6.7$ \cdot 10^{-27} $ & $1.0 \cdot 10^{-7}$ & $0.45$ \\ 
J1923+2515  & $528.0$ & $1.2$ & $9.1 \cdot 10^{-28} $ & $ $1.0$ \cdot 10^{-26} $ & $4.1 \cdot 10^{-8}$ & $0.15$ \\ 
J1925+1720  & $26.4$ & $5.1$ & $5.9 \cdot 10^{-26} $ & $ $7.0$ \cdot 10^{-26} $ & $4.8 \cdot 10^{-4}$ & $0.039$ \\ 
J1928+1746  & $29.1$ & $4.3$ & $8.2 \cdot 10^{-26} $ & $ $3.4$ \cdot 10^{-26} $ & $1.6 \cdot 10^{-4}$ & $0.54$ \\ 
J1933$-$6211  & $564.4$ & $0.7$ & $1.1 \cdot 10^{-27} $ & $ $6.9$ \cdot 10^{-27} $ & $1.3 \cdot 10^{-8}$ & $0.66$ \\ 
J1935+2025  & $25.0$ & $4.6$ & $1.5 \cdot 10^{-25} $ & $ $4.9$ \cdot 10^{-26} $ & $3.4 \cdot 10^{-4}$ & $0.66$ \\ 
J1939+2134  & $1283.9$ & $4.8$ & $1.4 \cdot 10^{-27} $ & $ $1.4$ \cdot 10^{-26} $ & $4.0 \cdot 10^{-8}$ & $0.59$ \\ 
J1943+2210  & $393.4$ & $6.8$ & $1.6 \cdot 10^{-28} $ & $ $6.4$ \cdot 10^{-27} $ & $2.6 \cdot 10^{-7}$ & $0.62$ \\ 
J1944+0907  & $385.7$ & $1.2$ & $7.4 \cdot 10^{-28} $ & $ $6.1$ \cdot 10^{-27} $ & $4.8 \cdot 10^{-8}$ & $0.75$ \\ 
J1946+3417  & $630.9$ & $6.9$ & $6.5 \cdot 10^{-29} $ & $ $9.3$ \cdot 10^{-27} $ & $1.5 \cdot 10^{-7}$ & $0.34$ \\ 
J1946$-$5403  & $737.8$ & $1.1$ & $7.0 \cdot 10^{-28} $ & $ $7.1$ \cdot 10^{-27} $ & $1.4 \cdot 10^{-8}$ & $0.89$ \\ 
J1949+3106  & $152.2$ & $7.5$ & $2.9 \cdot 10^{-28} $ & $ $5.8$ \cdot 10^{-27} $ & $1.8 \cdot 10^{-6}$ & $0.67$ \\ 
J1950+2414  & $464.6$ & $7.3$ & $2.4 \cdot 10^{-28} $ & $ $6.8$ \cdot 10^{-27} $ & $2.2 \cdot 10^{-7}$ & $0.67$ \\ 
J1952+3252  & $50.6$ & $3.0$ & $1.0 \cdot 10^{-25} $ & $ $1.3$ \cdot 10^{-26} $ & $1.4 \cdot 10^{-5}$ & $0.22$ \\ 
J1955+2527  & $410.4$ & $8.2$ & $1.5 \cdot 10^{-28} $ & $ $7.7$ \cdot 10^{-27} $ & $3.5 \cdot 10^{-7}$ & $0.36$ \\ 
J1955+2908  & $326.1$ & $6.3$ & $2.9 \cdot 10^{-28} $ & $ $6.6$ \cdot 10^{-27} $ & $3.7 \cdot 10^{-7}$ & $0.46$ \\ 
J2007+2722  & $81.6$ & $7.1$ & $7.1 \cdot 10^{-28} $ & $ $1.1$ \cdot 10^{-26} $ & $1.1 \cdot 10^{-5}$ & $0.037$ \\ 
J2010$-$1323  & $382.9$ & $1.2$ & $5.3 \cdot 10^{-28} $ & $ $8.4$ \cdot 10^{-27} $ & $6.3 \cdot 10^{-8}$ & $0.22$ \\ 
J2017+0603  & $690.6$ & $1.4$ & $9.6 \cdot 10^{-28} $ & $ $8.9$ \cdot 10^{-27} $ & $2.5 \cdot 10^{-8}$ & $0.6$ \\ 
J2019+2425  & $508.3$ & $1.2$ & $4.4 \cdot 10^{-28} $ & $ $5.4$ \cdot 10^{-26} $ & $2.3 \cdot 10^{-7}$ & $0.046$ \\ 
J2022+2534  & $755.9$ & - & - & $1.2 \cdot 10^{-26}$  & -  & $0.2$ \\ 
J2033+1734  & $336.2$ & $1.7$ & $5.5 \cdot 10^{-28} $ & $ $5.2$ \cdot 10^{-27} $ & $7.6 \cdot 10^{-8}$ & $0.92$ \\ 
J2039$-$3616  & $610.7$ & $1.7$ & $7.6 \cdot 10^{-28} $ & $ $7.6$ \cdot 10^{-27} $ & $3.3 \cdot 10^{-8}$ & $0.65$ \\ 
J2043+2740  & $20.8$ & $1.5$ & $6.3 \cdot 10^{-26} $ & $ $1.7$ \cdot 10^{-25} $ & $5.6 \cdot 10^{-4}$ & $0.32$ \\ 
J2045+3633  & $63.1$ & $5.6$ & $6.2 \cdot 10^{-28} $ & $ $9.1$ \cdot 10^{-27} $ & $1.2 \cdot 10^{-5}$ & $0.34$ \\ 
J2047+1053  & $466.6$ & $2.8$ & $6.4 \cdot 10^{-28} $ & $ $7.0$ \cdot 10^{-27} $ & $8.5 \cdot 10^{-8}$ & $0.69$ \\ 
J2053+4650  & $158.9$ & $3.8$ & $7.8 \cdot 10^{-28} $ & $ $4.6$ \cdot 10^{-27} $ & $6.6 \cdot 10^{-7}$ & $0.92$ \\ 
J2055+3829  & $957.3$ & $4.6$ & $1.1 \cdot 10^{-28} $ & $ $1.2$ \cdot 10^{-26} $ & $5.8 \cdot 10^{-8}$ & $0.36$ \\ 
J2124$-$3358  & $405.6$ & $0.4$ & $2.3 \cdot 10^{-27} $ & $ $8.0$ \cdot 10^{-27} $ & $2.0 \cdot 10^{-8}$ & $0.28$ \\ 
J2129$-$5721  & $536.7$ & $7.0$ & $2.6 \cdot 10^{-28} $ & $ $7.8$ \cdot 10^{-27} $ & $1.8 \cdot 10^{-7}$ & $0.41$ \\ 
J2144$-$5237  & $396.7$ & $1.7$ & $6.5 \cdot 10^{-28} $ & $ $6.0$ \cdot 10^{-27} $ & $6.0 \cdot 10^{-8}$ & $0.63$ \\ 
J2145$-$0750  & $124.6$ & $0.8$ & $1.3 \cdot 10^{-27} $ & $ $6.5$ \cdot 10^{-27} $ & $3.3 \cdot 10^{-7}$ & $0.52$ \\ 
J2150$-$0326  & $569.7$ & - & - & $8.9 \cdot 10^{-27}$  & -  & $0.46$ \\ 
J2205+6012  & $828.0$ & $3.5$ & $6.5 \cdot 10^{-28} $ & $ $8.9$ \cdot 10^{-27} $ & $4.3 \cdot 10^{-8}$ & $0.7$ \\ 
J2214+3000  & $641.2$ & $0.6$ & $2.7 \cdot 10^{-27} $ & $ $6.8$ \cdot 10^{-27} $ & $9.3 \cdot 10^{-9}$ & $0.92$ \\ 
J2222$-$0137  & $60.9$ & $0.3$ & $2.0 \cdot 10^{-27} $ & $ $7.2$ \cdot 10^{-27} $ & $5.0 \cdot 10^{-7}$ & $0.91$ \\ 
J2229+2643  & $671.6$ & $1.8$ & $3.1 \cdot 10^{-28} $ & $ $1.1$ \cdot 10^{-26} $ & $4.2 \cdot 10^{-8}$ & $0.19$ \\ 
J2229+6114  & $38.7$ & $3.0$ & $3.3 \cdot 10^{-25} $ & $ $1.3$ \cdot 10^{-26} $ & $2.4 \cdot 10^{-5}$ & $0.8$ \\ 
J2234+0611  & $559.2$ & $1.5$ & $8.2 \cdot 10^{-28} $ & $ $1.3$ \cdot 10^{-26} $ & $5.9 \cdot 10^{-8}$ & $0.071$ \\ 
J2234+0944  & $551.4$ & $0.8$ & $1.9 \cdot 10^{-27} $ & $ $7.6$ \cdot 10^{-27} $ & $1.9 \cdot 10^{-8}$ & $0.68$ \\ 
J2235+1506  & $33.5$ & $1.5$ & $6.5 \cdot 10^{-28} $ & $ $2.3$ \cdot 10^{-26} $ & $2.9 \cdot 10^{-5}$ & $0.53$ \\ 
J2236$-$5527  & $289.5$ & $2.0$ & $4.6 \cdot 10^{-28} $ & $ $5.0$ \cdot 10^{-27} $ & $1.2 \cdot 10^{-7}$ & $0.78$ \\ 
J2241$-$5236  & $914.6$ & $1.1$ & $1.2 \cdot 10^{-27} $ & $ $1.4$ \cdot 10^{-26} $ & $1.7 \cdot 10^{-8}$ & $0.097$ \\ 
J2256$-$1024  & $871.6$ & $1.3$ & $1.3 \cdot 10^{-27} $ & $ $1.6$ \cdot 10^{-26} $ & $2.7 \cdot 10^{-8}$ & $0.057$ \\ 
J2302+4442  & $385.2$ & $0.9$ & $1.5 \cdot 10^{-27} $ & $ $7.0$ \cdot 10^{-27} $ & $3.9 \cdot 10^{-8}$ & $0.39$ \\ 
J2317+1439  & $580.5$ & $2.2$ & $3.3 \cdot 10^{-28} $ & $ $7.5$ \cdot 10^{-27} $ & $4.6 \cdot 10^{-8}$ & $0.72$ \\ 
J2322+2057  & $415.9$ & $1.0$ & $6.5 \cdot 10^{-28} $ & $ $6.2$ \cdot 10^{-27} $ & $3.4 \cdot 10^{-8}$ & $0.77$ \\ 
J2322$-$2650  & $577.5$ & $0.2$ & $1.2 \cdot 10^{-27} $ & $ $7.0$ \cdot 10^{-27} $ & $4.6 \cdot 10^{-9}$ & $0.78$ \\

\hline
\end{longtable*}

\bibliography{biblio}

\providecommand{\noopsort}[1]{}\providecommand{\singleletter}[1]{#1}%
\begin{thebibliography}{43}%
\makeatletter
\providecommand \@ifxundefined [1]{%
 \@ifx{#1\undefined}
}%
\providecommand \@ifnum [1]{%
 \ifnum #1\expandafter \@firstoftwo
 \else \expandafter \@secondoftwo
 \fi
}%
\providecommand \@ifx [1]{%
 \ifx #1\expandafter \@firstoftwo
 \else \expandafter \@secondoftwo
 \fi
}%
\providecommand \natexlab [1]{#1}%
\providecommand \enquote  [1]{``#1''}%
\providecommand \bibnamefont  [1]{#1}%
\providecommand \bibfnamefont [1]{#1}%
\providecommand \citenamefont [1]{#1}%
\providecommand \href@noop [0]{\@secondoftwo}%
\providecommand \href [0]{\begingroup \@sanitize@url \@href}%
\providecommand \@href[1]{\@@startlink{#1}\@@href}%
\providecommand \@@href[1]{\endgroup#1\@@endlink}%
\providecommand \@sanitize@url [0]{\catcode `\\12\catcode `\$12\catcode
  `\&12\catcode `\#12\catcode `\^12\catcode `\_12\catcode `\%12\relax}%
\providecommand \@@startlink[1]{}%
\providecommand \@@endlink[0]{}%
\providecommand \url  [0]{\begingroup\@sanitize@url \@url }%
\providecommand \@url [1]{\endgroup\@href {#1}{\urlprefix }}%
\providecommand \urlprefix  [0]{URL }%
\providecommand \Eprint [0]{\href }%
\providecommand \doibase [0]{https://doi.org/}%
\providecommand \selectlanguage [0]{\@gobble}%
\providecommand \bibinfo  [0]{\@secondoftwo}%
\providecommand \bibfield  [0]{\@secondoftwo}%
\providecommand \translation [1]{[#1]}%
\providecommand \BibitemOpen [0]{}%
\providecommand \bibitemStop [0]{}%
\providecommand \bibitemNoStop [0]{.\EOS\space}%
\providecommand \EOS [0]{\spacefactor3000\relax}%
\providecommand \BibitemShut  [1]{\csname bibitem#1\endcsname}%
\let\auto@bib@innerbib\@empty
\bibitem [{\citenamefont {{The LIGO Scientific Collaboration}}(2015)}]{LIGO}%
  \BibitemOpen
  \bibfield  {author} {\bibinfo {author} {\bibnamefont {{The LIGO Scientific
  Collaboration}}},\ }\bibfield  {title} {\bibinfo {title} {Advanced ligo},\
  }\href {https://doi.org/10.1088/0264-9381/32/7/074001} {\bibfield  {journal}
  {\bibinfo  {journal} {Classical and Quantum Gravity}\ }\textbf {\bibinfo
  {volume} {32}},\ \bibinfo {pages} {074001} (\bibinfo {year}
  {2015})}\BibitemShut {NoStop}%
\bibitem [{\citenamefont {Acernese}\ and\ \citenamefont {al.}(2014)}]{virgo}%
  \BibitemOpen
  \bibfield  {author} {\bibinfo {author} {\bibfnamefont {F.}~\bibnamefont
  {Acernese}}\ and\ \bibinfo {author} {\bibnamefont {al.}},\ }\bibfield
  {title} {\bibinfo {title} {Advanced virgo: a second-generation
  interferometric gravitational wave detector},\ }\href
  {https://doi.org/10.1088/0264-9381/32/2/024001} {\bibfield  {journal}
  {\bibinfo  {journal} {Classical and Quantum Gravity}\ }\textbf {\bibinfo
  {volume} {32}},\ \bibinfo {pages} {024001} (\bibinfo {year}
  {2014})}\BibitemShut {NoStop}%
\bibitem [{\citenamefont {{The LIGO Scientific Collaboration and The Virgo
  Collaboration}}(2014)}]{1era}%
  \BibitemOpen
  \bibfield  {author} {\bibinfo {author} {\bibnamefont {{The LIGO Scientific
  Collaboration and The Virgo Collaboration}}},\ }\bibfield  {title} {\bibinfo
  {title} {Gravitational waves from known pulsars: Results from the initial
  detector era},\ }\href {https://doi.org/10.1088/0004-637X/785/2/119}
  {\bibfield  {journal} {\bibinfo  {journal} {The Astrophysical Journal}\
  }\textbf {\bibinfo {volume} {785}},\ \bibinfo {pages} {119} (\bibinfo {year}
  {2014})}\BibitemShut {NoStop}%
\bibitem [{\citenamefont {Glampedakis}\ and\ \citenamefont
  {Gualtieri}(2018)}]{review}%
  \BibitemOpen
  \bibfield  {author} {\bibinfo {author} {\bibfnamefont {K.}~\bibnamefont
  {Glampedakis}}\ and\ \bibinfo {author} {\bibfnamefont {L.}~\bibnamefont
  {Gualtieri}},\ }\bibinfo {title} {Gravitational waves from single neutron
  stars: An advanced detector era survey},\ in\ \href
  {https://doi.org/10.1007/978-3-319-97616-7_12} {\emph {\bibinfo {booktitle}
  {The Physics and Astrophysics of Neutron Stars}}},\ \bibinfo {editor} {edited
  by\ \bibinfo {editor} {\bibfnamefont {L.}~\bibnamefont {Rezzolla}}, \bibinfo
  {editor} {\bibfnamefont {P.}~\bibnamefont {Pizzochero}}, \bibinfo {editor}
  {\bibfnamefont {D.~I.}\ \bibnamefont {Jones}}, \bibinfo {editor}
  {\bibfnamefont {N.}~\bibnamefont {Rea}},\ and\ \bibinfo {editor}
  {\bibfnamefont {I.}~\bibnamefont {Vida{\~{n}}a}}}\ (\bibinfo  {publisher}
  {Springer International Publishing},\ \bibinfo {address} {Cham},\ \bibinfo
  {year} {2018})\BibitemShut {NoStop}%
\bibitem [{\citenamefont {Jaranowski}\ \emph {et~al.}(1998)\citenamefont
  {Jaranowski}, \citenamefont {Kr\'olak},\ and\ \citenamefont {Schutz}}]{JKS}%
  \BibitemOpen
  \bibfield  {author} {\bibinfo {author} {\bibfnamefont {P.}~\bibnamefont
  {Jaranowski}}, \bibinfo {author} {\bibfnamefont {A.}~\bibnamefont
  {Kr\'olak}},\ and\ \bibinfo {author} {\bibfnamefont {B.~F.}\ \bibnamefont
  {Schutz}},\ }\bibfield  {title} {\bibinfo {title} {{Data analysis of
  gravitational-wave signals from spinning neutron stars: The signal and its
  detection}},\ }\href {https://doi.org/10.1103/PhysRevD.58.063001} {\bibfield
  {journal} {\bibinfo  {journal} {Phys. Rev. D}\ }\textbf {\bibinfo {volume}
  {58}},\ \bibinfo {pages} {063001} (\bibinfo {year} {1998})}\BibitemShut
  {NoStop}%
\bibitem [{\citenamefont {Jones}(2021)}]{cwemission}%
  \BibitemOpen
  \bibfield  {author} {\bibinfo {author} {\bibfnamefont {D.~I.}\ \bibnamefont
  {Jones}},\ }\href@noop {} {\bibinfo {title} {{Learning from the Frequency
  Content of Continuous Gravitational Wave Signals}}} (\bibinfo {year}
  {2021}),\ \Eprint {https://arxiv.org/abs/2111.08561} {arXiv:2111.08561
  [astro-ph.HE]} \BibitemShut {NoStop}%
\bibitem [{\citenamefont {Piccinni}(2022)}]{ornella}%
  \BibitemOpen
  \bibfield  {author} {\bibinfo {author} {\bibfnamefont {O.~J.}\ \bibnamefont
  {Piccinni}},\ }\bibfield  {title} {\bibinfo {title} {Status and perspectives
  of continuous gravitational wave searches},\ }\bibfield  {journal} {\bibinfo
  {journal} {Galaxies}\ }\textbf {\bibinfo {volume} {10}},\ \href
  {https://doi.org/10.3390/galaxies10030072} {10.3390/galaxies10030072}
  (\bibinfo {year} {2022})\BibitemShut {NoStop}%
\bibitem [{\citenamefont {et~al.}(2017)}]{firstsearch}%
  \BibitemOpen
  \bibfield  {author} {\bibinfo {author} {\bibfnamefont {B.~P.~A.}\
  \bibnamefont {et~al.}},\ }\bibfield  {title} {\bibinfo {title} {First search
  for gravitational waves from known pulsars with advanced ligo},\ }\href
  {https://doi.org/10.3847/1538-4357/aa677f} {\bibfield  {journal} {\bibinfo
  {journal} {The Astrophysical Journal}\ }\textbf {\bibinfo {volume} {839}},\
  \bibinfo {pages} {12} (\bibinfo {year} {2017})}\BibitemShut {NoStop}%
\bibitem [{\citenamefont {{The LIGO Scientific Collaboration and the Virgo
  Collaboration and the KAGRA Collaboration }}(2022)}]{O3targ}%
  \BibitemOpen
  \bibfield  {author} {\bibinfo {author} {\bibnamefont {{The LIGO Scientific
  Collaboration and the Virgo Collaboration and the KAGRA Collaboration }}},\
  }\bibfield  {title} {\bibinfo {title} {Searches for gravitational waves from
  known pulsars at two harmonics in the second and third ligo-virgo observing
  runs},\ }\href@noop {} {\bibfield  {journal} {\bibinfo  {journal} {The
  Astrophysical Journal Letters}\ }\textbf {\bibinfo {volume} {935}} (\bibinfo
  {year} {2022})}\BibitemShut {NoStop}%
\bibitem [{\citenamefont {Astone}\ \emph {et~al.}(2010)\citenamefont {Astone},
  \citenamefont {D'Antonio}, \citenamefont {Frasca},\ and\ \citenamefont
  {Palomba}}]{2010}%
  \BibitemOpen
  \bibfield  {author} {\bibinfo {author} {\bibfnamefont {P.}~\bibnamefont
  {Astone}}, \bibinfo {author} {\bibfnamefont {S.}~\bibnamefont {D'Antonio}},
  \bibinfo {author} {\bibfnamefont {S.}~\bibnamefont {Frasca}},\ and\ \bibinfo
  {author} {\bibfnamefont {C.}~\bibnamefont {Palomba}},\ }\bibfield  {title}
  {\bibinfo {title} {A method for detection of known sources of continuous
  gravitational wave signals in non-stationary data},\ }\href
  {https://doi.org/10.1088/0264-9381/27/19/194016} {\bibfield  {journal}
  {\bibinfo  {journal} {Classical and Quantum Gravity}\ }\textbf {\bibinfo
  {volume} {27}},\ \bibinfo {pages} {194016} (\bibinfo {year}
  {2010})}\BibitemShut {NoStop}%
\bibitem [{\citenamefont {Astone}\ \emph {et~al.}(2014)\citenamefont {Astone},
  \citenamefont {Colla}, \citenamefont {D'Antonio}, \citenamefont {Frasca},
  \citenamefont {Palomba},\ and\ \citenamefont {Serafinelli}}]{2014}%
  \BibitemOpen
  \bibfield  {author} {\bibinfo {author} {\bibfnamefont {P.}~\bibnamefont
  {Astone}}, \bibinfo {author} {\bibfnamefont {A.}~\bibnamefont {Colla}},
  \bibinfo {author} {\bibfnamefont {S.}~\bibnamefont {D'Antonio}}, \bibinfo
  {author} {\bibfnamefont {S.}~\bibnamefont {Frasca}}, \bibinfo {author}
  {\bibfnamefont {C.}~\bibnamefont {Palomba}},\ and\ \bibinfo {author}
  {\bibfnamefont {R.}~\bibnamefont {Serafinelli}},\ }\bibfield  {title}
  {\bibinfo {title} {Method for narrow-band search of continuous gravitational
  wave signals},\ }\href {https://doi.org/10.1103/PhysRevD.89.062008}
  {\bibfield  {journal} {\bibinfo  {journal} {Phys. Rev. D}\ }\textbf {\bibinfo
  {volume} {89}},\ \bibinfo {pages} {062008} (\bibinfo {year}
  {2014})}\BibitemShut {NoStop}%
\bibitem [{\citenamefont {Pitkin}\ \emph {et~al.}(2018)\citenamefont {Pitkin},
  \citenamefont {Messenger},\ and\ \citenamefont {Fan}}]{ensbayes}%
  \BibitemOpen
  \bibfield  {author} {\bibinfo {author} {\bibfnamefont {M.}~\bibnamefont
  {Pitkin}}, \bibinfo {author} {\bibfnamefont {C.}~\bibnamefont {Messenger}},\
  and\ \bibinfo {author} {\bibfnamefont {X.}~\bibnamefont {Fan}},\ }\bibfield
  {title} {\bibinfo {title} {{Hierarchical Bayesian method for detecting
  continuous gravitational waves from an ensemble of pulsars}},\ }\href
  {https://doi.org/10.1103/PhysRevD.98.063001} {\bibfield  {journal} {\bibinfo
  {journal} {Phys. Rev. D}\ }\textbf {\bibinfo {volume} {98}},\ \bibinfo
  {pages} {063001} (\bibinfo {year} {2018})}\BibitemShut {NoStop}%
\bibitem [{\citenamefont {Fan}\ \emph {et~al.}(2016)\citenamefont {Fan},
  \citenamefont {Chen},\ and\ \citenamefont {Messenger}}]{Fstat2}%
  \BibitemOpen
  \bibfield  {author} {\bibinfo {author} {\bibfnamefont {X.}~\bibnamefont
  {Fan}}, \bibinfo {author} {\bibfnamefont {Y.}~\bibnamefont {Chen}},\ and\
  \bibinfo {author} {\bibfnamefont {C.}~\bibnamefont {Messenger}},\ }\bibfield
  {title} {\bibinfo {title} {Method to detect gravitational waves from an
  ensemble of known pulsars},\ }\href
  {https://doi.org/10.1103/PhysRevD.94.084029} {\bibfield  {journal} {\bibinfo
  {journal} {Physical Review D}\ }\textbf {\bibinfo {volume} {94}} (\bibinfo
  {year} {2016})}\BibitemShut {NoStop}%
\bibitem [{\citenamefont {Buono}\ \emph {et~al.}(2021)\citenamefont {Buono},
  \citenamefont {De~Rosa}, \citenamefont {D'Onofrio}, \citenamefont {Errico},
  \citenamefont {Palomba}, \citenamefont {Piccinni},\ and\ \citenamefont
  {Sequino}}]{mio}%
  \BibitemOpen
  \bibfield  {author} {\bibinfo {author} {\bibfnamefont {M.}~\bibnamefont
  {Buono}}, \bibinfo {author} {\bibfnamefont {R.}~\bibnamefont {De~Rosa}},
  \bibinfo {author} {\bibfnamefont {L.}~\bibnamefont {D'Onofrio}}, \bibinfo
  {author} {\bibfnamefont {L.}~\bibnamefont {Errico}}, \bibinfo {author}
  {\bibfnamefont {C.}~\bibnamefont {Palomba}}, \bibinfo {author} {\bibfnamefont
  {O.}~\bibnamefont {Piccinni}},\ and\ \bibinfo {author} {\bibfnamefont
  {V.}~\bibnamefont {Sequino}},\ }\bibfield  {title} {\bibinfo {title} {A
  method for detecting continuous gravitational wave signals from an ensemble
  of known pulsars},\ }\bibfield  {journal} {\bibinfo  {journal} {Classical and
  Quantum Gravity}\ }\textbf {\bibinfo {volume} {38}},\ \href
  {https://doi.org/10.1088/1361-6382/abf1c0} {10.1088/1361-6382/abf1c0}
  (\bibinfo {year} {2021})\BibitemShut {NoStop}%
\bibitem [{\citenamefont {D'Onofrio}\ \emph {et~al.}(2022)\citenamefont
  {D'Onofrio}, \citenamefont {De~Rosa}, \citenamefont {Errico}, \citenamefont
  {Palomba}, \citenamefont {Sequino},\ and\ \citenamefont {Trozzo}}]{mio2}%
  \BibitemOpen
  \bibfield  {author} {\bibinfo {author} {\bibfnamefont {L.}~\bibnamefont
  {D'Onofrio}}, \bibinfo {author} {\bibfnamefont {R.}~\bibnamefont {De~Rosa}},
  \bibinfo {author} {\bibfnamefont {L.}~\bibnamefont {Errico}}, \bibinfo
  {author} {\bibfnamefont {C.}~\bibnamefont {Palomba}}, \bibinfo {author}
  {\bibfnamefont {V.}~\bibnamefont {Sequino}},\ and\ \bibinfo {author}
  {\bibfnamefont {L.}~\bibnamefont {Trozzo}},\ }\bibfield  {title} {\bibinfo
  {title} {{5n-vector ensemble method for detecting gravitational waves from
  known pulsars}},\ }\href {https://doi.org/10.1103/PhysRevD.105.063012}
  {\bibfield  {journal} {\bibinfo  {journal} {Phys. Rev. D}\ }\textbf {\bibinfo
  {volume} {105}},\ \bibinfo {pages} {063012} (\bibinfo {year}
  {2022})}\BibitemShut {NoStop}%
\bibitem [{\citenamefont {Piccinni}\ \emph {et~al.}(2018)\citenamefont
  {Piccinni}, \citenamefont {Astone}, \citenamefont {D’Antonio},
  \citenamefont {Frasca}, \citenamefont {Intini}, \citenamefont {Leaci},
  \citenamefont {Mastrogiovanni}, \citenamefont {Miller}, \citenamefont
  {Palomba},\ and\ \citenamefont {Singhal}}]{2019}%
  \BibitemOpen
  \bibfield  {author} {\bibinfo {author} {\bibfnamefont {O.~J.}\ \bibnamefont
  {Piccinni}}, \bibinfo {author} {\bibfnamefont {P.}~\bibnamefont {Astone}},
  \bibinfo {author} {\bibfnamefont {S.}~\bibnamefont {D’Antonio}}, \bibinfo
  {author} {\bibfnamefont {S.}~\bibnamefont {Frasca}}, \bibinfo {author}
  {\bibfnamefont {G.}~\bibnamefont {Intini}}, \bibinfo {author} {\bibfnamefont
  {P.}~\bibnamefont {Leaci}}, \bibinfo {author} {\bibfnamefont
  {S.}~\bibnamefont {Mastrogiovanni}}, \bibinfo {author} {\bibfnamefont
  {A.~L.}\ \bibnamefont {Miller}}, \bibinfo {author} {\bibfnamefont
  {C.}~\bibnamefont {Palomba}},\ and\ \bibinfo {author} {\bibfnamefont
  {A.}~\bibnamefont {Singhal}},\ }\bibfield  {title} {\bibinfo {title} {A new
  data analysis framework for the search of continuous gravitational wave
  signals},\ }\href@noop {} {\bibfield  {journal} {\bibinfo  {journal}
  {Classical and Quantum Gravity}\ } (\bibinfo {year} {2018})}\BibitemShut
  {NoStop}%
\bibitem [{Note1()}]{Note1}%
  \BibitemOpen
  \bibinfo {note} {\protect \textit {Observation time} refers to the amount of
  data; that is, the time span not considering the periods where the detector
  was in NO-science mode}\BibitemShut {NoStop}%
\bibitem [{\citenamefont {{The LIGO and Virgo Collaboration}}(2019)}]{targO2}%
  \BibitemOpen
  \bibfield  {author} {\bibinfo {author} {\bibnamefont {{The LIGO and Virgo
  Collaboration}}},\ }\bibfield  {title} {\bibinfo {title} {Searches for
  gravitational waves from known pulsars at two harmonics in 2015-2017 {LIGO}
  data},\ }\href {https://doi.org/10.3847/1538-4357/ab20cb} {\bibfield
  {journal} {\bibinfo  {journal} {The Astrophysical Journal}\ }\textbf
  {\bibinfo {volume} {879}},\ \bibinfo {pages} {10} (\bibinfo {year}
  {2019})}\BibitemShut {NoStop}%
\bibitem [{\citenamefont {Thrane}\ and\ \citenamefont
  {Talbot}(2019)}]{hierarc}%
  \BibitemOpen
  \bibfield  {author} {\bibinfo {author} {\bibfnamefont {E.}~\bibnamefont
  {Thrane}}\ and\ \bibinfo {author} {\bibfnamefont {C.}~\bibnamefont
  {Talbot}},\ }\bibfield  {title} {\bibinfo {title} {An introduction to
  bayesian inference in gravitational-wave astronomy: Parameter estimation,
  model selection, and hierarchical models},\ }\href
  {https://doi.org/10.1017/pasa.2019.2} {\bibfield  {journal} {\bibinfo
  {journal} {Publications of the Astronomical Society of Australia}\ }\textbf
  {\bibinfo {volume} {36}},\ \bibinfo {pages} {e010} (\bibinfo {year}
  {2019})}\BibitemShut {NoStop}%
\bibitem [{gwd()}]{gwdata}%
  \BibitemOpen
  \href@noop {} {\bibinfo {title} {{Gravitational Wave Open Science Center}}},\
  \bibinfo {howpublished} {\url{https://www.gw-openscience.
  org/O3/index/}}\BibitemShut {NoStop}%
\bibitem [{Note2()}]{Note2}%
  \BibitemOpen
  \bibinfo {note} {The ephemerides and the rotational/orbital parameters may be
  obtainable on request at the discretion of the individual EM pulsar groups
  from the above-mentioned observatories.}\BibitemShut {Stop}%
\bibitem [{\citenamefont {{Amiri}}\ \emph {et~al.}(2021)\citenamefont
  {{Amiri}}, \citenamefont {{Bandura}}, \citenamefont {{Boyle}}, \citenamefont
  {{Brar}}, \citenamefont {{Cliche}}, \citenamefont {{Crowter}}, \citenamefont
  {{Cubranic}}, \citenamefont {{Demorest}}, \citenamefont {{Denman}},
  \citenamefont {{Dobbs}}, \citenamefont {{Dong}}, \citenamefont {{Fandino}},
  \citenamefont {{Fonseca}}, \citenamefont {{Good}}, \citenamefont {{Halpern}},
  \citenamefont {{Hill}}, \citenamefont {{H{\"o}fer}}, \citenamefont {{Kaspi}},
  \citenamefont {{Landecker}}, \citenamefont {{Leung}}, \citenamefont {{Lin}},
  \citenamefont {{Luo}}, \citenamefont {{Masui}}, \citenamefont {{McKee}},
  \citenamefont {{Mena-Parra}}, \citenamefont {{Meyers}}, \citenamefont
  {{Michilli}}, \citenamefont {{Naidu}}, \citenamefont {{Newburgh}},
  \citenamefont {{Ng}}, \citenamefont {{Patel}}, \citenamefont
  {{Pinsonneault-Marotte}}, \citenamefont {{Ransom}}, \citenamefont {{Renard}},
  \citenamefont {{Scholz}}, \citenamefont {{Shaw}}, \citenamefont {{Sikora}},
  \citenamefont {{Stairs}}, \citenamefont {{Tan}}, \citenamefont {{Tendulkar}},
  \citenamefont {{Tretyakov}}, \citenamefont {{Vanderlinde}}, \citenamefont
  {{Wang}},\ and\ \citenamefont {{Wang}}}]{2021ApJS..255....5A}%
  \BibitemOpen
  \bibfield  {author} {\bibinfo {author} {\bibfnamefont {M.}~\bibnamefont
  {{Amiri}}}, \bibinfo {author} {\bibfnamefont {K.~M.}\ \bibnamefont
  {{Bandura}}}, \bibinfo {author} {\bibfnamefont {P.~J.}\ \bibnamefont
  {{Boyle}}}, \bibinfo {author} {\bibfnamefont {C.}~\bibnamefont {{Brar}}},
  \bibinfo {author} {\bibfnamefont {J.~F.}\ \bibnamefont {{Cliche}}}, \bibinfo
  {author} {\bibfnamefont {K.}~\bibnamefont {{Crowter}}}, \bibinfo {author}
  {\bibfnamefont {D.}~\bibnamefont {{Cubranic}}}, \bibinfo {author}
  {\bibfnamefont {P.~B.}\ \bibnamefont {{Demorest}}}, \bibinfo {author}
  {\bibfnamefont {N.~T.}\ \bibnamefont {{Denman}}}, \bibinfo {author}
  {\bibfnamefont {M.}~\bibnamefont {{Dobbs}}}, \bibinfo {author} {\bibfnamefont
  {F.~Q.}\ \bibnamefont {{Dong}}}, \bibinfo {author} {\bibfnamefont
  {M.}~\bibnamefont {{Fandino}}}, \bibinfo {author} {\bibfnamefont
  {E.}~\bibnamefont {{Fonseca}}}, \bibinfo {author} {\bibfnamefont {D.~C.}\
  \bibnamefont {{Good}}}, \bibinfo {author} {\bibfnamefont {M.}~\bibnamefont
  {{Halpern}}}, \bibinfo {author} {\bibfnamefont {A.~S.}\ \bibnamefont
  {{Hill}}}, \bibinfo {author} {\bibfnamefont {C.}~\bibnamefont {{H{\"o}fer}}},
  \bibinfo {author} {\bibfnamefont {V.~M.}\ \bibnamefont {{Kaspi}}}, \bibinfo
  {author} {\bibfnamefont {T.~L.}\ \bibnamefont {{Landecker}}}, \bibinfo
  {author} {\bibfnamefont {C.}~\bibnamefont {{Leung}}}, \bibinfo {author}
  {\bibfnamefont {H.~H.}\ \bibnamefont {{Lin}}}, \bibinfo {author}
  {\bibfnamefont {J.}~\bibnamefont {{Luo}}}, \bibinfo {author} {\bibfnamefont
  {K.~W.}\ \bibnamefont {{Masui}}}, \bibinfo {author} {\bibfnamefont {J.~W.}\
  \bibnamefont {{McKee}}}, \bibinfo {author} {\bibfnamefont {J.}~\bibnamefont
  {{Mena-Parra}}}, \bibinfo {author} {\bibfnamefont {B.~W.}\ \bibnamefont
  {{Meyers}}}, \bibinfo {author} {\bibfnamefont {D.}~\bibnamefont
  {{Michilli}}}, \bibinfo {author} {\bibfnamefont {A.}~\bibnamefont {{Naidu}}},
  \bibinfo {author} {\bibfnamefont {L.}~\bibnamefont {{Newburgh}}}, \bibinfo
  {author} {\bibfnamefont {C.}~\bibnamefont {{Ng}}}, \bibinfo {author}
  {\bibfnamefont {C.}~\bibnamefont {{Patel}}}, \bibinfo {author} {\bibfnamefont
  {T.}~\bibnamefont {{Pinsonneault-Marotte}}}, \bibinfo {author} {\bibfnamefont
  {S.~M.}\ \bibnamefont {{Ransom}}}, \bibinfo {author} {\bibfnamefont
  {A.}~\bibnamefont {{Renard}}}, \bibinfo {author} {\bibfnamefont
  {P.}~\bibnamefont {{Scholz}}}, \bibinfo {author} {\bibfnamefont {J.~R.}\
  \bibnamefont {{Shaw}}}, \bibinfo {author} {\bibfnamefont {A.~E.}\
  \bibnamefont {{Sikora}}}, \bibinfo {author} {\bibfnamefont {I.~H.}\
  \bibnamefont {{Stairs}}}, \bibinfo {author} {\bibfnamefont {C.~M.}\
  \bibnamefont {{Tan}}}, \bibinfo {author} {\bibfnamefont {S.~P.}\ \bibnamefont
  {{Tendulkar}}}, \bibinfo {author} {\bibfnamefont {I.}~\bibnamefont
  {{Tretyakov}}}, \bibinfo {author} {\bibfnamefont {K.}~\bibnamefont
  {{Vanderlinde}}}, \bibinfo {author} {\bibfnamefont {H.}~\bibnamefont
  {{Wang}}},\ and\ \bibinfo {author} {\bibfnamefont {X.}~\bibnamefont
  {{Wang}}},\ }\bibfield  {title} {\bibinfo {title} {{The CHIME Pulsar Project:
  System Overview}},\ }\href {https://doi.org/10.3847/1538-4365/abfdcb}
  {\bibfield  {journal} {\bibinfo  {journal} {Astrophysical Journal,
  Supplement}\ }\textbf {\bibinfo {volume} {255}},\ \bibinfo {eid} {5}
  (\bibinfo {year} {2021})},\ \Eprint {https://arxiv.org/abs/2008.05681}
  {arXiv:2008.05681 [astro-ph.IM]} \BibitemShut {NoStop}%
\bibitem [{\citenamefont {{Bailes}}\ \emph {et~al.}(2020)\citenamefont
  {{Bailes}}, \citenamefont {{Jameson}}, \citenamefont {{Abbate}},
  \citenamefont {{Barr}}, \citenamefont {{Bhat}}, \citenamefont {{Bondonneau}},
  \citenamefont {{Burgay}}, \citenamefont {{Buchner}}, \citenamefont
  {{Camilo}}, \citenamefont {{Champion}}, \citenamefont {{Cognard}},
  \citenamefont {{Demorest}}, \citenamefont {{Freire}}, \citenamefont
  {{Gautam}}, \citenamefont {{Geyer}}, \citenamefont {{Griessmeier}},
  \citenamefont {{Guillemot}}, \citenamefont {{Hu}}, \citenamefont
  {{Jankowski}}, \citenamefont {{Johnston}}, \citenamefont {{Karastergiou}},
  \citenamefont {{Karuppusamy}}, \citenamefont {{Kaur}}, \citenamefont
  {{Keith}}, \citenamefont {{Kramer}}, \citenamefont {{van Leeuwen}},
  \citenamefont {{Lower}}, \citenamefont {{Maan}}, \citenamefont
  {{McLaughlin}}, \citenamefont {{Meyers}}, \citenamefont {{Os{\l}owski}},
  \citenamefont {{Oswald}}, \citenamefont {{Parthasarathy}}, \citenamefont
  {{Pennucci}}, \citenamefont {{Posselt}}, \citenamefont {{Possenti}},
  \citenamefont {{Ransom}}, \citenamefont {{Reardon}}, \citenamefont
  {{Ridolfi}}, \citenamefont {{Schollar}}, \citenamefont {{Serylak}},
  \citenamefont {{Shaifullah}}, \citenamefont {{Shamohammadi}}, \citenamefont
  {{Shannon}}, \citenamefont {{Sobey}}, \citenamefont {{Song}}, \citenamefont
  {{Spiewak}}, \citenamefont {{Stairs}}, \citenamefont {{Stappers}},
  \citenamefont {{van Straten}}, \citenamefont {{Szary}}, \citenamefont
  {{Theureau}}, \citenamefont {{Venkatraman Krishnan}}, \citenamefont
  {{Weltevrede}}, \citenamefont {{Wex}}, \citenamefont {{Abbott}},
  \citenamefont {{Adams}}, \citenamefont {{Burger}}, \citenamefont
  {{Gamatham}}, \citenamefont {{Gouws}}, \citenamefont {{Horn}}, \citenamefont
  {{Hugo}}, \citenamefont {{Joubert}}, \citenamefont {{Manley}}, \citenamefont
  {{McAlpine}}, \citenamefont {{Passmoor}}, \citenamefont {{Peens-Hough}},
  \citenamefont {{Ramudzuli}}, \citenamefont {{Rust}}, \citenamefont {{Salie}},
  \citenamefont {{Schwardt}}, \citenamefont {{Siebrits}}, \citenamefont {{Van
  Tonder}}, \citenamefont {{Van Tonder}},\ and\ \citenamefont
  {{Welz}}}]{MeerTime}%
  \BibitemOpen
  \bibfield  {author} {\bibinfo {author} {\bibfnamefont {M.}~\bibnamefont
  {{Bailes}}}, \bibinfo {author} {\bibfnamefont {A.}~\bibnamefont {{Jameson}}},
  \bibinfo {author} {\bibfnamefont {F.}~\bibnamefont {{Abbate}}}, \bibinfo
  {author} {\bibfnamefont {E.~D.}\ \bibnamefont {{Barr}}}, \bibinfo {author}
  {\bibfnamefont {N.~D.~R.}\ \bibnamefont {{Bhat}}}, \bibinfo {author}
  {\bibfnamefont {L.}~\bibnamefont {{Bondonneau}}}, \bibinfo {author}
  {\bibfnamefont {M.}~\bibnamefont {{Burgay}}}, \bibinfo {author}
  {\bibfnamefont {S.~J.}\ \bibnamefont {{Buchner}}}, \bibinfo {author}
  {\bibfnamefont {F.}~\bibnamefont {{Camilo}}}, \bibinfo {author}
  {\bibfnamefont {D.~J.}\ \bibnamefont {{Champion}}}, \bibinfo {author}
  {\bibfnamefont {I.}~\bibnamefont {{Cognard}}}, \bibinfo {author}
  {\bibfnamefont {P.~B.}\ \bibnamefont {{Demorest}}}, \bibinfo {author}
  {\bibfnamefont {P.~C.~C.}\ \bibnamefont {{Freire}}}, \bibinfo {author}
  {\bibfnamefont {T.}~\bibnamefont {{Gautam}}}, \bibinfo {author}
  {\bibfnamefont {M.}~\bibnamefont {{Geyer}}}, \bibinfo {author} {\bibfnamefont
  {J.~M.}\ \bibnamefont {{Griessmeier}}}, \bibinfo {author} {\bibfnamefont
  {L.}~\bibnamefont {{Guillemot}}}, \bibinfo {author} {\bibfnamefont
  {H.}~\bibnamefont {{Hu}}}, \bibinfo {author} {\bibfnamefont {F.}~\bibnamefont
  {{Jankowski}}}, \bibinfo {author} {\bibfnamefont {S.}~\bibnamefont
  {{Johnston}}}, \bibinfo {author} {\bibfnamefont {A.}~\bibnamefont
  {{Karastergiou}}}, \bibinfo {author} {\bibfnamefont {R.}~\bibnamefont
  {{Karuppusamy}}}, \bibinfo {author} {\bibfnamefont {D.}~\bibnamefont
  {{Kaur}}}, \bibinfo {author} {\bibfnamefont {M.~J.}\ \bibnamefont {{Keith}}},
  \bibinfo {author} {\bibfnamefont {M.}~\bibnamefont {{Kramer}}}, \bibinfo
  {author} {\bibfnamefont {J.}~\bibnamefont {{van Leeuwen}}}, \bibinfo {author}
  {\bibfnamefont {M.~E.}\ \bibnamefont {{Lower}}}, \bibinfo {author}
  {\bibfnamefont {Y.}~\bibnamefont {{Maan}}}, \bibinfo {author} {\bibfnamefont
  {M.~A.}\ \bibnamefont {{McLaughlin}}}, \bibinfo {author} {\bibfnamefont
  {B.~W.}\ \bibnamefont {{Meyers}}}, \bibinfo {author} {\bibfnamefont
  {S.}~\bibnamefont {{Os{\l}owski}}}, \bibinfo {author} {\bibfnamefont {L.~S.}\
  \bibnamefont {{Oswald}}}, \bibinfo {author} {\bibfnamefont {A.}~\bibnamefont
  {{Parthasarathy}}}, \bibinfo {author} {\bibfnamefont {T.}~\bibnamefont
  {{Pennucci}}}, \bibinfo {author} {\bibfnamefont {B.}~\bibnamefont
  {{Posselt}}}, \bibinfo {author} {\bibfnamefont {A.}~\bibnamefont
  {{Possenti}}}, \bibinfo {author} {\bibfnamefont {S.~M.}\ \bibnamefont
  {{Ransom}}}, \bibinfo {author} {\bibfnamefont {D.~J.}\ \bibnamefont
  {{Reardon}}}, \bibinfo {author} {\bibfnamefont {A.}~\bibnamefont
  {{Ridolfi}}}, \bibinfo {author} {\bibfnamefont {C.~T.~G.}\ \bibnamefont
  {{Schollar}}}, \bibinfo {author} {\bibfnamefont {M.}~\bibnamefont
  {{Serylak}}}, \bibinfo {author} {\bibfnamefont {G.}~\bibnamefont
  {{Shaifullah}}}, \bibinfo {author} {\bibfnamefont {M.}~\bibnamefont
  {{Shamohammadi}}}, \bibinfo {author} {\bibfnamefont {R.~M.}\ \bibnamefont
  {{Shannon}}}, \bibinfo {author} {\bibfnamefont {C.}~\bibnamefont {{Sobey}}},
  \bibinfo {author} {\bibfnamefont {X.}~\bibnamefont {{Song}}}, \bibinfo
  {author} {\bibfnamefont {R.}~\bibnamefont {{Spiewak}}}, \bibinfo {author}
  {\bibfnamefont {I.~H.}\ \bibnamefont {{Stairs}}}, \bibinfo {author}
  {\bibfnamefont {B.~W.}\ \bibnamefont {{Stappers}}}, \bibinfo {author}
  {\bibfnamefont {W.}~\bibnamefont {{van Straten}}}, \bibinfo {author}
  {\bibfnamefont {A.}~\bibnamefont {{Szary}}}, \bibinfo {author} {\bibfnamefont
  {G.}~\bibnamefont {{Theureau}}}, \bibinfo {author} {\bibfnamefont
  {V.}~\bibnamefont {{Venkatraman Krishnan}}}, \bibinfo {author} {\bibfnamefont
  {P.}~\bibnamefont {{Weltevrede}}}, \bibinfo {author} {\bibfnamefont
  {N.}~\bibnamefont {{Wex}}}, \bibinfo {author} {\bibfnamefont {T.~D.}\
  \bibnamefont {{Abbott}}}, \bibinfo {author} {\bibfnamefont {G.~B.}\
  \bibnamefont {{Adams}}}, \bibinfo {author} {\bibfnamefont {J.~P.}\
  \bibnamefont {{Burger}}}, \bibinfo {author} {\bibfnamefont {R.~R.~G.}\
  \bibnamefont {{Gamatham}}}, \bibinfo {author} {\bibfnamefont
  {M.}~\bibnamefont {{Gouws}}}, \bibinfo {author} {\bibfnamefont {D.~M.}\
  \bibnamefont {{Horn}}}, \bibinfo {author} {\bibfnamefont {B.}~\bibnamefont
  {{Hugo}}}, \bibinfo {author} {\bibfnamefont {A.~F.}\ \bibnamefont
  {{Joubert}}}, \bibinfo {author} {\bibfnamefont {J.~R.}\ \bibnamefont
  {{Manley}}}, \bibinfo {author} {\bibfnamefont {K.}~\bibnamefont
  {{McAlpine}}}, \bibinfo {author} {\bibfnamefont {S.~S.}\ \bibnamefont
  {{Passmoor}}}, \bibinfo {author} {\bibfnamefont {A.}~\bibnamefont
  {{Peens-Hough}}}, \bibinfo {author} {\bibfnamefont {Z.~R.}\ \bibnamefont
  {{Ramudzuli}}}, \bibinfo {author} {\bibfnamefont {A.}~\bibnamefont {{Rust}}},
  \bibinfo {author} {\bibfnamefont {S.}~\bibnamefont {{Salie}}}, \bibinfo
  {author} {\bibfnamefont {L.~C.}\ \bibnamefont {{Schwardt}}}, \bibinfo
  {author} {\bibfnamefont {R.}~\bibnamefont {{Siebrits}}}, \bibinfo {author}
  {\bibfnamefont {G.}~\bibnamefont {{Van Tonder}}}, \bibinfo {author}
  {\bibfnamefont {V.}~\bibnamefont {{Van Tonder}}},\ and\ \bibinfo {author}
  {\bibfnamefont {M.~G.}\ \bibnamefont {{Welz}}},\ }\bibfield  {title}
  {\bibinfo {title} {{The MeerKAT telescope as a pulsar facility: System
  verification and early science results from MeerTime}},\ }\href
  {https://doi.org/10.1017/pasa.2020.19} {\bibfield  {journal} {\bibinfo
  {journal} {Publications of the Astron. Soc. of Australia}\ }\textbf {\bibinfo
  {volume} {37}},\ \bibinfo {eid} {e028} (\bibinfo {year} {2020})},\ \Eprint
  {https://arxiv.org/abs/2005.14366} {arXiv:2005.14366 [astro-ph.IM]}
  \BibitemShut {NoStop}%
\bibitem [{\citenamefont {{Jankowski}}\ \emph {et~al.}(2019)\citenamefont
  {{Jankowski}}, \citenamefont {{Bailes}}, \citenamefont {{van Straten}},
  \citenamefont {{Keane}}, \citenamefont {{Flynn}}, \citenamefont {{Barr}},
  \citenamefont {{Bateman}}, \citenamefont {{Bhandari}}, \citenamefont
  {{Caleb}}, \citenamefont {{Campbell-Wilson}}, \citenamefont {{Farah}},
  \citenamefont {{Green}}, \citenamefont {{Hunstead}}, \citenamefont
  {{Jameson}}, \citenamefont {{Os{\l}owski}}, \citenamefont {{Parthasarathy}},
  \citenamefont {{Rosado}},\ and\ \citenamefont {{Venkatraman
  Krishnan}}}]{2019MNRAS.484.3691J}%
  \BibitemOpen
  \bibfield  {author} {\bibinfo {author} {\bibfnamefont {F.}~\bibnamefont
  {{Jankowski}}}, \bibinfo {author} {\bibfnamefont {M.}~\bibnamefont
  {{Bailes}}}, \bibinfo {author} {\bibfnamefont {W.}~\bibnamefont {{van
  Straten}}}, \bibinfo {author} {\bibfnamefont {E.~F.}\ \bibnamefont
  {{Keane}}}, \bibinfo {author} {\bibfnamefont {C.}~\bibnamefont {{Flynn}}},
  \bibinfo {author} {\bibfnamefont {E.~D.}\ \bibnamefont {{Barr}}}, \bibinfo
  {author} {\bibfnamefont {T.}~\bibnamefont {{Bateman}}}, \bibinfo {author}
  {\bibfnamefont {S.}~\bibnamefont {{Bhandari}}}, \bibinfo {author}
  {\bibfnamefont {M.}~\bibnamefont {{Caleb}}}, \bibinfo {author} {\bibfnamefont
  {D.}~\bibnamefont {{Campbell-Wilson}}}, \bibinfo {author} {\bibfnamefont
  {W.}~\bibnamefont {{Farah}}}, \bibinfo {author} {\bibfnamefont {A.~J.}\
  \bibnamefont {{Green}}}, \bibinfo {author} {\bibfnamefont {R.~W.}\
  \bibnamefont {{Hunstead}}}, \bibinfo {author} {\bibfnamefont
  {A.}~\bibnamefont {{Jameson}}}, \bibinfo {author} {\bibfnamefont
  {S.}~\bibnamefont {{Os{\l}owski}}}, \bibinfo {author} {\bibfnamefont
  {A.}~\bibnamefont {{Parthasarathy}}}, \bibinfo {author} {\bibfnamefont
  {P.~A.}\ \bibnamefont {{Rosado}}},\ and\ \bibinfo {author} {\bibfnamefont
  {V.}~\bibnamefont {{Venkatraman Krishnan}}},\ }\bibfield  {title} {\bibinfo
  {title} {{The UTMOST pulsar timing programme I: Overview and first
  results}},\ }\href {https://doi.org/10.1093/mnras/sty3390} {\bibfield
  {journal} {\bibinfo  {journal} {Monthly Notices of the RAS}\ }\textbf
  {\bibinfo {volume} {484}},\ \bibinfo {pages} {3691} (\bibinfo {year}
  {2019})},\ \Eprint {https://arxiv.org/abs/1812.04038} {arXiv:1812.04038
  [astro-ph.HE]} \BibitemShut {NoStop}%
\bibitem [{\citenamefont {{Lower}}\ \emph {et~al.}(2020)\citenamefont
  {{Lower}}, \citenamefont {{Bailes}}, \citenamefont {{Shannon}}, \citenamefont
  {{Johnston}}, \citenamefont {{Flynn}}, \citenamefont {{Os{\l}owski}},
  \citenamefont {{Gupta}}, \citenamefont {{Farah}}, \citenamefont {{Bateman}},
  \citenamefont {{Green}}, \citenamefont {{Hunstead}}, \citenamefont
  {{Jameson}}, \citenamefont {{Jankowski}}, \citenamefont {{Parthasarathy}},
  \citenamefont {{Price}}, \citenamefont {{Sutherland}}, \citenamefont
  {{Temby}},\ and\ \citenamefont {{Venkatraman
  Krishnan}}}]{2020MNRAS.494..228L}%
  \BibitemOpen
  \bibfield  {author} {\bibinfo {author} {\bibfnamefont {M.~E.}\ \bibnamefont
  {{Lower}}}, \bibinfo {author} {\bibfnamefont {M.}~\bibnamefont {{Bailes}}},
  \bibinfo {author} {\bibfnamefont {R.~M.}\ \bibnamefont {{Shannon}}}, \bibinfo
  {author} {\bibfnamefont {S.}~\bibnamefont {{Johnston}}}, \bibinfo {author}
  {\bibfnamefont {C.}~\bibnamefont {{Flynn}}}, \bibinfo {author} {\bibfnamefont
  {S.}~\bibnamefont {{Os{\l}owski}}}, \bibinfo {author} {\bibfnamefont
  {V.}~\bibnamefont {{Gupta}}}, \bibinfo {author} {\bibfnamefont
  {W.}~\bibnamefont {{Farah}}}, \bibinfo {author} {\bibfnamefont
  {T.}~\bibnamefont {{Bateman}}}, \bibinfo {author} {\bibfnamefont {A.~J.}\
  \bibnamefont {{Green}}}, \bibinfo {author} {\bibfnamefont {R.}~\bibnamefont
  {{Hunstead}}}, \bibinfo {author} {\bibfnamefont {A.}~\bibnamefont
  {{Jameson}}}, \bibinfo {author} {\bibfnamefont {F.}~\bibnamefont
  {{Jankowski}}}, \bibinfo {author} {\bibfnamefont {A.}~\bibnamefont
  {{Parthasarathy}}}, \bibinfo {author} {\bibfnamefont {D.~C.}\ \bibnamefont
  {{Price}}}, \bibinfo {author} {\bibfnamefont {A.}~\bibnamefont
  {{Sutherland}}}, \bibinfo {author} {\bibfnamefont {D.}~\bibnamefont
  {{Temby}}},\ and\ \bibinfo {author} {\bibfnamefont {V.}~\bibnamefont
  {{Venkatraman Krishnan}}},\ }\bibfield  {title} {\bibinfo {title} {{The
  UTMOST pulsar timing programme - II. Timing noise across the pulsar
  population}},\ }\href {https://doi.org/10.1093/mnras/staa615} {\bibfield
  {journal} {\bibinfo  {journal} {Monthly Notices of the RAS}\ }\textbf
  {\bibinfo {volume} {494}},\ \bibinfo {pages} {228} (\bibinfo {year}
  {2020})},\ \Eprint {https://arxiv.org/abs/2002.12481} {arXiv:2002.12481
  [astro-ph.HE]} \BibitemShut {NoStop}%
\bibitem [{\citenamefont {{Nice}}\ \emph {et~al.}(2015)\citenamefont {{Nice}},
  \citenamefont {{Demorest}}, \citenamefont {{Stairs}}, \citenamefont
  {{Manchester}}, \citenamefont {{Taylor}}, \citenamefont {{Peters}},
  \citenamefont {{Weisberg}}, \citenamefont {{Irwin}}, \citenamefont {{Wex}},\
  and\ \citenamefont {{Huang}}}]{2015ascl.soft09002N}%
  \BibitemOpen
  \bibfield  {author} {\bibinfo {author} {\bibfnamefont {D.}~\bibnamefont
  {{Nice}}}, \bibinfo {author} {\bibfnamefont {P.}~\bibnamefont {{Demorest}}},
  \bibinfo {author} {\bibfnamefont {I.}~\bibnamefont {{Stairs}}}, \bibinfo
  {author} {\bibfnamefont {R.}~\bibnamefont {{Manchester}}}, \bibinfo {author}
  {\bibfnamefont {J.}~\bibnamefont {{Taylor}}}, \bibinfo {author}
  {\bibfnamefont {W.}~\bibnamefont {{Peters}}}, \bibinfo {author}
  {\bibfnamefont {J.}~\bibnamefont {{Weisberg}}}, \bibinfo {author}
  {\bibfnamefont {A.}~\bibnamefont {{Irwin}}}, \bibinfo {author} {\bibfnamefont
  {N.}~\bibnamefont {{Wex}}},\ and\ \bibinfo {author} {\bibfnamefont
  {Y.}~\bibnamefont {{Huang}}},\ }\href@noop {} {\bibinfo {title} {{Tempo:
  Pulsar timing data analysis}}} (\bibinfo {year} {2015}),\ \Eprint
  {https://arxiv.org/abs/1509.002} {ascl:1509.002} \BibitemShut {NoStop}%
\bibitem [{\citenamefont {{Edwards}}\ \emph {et~al.}(2006)\citenamefont
  {{Edwards}}, \citenamefont {{Hobbs}},\ and\ \citenamefont
  {{Manchester}}}]{tempo22}%
  \BibitemOpen
  \bibfield  {author} {\bibinfo {author} {\bibfnamefont {R.~T.}\ \bibnamefont
  {{Edwards}}}, \bibinfo {author} {\bibfnamefont {G.~B.}\ \bibnamefont
  {{Hobbs}}},\ and\ \bibinfo {author} {\bibfnamefont {R.~N.}\ \bibnamefont
  {{Manchester}}},\ }\bibfield  {title} {\bibinfo {title} {{TEMPO2, a new
  pulsar timing package - II. The timing model and precision estimates}},\
  }\href {https://doi.org/10.1111/j.1365-2966.2006.10870.x} {\bibfield
  {journal} {\bibinfo  {journal} {Monthly Notices of the RAS}\ }\textbf
  {\bibinfo {volume} {372}},\ \bibinfo {pages} {1549} (\bibinfo {year}
  {2006})},\ \Eprint {https://arxiv.org/abs/astro-ph/0607664}
  {arXiv:astro-ph/0607664 [astro-ph]} \BibitemShut {NoStop}%
\bibitem [{\citenamefont {{Hobbs}}\ \emph {et~al.}(2006)\citenamefont
  {{Hobbs}}, \citenamefont {{Edwards}},\ and\ \citenamefont
  {{Manchester}}}]{tempo21}%
  \BibitemOpen
  \bibfield  {author} {\bibinfo {author} {\bibfnamefont {G.}~\bibnamefont
  {{Hobbs}}}, \bibinfo {author} {\bibfnamefont {R.}~\bibnamefont {{Edwards}}},\
  and\ \bibinfo {author} {\bibfnamefont {R.}~\bibnamefont {{Manchester}}},\
  }\bibfield  {title} {\bibinfo {title} {{TEMPO2: a New Pulsar Timing
  Package}},\ }\href@noop {} {\bibfield  {journal} {\bibinfo  {journal}
  {Chinese Journal of Astronomy and Astrophysics Supplement}\ }\textbf
  {\bibinfo {volume} {6}},\ \bibinfo {pages} {189} (\bibinfo {year}
  {2006})}\BibitemShut {NoStop}%
\bibitem [{\citenamefont {{Hobbs}}\ \emph {et~al.}(2009)\citenamefont
  {{Hobbs}}, \citenamefont {{Jenet}}, \citenamefont {{Lee}}, \citenamefont
  {{Verbiest}}, \citenamefont {{Yardley}}, \citenamefont {{Manchester}},
  \citenamefont {{Lommen}}, \citenamefont {{Coles}}, \citenamefont
  {{Edwards}},\ and\ \citenamefont {{Shettigara}}}]{tempo23}%
  \BibitemOpen
  \bibfield  {author} {\bibinfo {author} {\bibfnamefont {G.}~\bibnamefont
  {{Hobbs}}}, \bibinfo {author} {\bibfnamefont {F.}~\bibnamefont {{Jenet}}},
  \bibinfo {author} {\bibfnamefont {K.~J.}\ \bibnamefont {{Lee}}}, \bibinfo
  {author} {\bibfnamefont {J.~P.~W.}\ \bibnamefont {{Verbiest}}}, \bibinfo
  {author} {\bibfnamefont {D.}~\bibnamefont {{Yardley}}}, \bibinfo {author}
  {\bibfnamefont {R.}~\bibnamefont {{Manchester}}}, \bibinfo {author}
  {\bibfnamefont {A.}~\bibnamefont {{Lommen}}}, \bibinfo {author}
  {\bibfnamefont {W.}~\bibnamefont {{Coles}}}, \bibinfo {author} {\bibfnamefont
  {R.}~\bibnamefont {{Edwards}}},\ and\ \bibinfo {author} {\bibfnamefont
  {C.}~\bibnamefont {{Shettigara}}},\ }\bibfield  {title} {\bibinfo {title}
  {{TEMPO2: a new pulsar timing package - III. Gravitational wave
  simulation}},\ }\href {https://doi.org/10.1111/j.1365-2966.2009.14391.x}
  {\bibfield  {journal} {\bibinfo  {journal} {Monthly Notices of the RAS}\
  }\textbf {\bibinfo {volume} {394}},\ \bibinfo {pages} {1945} (\bibinfo {year}
  {2009})},\ \Eprint {https://arxiv.org/abs/0901.0592} {arXiv:0901.0592
  [astro-ph.SR]} \BibitemShut {NoStop}%
\bibitem [{atn()}]{atnf}%
  \BibitemOpen
  \href@noop {} {\bibinfo {title} {{ATNF Pulsar Catalogue}, catalogue version
  1.61}},\ \bibinfo {howpublished}
  {\url{https://www.atnf.csiro.au/research/pulsar/psrcat/}}\BibitemShut
  {NoStop}%
\bibitem [{\citenamefont {{Yao}}\ \emph {et~al.}(2017)\citenamefont {{Yao}},
  \citenamefont {{Manchester}},\ and\ \citenamefont {{Wang}}}]{pulsarYao}%
  \BibitemOpen
  \bibfield  {author} {\bibinfo {author} {\bibfnamefont {J.~M.}\ \bibnamefont
  {{Yao}}}, \bibinfo {author} {\bibfnamefont {R.~N.}\ \bibnamefont
  {{Manchester}}},\ and\ \bibinfo {author} {\bibfnamefont {N.}~\bibnamefont
  {{Wang}}},\ }\bibfield  {title} {\bibinfo {title} {{A New Electron-density
  Model for Estimation of Pulsar and FRB Distances}},\ }\href
  {https://doi.org/10.3847/1538-4357/835/1/29} {\bibfield  {journal} {\bibinfo
  {journal} {Astrophysical Journal}\ }\textbf {\bibinfo {volume} {835}},\
  \bibinfo {eid} {29} (\bibinfo {year} {2017})},\ \Eprint
  {https://arxiv.org/abs/1610.09448} {arXiv:1610.09448 [astro-ph.GA]}
  \BibitemShut {NoStop}%
\bibitem [{\citenamefont {Cordes}\ and\ \citenamefont {Lazio}(2002)}]{NE2001}%
  \BibitemOpen
  \bibfield  {author} {\bibinfo {author} {\bibfnamefont {J.~M.}\ \bibnamefont
  {Cordes}}\ and\ \bibinfo {author} {\bibfnamefont {T.~J.~W.}\ \bibnamefont
  {Lazio}},\ }\href {https://doi.org/10.48550/ARXIV.ASTRO-PH/0207156} {\bibinfo
  {title} {Ne2001.i. a new model for the galactic distribution of free
  electrons and its fluctuations}} (\bibinfo {year} {2002})\BibitemShut
  {NoStop}%
\bibitem [{\citenamefont {{Haskell}}\ and\ \citenamefont
  {{Melatos}}(2015)}]{glitch}%
  \BibitemOpen
  \bibfield  {author} {\bibinfo {author} {\bibfnamefont {B.}~\bibnamefont
  {{Haskell}}}\ and\ \bibinfo {author} {\bibfnamefont {A.}~\bibnamefont
  {{Melatos}}},\ }\bibfield  {title} {\bibinfo {title} {{Models of pulsar
  glitches}},\ }\href {https://doi.org/10.1142/S0218271815300086} {\bibfield
  {journal} {\bibinfo  {journal} {International Journal of Modern Physics D}\
  }\textbf {\bibinfo {volume} {24}},\ \bibinfo {eid} {1530008} (\bibinfo {year}
  {2015})},\ \Eprint {https://arxiv.org/abs/1502.07062} {arXiv:1502.07062
  [astro-ph.SR]} \BibitemShut {NoStop}%
\bibitem [{\citenamefont {Leaci}\ and\ \citenamefont {Prix}(2015)}]{deltabin}%
  \BibitemOpen
  \bibfield  {author} {\bibinfo {author} {\bibfnamefont {P.}~\bibnamefont
  {Leaci}}\ and\ \bibinfo {author} {\bibfnamefont {R.}~\bibnamefont {Prix}},\
  }\bibfield  {title} {\bibinfo {title} {Directed searches for continuous
  gravitational waves from binary systems: Parameter-space metrics and optimal
  scorpius x-1 sensitivity},\ }\href
  {https://doi.org/10.1103/PhysRevD.91.102003} {\bibfield  {journal} {\bibinfo
  {journal} {Phys. Rev. D}\ }\textbf {\bibinfo {volume} {91}},\ \bibinfo
  {pages} {102003} (\bibinfo {year} {2015})}\BibitemShut {NoStop}%
\bibitem [{\citenamefont {Condon}\ and\ \citenamefont {Ransom}(2016)}]{ppdot}%
  \BibitemOpen
  \bibfield  {author} {\bibinfo {author} {\bibfnamefont {J.}~\bibnamefont
  {Condon}}\ and\ \bibinfo {author} {\bibfnamefont {S.}~\bibnamefont
  {Ransom}},\ }\href {https://books.google.it/books?id=vWWYDwAAQBAJ} {\emph
  {\bibinfo {title} {Essential Radio Astronomy}}},\ Princeton Series in Modern
  Observational Astronomy\ (\bibinfo  {publisher} {Princeton University
  Press},\ \bibinfo {year} {2016})\BibitemShut {NoStop}%
\bibitem [{\citenamefont {De~Lillo}\ \emph {et~al.}(2022)\citenamefont
  {De~Lillo}, \citenamefont {Suresh},\ and\ \citenamefont {Miller}}]{delillo}%
  \BibitemOpen
  \bibfield  {author} {\bibinfo {author} {\bibfnamefont {F.}~\bibnamefont
  {De~Lillo}}, \bibinfo {author} {\bibfnamefont {J.}~\bibnamefont {Suresh}},\
  and\ \bibinfo {author} {\bibfnamefont {A.~L.}\ \bibnamefont {Miller}},\
  }\bibfield  {title} {\bibinfo {title} {{Stochastic gravitational-wave
  background searches and constraints on neutron-star ellipticity}},\ }\href
  {https://doi.org/10.1093/mnras/stac984} {\bibfield  {journal} {\bibinfo
  {journal} {Mon. Not. Roy. Astron. Soc.}\ }\textbf {\bibinfo {volume} {513}},\
  \bibinfo {pages} {1105} (\bibinfo {year} {2022})},\ \Eprint
  {https://arxiv.org/abs/2203.03536} {arXiv:2203.03536 [gr-qc]} \BibitemShut
  {NoStop}%
\bibitem [{\citenamefont {Agarwal}\ \emph {et~al.}(2022)\citenamefont
  {Agarwal}, \citenamefont {Suresh}, \citenamefont {Mandic}, \citenamefont
  {Matas},\ and\ \citenamefont {Regimbau}}]{deep}%
  \BibitemOpen
  \bibfield  {author} {\bibinfo {author} {\bibfnamefont {D.}~\bibnamefont
  {Agarwal}}, \bibinfo {author} {\bibfnamefont {J.}~\bibnamefont {Suresh}},
  \bibinfo {author} {\bibfnamefont {V.}~\bibnamefont {Mandic}}, \bibinfo
  {author} {\bibfnamefont {A.}~\bibnamefont {Matas}},\ and\ \bibinfo {author}
  {\bibfnamefont {T.}~\bibnamefont {Regimbau}},\ }\bibfield  {title} {\bibinfo
  {title} {Targeted search for the stochastic gravitational-wave background
  from the galactic millisecond pulsar population},\ }\href
  {https://doi.org/10.1103/PhysRevD.106.043019} {\bibfield  {journal} {\bibinfo
   {journal} {Phys. Rev. D}\ }\textbf {\bibinfo {volume} {106}},\ \bibinfo
  {pages} {043019} (\bibinfo {year} {2022})}\BibitemShut {NoStop}%
\bibitem [{\citenamefont {et~al.}(2022)}]{O3allsky}%
  \BibitemOpen
  \bibfield  {author} {\bibinfo {author} {\bibfnamefont {B.~P.~A.}\
  \bibnamefont {et~al.}},\ }\bibfield  {title} {\bibinfo {title} {All-sky
  search for continuous gravitational waves from isolated neutron stars using
  advanced ligo and advanced virgo o3 data},\ }\href
  {https://doi.org/10.1103/PhysRevD.106.102008} {\bibfield  {journal} {\bibinfo
   {journal} {Phys. Rev. D}\ }\textbf {\bibinfo {volume} {106}},\ \bibinfo
  {pages} {102008} (\bibinfo {year} {2022})}\BibitemShut {NoStop}%
\bibitem [{\citenamefont {Woan}\ \emph {et~al.}(2018)\citenamefont {Woan},
  \citenamefont {Pitkin}, \citenamefont {Haskell}, \citenamefont {Jones},\ and\
  \citenamefont {Lasky}}]{min_ell}%
  \BibitemOpen
  \bibfield  {author} {\bibinfo {author} {\bibfnamefont {G.}~\bibnamefont
  {Woan}}, \bibinfo {author} {\bibfnamefont {M.~D.}\ \bibnamefont {Pitkin}},
  \bibinfo {author} {\bibfnamefont {B.}~\bibnamefont {Haskell}}, \bibinfo
  {author} {\bibfnamefont {D.~I.}\ \bibnamefont {Jones}},\ and\ \bibinfo
  {author} {\bibfnamefont {P.~D.}\ \bibnamefont {Lasky}},\ }\bibfield  {title}
  {\bibinfo {title} {Evidence for a minimum ellipticity in millisecond
  pulsars},\ }\href {https://doi.org/10.3847/2041-8213/aad86a} {\bibfield
  {journal} {\bibinfo  {journal} {The Astrophysical Journal Letters}\ }\textbf
  {\bibinfo {volume} {863}},\ \bibinfo {pages} {L40} (\bibinfo {year}
  {2018})}\BibitemShut {NoStop}%
\bibitem [{\citenamefont {Astone}\ \emph {et~al.}(2012)\citenamefont {Astone},
  \citenamefont {Colla}, \citenamefont {D'Antonio}, \citenamefont {Frasca},\
  and\ \citenamefont {Palomba}}]{5n}%
  \BibitemOpen
  \bibfield  {author} {\bibinfo {author} {\bibfnamefont {P.}~\bibnamefont
  {Astone}}, \bibinfo {author} {\bibfnamefont {A.}~\bibnamefont {Colla}},
  \bibinfo {author} {\bibfnamefont {S.}~\bibnamefont {D'Antonio}}, \bibinfo
  {author} {\bibfnamefont {S.}~\bibnamefont {Frasca}},\ and\ \bibinfo {author}
  {\bibfnamefont {C.}~\bibnamefont {Palomba}},\ }\bibfield  {title} {\bibinfo
  {title} {Coherent search of continuous gravitational wave signals: Extension
  of the 5-vectors method to a network of detectors},\ }\bibfield  {journal}
  {\bibinfo  {journal} {Journal of Physics: Conference Series}\ }\textbf
  {\bibinfo {volume} {363}},\ \href
  {https://doi.org/10.1088/1742-6596/363/1/012038}
  {10.1088/1742-6596/363/1/012038} (\bibinfo {year} {2012})\BibitemShut
  {NoStop}%
\bibitem [{\citenamefont {Biwer}\ \emph {et~al.}(2017)\citenamefont {Biwer},
  \citenamefont {Barker}, \citenamefont {Batch}, \citenamefont {Betzwieser},
  \citenamefont {Fisher}, \citenamefont {Goetz}, \citenamefont {Kandhasamy},
  \citenamefont {Karki}, \citenamefont {Kissel}, \citenamefont {Lundgren},
  \citenamefont {Macleod}, \citenamefont {Mullavey}, \citenamefont {Riles},
  \citenamefont {Rollins}, \citenamefont {Thorne}, \citenamefont {Thrane},
  \citenamefont {Abbott}, \citenamefont {Allen}, \citenamefont {Brown},
  \citenamefont {Charlton}, \citenamefont {Crowder}, \citenamefont {Fritschel},
  \citenamefont {Kanner}, \citenamefont {Landry}, \citenamefont {Lazzaro},
  \citenamefont {Millhouse}, \citenamefont {Pitkin}, \citenamefont {Savage},
  \citenamefont {Shawhan}, \citenamefont {Shoemaker}, \citenamefont {Smith},
  \citenamefont {Sun}, \citenamefont {Veitch}, \citenamefont {Vitale},
  \citenamefont {Weinstein}, \citenamefont {Cornish}, \citenamefont {Essick},
  \citenamefont {Fays}, \citenamefont {Katsavounidis}, \citenamefont {Lange},
  \citenamefont {Littenberg}, \citenamefont {Lynch}, \citenamefont {Meyers},
  \citenamefont {Pannarale}, \citenamefont {Prix}, \citenamefont
  {O'Shaughnessy},\ and\ \citenamefont {Sigg}}]{HI}%
  \BibitemOpen
  \bibfield  {author} {\bibinfo {author} {\bibfnamefont {C.}~\bibnamefont
  {Biwer}}, \bibinfo {author} {\bibfnamefont {D.}~\bibnamefont {Barker}},
  \bibinfo {author} {\bibfnamefont {J.~C.}\ \bibnamefont {Batch}}, \bibinfo
  {author} {\bibfnamefont {J.}~\bibnamefont {Betzwieser}}, \bibinfo {author}
  {\bibfnamefont {R.~P.}\ \bibnamefont {Fisher}}, \bibinfo {author}
  {\bibfnamefont {E.}~\bibnamefont {Goetz}}, \bibinfo {author} {\bibfnamefont
  {S.}~\bibnamefont {Kandhasamy}}, \bibinfo {author} {\bibfnamefont
  {S.}~\bibnamefont {Karki}}, \bibinfo {author} {\bibfnamefont {J.~S.}\
  \bibnamefont {Kissel}}, \bibinfo {author} {\bibfnamefont {A.~P.}\
  \bibnamefont {Lundgren}}, \bibinfo {author} {\bibfnamefont {D.~M.}\
  \bibnamefont {Macleod}}, \bibinfo {author} {\bibfnamefont {A.}~\bibnamefont
  {Mullavey}}, \bibinfo {author} {\bibfnamefont {K.}~\bibnamefont {Riles}},
  \bibinfo {author} {\bibfnamefont {J.~G.}\ \bibnamefont {Rollins}}, \bibinfo
  {author} {\bibfnamefont {K.~A.}\ \bibnamefont {Thorne}}, \bibinfo {author}
  {\bibfnamefont {E.}~\bibnamefont {Thrane}}, \bibinfo {author} {\bibfnamefont
  {T.~D.}\ \bibnamefont {Abbott}}, \bibinfo {author} {\bibfnamefont
  {B.}~\bibnamefont {Allen}}, \bibinfo {author} {\bibfnamefont {D.~A.}\
  \bibnamefont {Brown}}, \bibinfo {author} {\bibfnamefont {P.}~\bibnamefont
  {Charlton}}, \bibinfo {author} {\bibfnamefont {S.~G.}\ \bibnamefont
  {Crowder}}, \bibinfo {author} {\bibfnamefont {P.}~\bibnamefont {Fritschel}},
  \bibinfo {author} {\bibfnamefont {J.~B.}\ \bibnamefont {Kanner}}, \bibinfo
  {author} {\bibfnamefont {M.}~\bibnamefont {Landry}}, \bibinfo {author}
  {\bibfnamefont {C.}~\bibnamefont {Lazzaro}}, \bibinfo {author} {\bibfnamefont
  {M.}~\bibnamefont {Millhouse}}, \bibinfo {author} {\bibfnamefont
  {M.}~\bibnamefont {Pitkin}}, \bibinfo {author} {\bibfnamefont {R.~L.}\
  \bibnamefont {Savage}}, \bibinfo {author} {\bibfnamefont {P.}~\bibnamefont
  {Shawhan}}, \bibinfo {author} {\bibfnamefont {D.~H.}\ \bibnamefont
  {Shoemaker}}, \bibinfo {author} {\bibfnamefont {J.~R.}\ \bibnamefont
  {Smith}}, \bibinfo {author} {\bibfnamefont {L.}~\bibnamefont {Sun}}, \bibinfo
  {author} {\bibfnamefont {J.}~\bibnamefont {Veitch}}, \bibinfo {author}
  {\bibfnamefont {S.}~\bibnamefont {Vitale}}, \bibinfo {author} {\bibfnamefont
  {A.~J.}\ \bibnamefont {Weinstein}}, \bibinfo {author} {\bibfnamefont
  {N.}~\bibnamefont {Cornish}}, \bibinfo {author} {\bibfnamefont {R.~C.}\
  \bibnamefont {Essick}}, \bibinfo {author} {\bibfnamefont {M.}~\bibnamefont
  {Fays}}, \bibinfo {author} {\bibfnamefont {E.}~\bibnamefont {Katsavounidis}},
  \bibinfo {author} {\bibfnamefont {J.}~\bibnamefont {Lange}}, \bibinfo
  {author} {\bibfnamefont {T.~B.}\ \bibnamefont {Littenberg}}, \bibinfo
  {author} {\bibfnamefont {R.}~\bibnamefont {Lynch}}, \bibinfo {author}
  {\bibfnamefont {P.~M.}\ \bibnamefont {Meyers}}, \bibinfo {author}
  {\bibfnamefont {F.}~\bibnamefont {Pannarale}}, \bibinfo {author}
  {\bibfnamefont {R.}~\bibnamefont {Prix}}, \bibinfo {author} {\bibfnamefont
  {R.}~\bibnamefont {O'Shaughnessy}},\ and\ \bibinfo {author} {\bibfnamefont
  {D.}~\bibnamefont {Sigg}},\ }\bibfield  {title} {\bibinfo {title} {Validating
  gravitational-wave detections: The advanced ligo hardware injection system},\
  }\href {https://doi.org/10.1103/PhysRevD.95.062002} {\bibfield  {journal}
  {\bibinfo  {journal} {Phys. Rev. D}\ }\textbf {\bibinfo {volume} {95}},\
  \bibinfo {pages} {062002} (\bibinfo {year} {2017})}\BibitemShut {NoStop}%
\bibitem [{HIl()}]{HIlist}%
  \BibitemOpen
  \href@noop {} {\bibinfo {title} {{O}3 continuous wave hardware injections}},\
  \bibinfo {howpublished} {\url{https://www.gw-openscience.org/O3/o3_inj/}},\
  \bibinfo {note} {accessed: 2022-07-20}\BibitemShut {NoStop}%
\bibitem [{Note3()}]{Note3}%
  \BibitemOpen
  \bibinfo {note} {The coherence is a number between 0 and 1 that measures the
  resemblance between the shape of the expected signal and the data.
  Differently form the p-value, the noise distribution of the coherence is not
  flat between 0 and 1 (see \cite {2010}). It is a supplemental parameter to
  show the significance of the detection.}\BibitemShut {Stop}%
\end{thebibliography}%

\end{document}